\documentclass[aps,prev,twocolumn,preprintnumbers,floatfix,nofootinbib]{revtex4-1}
\pdfoutput=1
\usepackage{graphicx}
\usepackage{bm}
\usepackage{times}
\usepackage{slashed}
\usepackage{color}
\usepackage{xcolor, hyperref}
\usepackage{aas_macros}

\usepackage{slashed}
\usepackage{lipsum}
\usepackage{subfigure}
\usepackage{multirow}
\usepackage{amsmath}
\usepackage{array} 
\usepackage{varwidth} 
\usepackage{tabularx}
\newcolumntype{L}{>{$}l<{$}} 

\newcommand{\be}{\begin{equation}}
\newcommand{\ee}{\end{equation}}
\newcommand{\bea}{\begin{eqnarray}}
\newcommand{\eea}{\end{eqnarray}}
\newcommand{\ergs}{\text{ergs}/\text{s}}

\newcommand{\commentOut}[1]{}

\DeclareMathOperator*{\argmax}{arg\,max}

\begin{document}

\title{Gamma-Rays from Star Forming Activity Appear to Outshine Misaligned Active Galactic Nuclei}
\author{Carlos Blanco}
\email{carlos.blanco@fysik.su.se, ORCID: orcid.org/0000-0001-8971-834X}
\affiliation{Stockholm University and The Oskar Klein Centre for Cosmoparticle Physics,  Alba Nova, 10691 Stockholm, Sweden}
\affiliation{Princeton University, Department of Physics, Princeton, NJ 08544}
\author{Tim Linden}
\email{linden@fysik.su.se, ORCID: orcid.org/0000-0001-9888-0971}
\affiliation{Stockholm University and The Oskar Klein Centre for Cosmoparticle Physics,  Alba Nova, 10691 Stockholm, Sweden}

\begin{abstract}
The total extragalactic $\gamma$-ray flux provides a powerful probe into the origin and evolution of the highest energy processes in our universe. An important component of this emission is the isotropic $\gamma$-ray background (IGRB), composed of sources that cannot be individually resolved by current experiments.  Previous studies have determined that the IGRB can be dominated by either misaligned active galactic nuclei (mAGN) or star-forming galaxies (SFGs). However, these analyses are limited, because they have utilized binary source classifications and examined only one source class at a time. We perform the first combined joint-likelihood analysis that simultaneously correlates the $\gamma$-ray luminosity of extragalactic objects with both star-formation and mAGN activity. We find that SFGs produce 48$^{+33}_{-20}$\% of the total IGRB at~1 GeV and 56$^{+40}_{-23}$\% of the total IGRB at 10 GeV. The contribution of mAGN is more uncertain, but can also be significant. Future work to quantify the radio and infrared properties of nearby galaxies could significantly improve these constraints.
\end{abstract}

\maketitle

\section{Introduction}

The extragalactic GeV $\gamma$-ray emission in our universe is surprisingly well-balanced. Models indicate that multiple source classes meaningfully contribute ($\gtrsim$10\%) to the total \mbox{$\gamma$-ray} luminosity of our universe. Moreover, the extragalactic $\gamma$-ray luminosity (EGRB) is relatively equally divided into emission between a handful of extremely bright point sources~\cite{Ajello:2015mfa, Lisanti:2016jub, DiMauro:2017ing, Manconi:2019ynl, 2020ApJ...896....6M}, and an underlying sea of dim sources, known collectively as the isotropic $\gamma$-ray background (IGRB)~\cite{Ackermann:2012uf, dimauro:2013xta, Ackermann:2014usa, Hooper:2016gjy, Linden:2016fdd, Blanco2017, Komis:2017jta, Stecker:2019ybn}. While the latter coincidence depends on the sensitivity of the instrument (in this case the Fermi-LAT), population models and statistical studies~\cite{Ackermann:2012uf, Zechlin:2015wdz, Fornasa:2016ohl, Zechlin:2016pme, Ando:2017alx,  Ackermann:2018wlo} indicate that the distinction is also physically meaningful, with significant and distinct contributions from very bright and very dim sources. 

While the EGRB is dominated by a handful of extremely luminous blazars (again, divided relatively equally between BL Lac objects and Flat Spectrum Radio Quasars~\cite{Ajello:2015mfa, Lisanti:2016jub, Zechlin:2016pme}, the IGRB includes only a $\sim$20\% contribution from on-axis blazar jets~\cite{Ackermann:2012uf, Fornasa:2016ohl}. Models instead indicate that the IGRB is produced either by off-axis emission from supermassive black holes (hereafter, misaligned AGN (mAGN)), or by cosmic-rays accelerated in the supernova explosions of star-forming galaxies (SFGs). The typical $\gamma$-ray luminosities of these objects has profound implications for the energetics of mAGN and SFG activity, and thus their relative importance in shaping galaxy evolution~\cite{Socrates:2006dv, Kormendy:2013dxa, 2015MNRAS.452..575S,  Pakmor:2016zbu, Wiener:2016zcr, 2017A&A...602A.123L}.

Because most SFGs and mAGN do not have \mbox{$\gamma$-ray} fluxes that are individually detectable, statistical methods are used to determine their contribution to the IGRB. However, these studies have obtained disparate results. Work by Ref.~\cite{dimauro:2013xta} found that mAGN produce $\sim$10\% of the total IGRB intensity, while a subsequent joint-likelihood analysis by Ref.~\cite{Hooper:2016gjy} determined their mean contribution to be $\sim$77\%. An analysis of the brightest SFGs by the Fermi-LAT collaboration found that they produce 4-23\% of the IGRB~\cite{2012ApJ...755..164A}, while a study of the combined emission from more than 600 SFGs found their contribution to be roughly 60\%~\cite{Linden:2016fdd}. More recent work has also provided inconsistent results~\cite{Stecker:2019ybn, 2020A&A...641A.147K, Ajello:2020zna}, and the totality of the research definitively shows only that \emph{either} mAGN or SFGs (or \emph{both}, or \emph{neither}) dominate the intensity of IGRB. 

Despite differing results, most studies employ a similar methodology. The $\gamma$-ray flux from the brightest mAGN or SFGs is compared with a multiwavelength tracer of the mAGN or SFG luminosity. The derived correlation between the $\gamma$-ray and multiwavelength luminosity is then extrapolated down to systems with smaller fluxes, where the \mbox{$\gamma$-ray} data is unconstrained. Using models for the luminosity and redshift evolution of each source class, the total $\gamma$-ray emission is then calculated. For mAGN, studies have typically correlated the $\gamma$-ray luminosity with the 5~GHz core radio luminosity, which serves as a proxy for supermassive black hole activity. For SFGs, studies have used the FIR \mbox{(8-1000 $\mu$m)} flux as a tracer of the star-formation and subsequent supernova rate.

A significant weakness of these studies is that each has independently investigated only mAGN or SFG activity, and has further assigned binary classifications to observed mAGN and SFGs, assuming that the entirety of their $\gamma$-ray flux is due to a single effect\footnote{Throughout this work, we use SFG and mAGN to correspond to the primary classification of a given galaxy, and ``SF activity" and ``mAGN activity" to relate to the physical process which produces $\gamma$-ray emission.}. However, reality is much trickier. NGC 1068, NGC 4945 and Circinus, traditionally described as SFGs, are also Seyfert galaxies. Essentially every mAGN galaxy is also undergoing star-formation activity. At best, this binary classification produces ``double-counting", as $\gamma$-ray emission can be prescribed to both mAGN and SFGs inside the correlation analysis, and then extrapolated to other systems. At worst, such misidentifications could significantly affect the calculated $\gamma$-ray correlations, producing large systematic errors which will be magnified by the subsequent extrapolations.

In this paper, we perform the first joint-likelihood analysis of $\gamma$-ray fluxes from a population of SFGs and mAGN that simultaneously accounts for emission from both supernovae (tracing the FIR intensity) and proton acceleration from a central black hole (tracing the 5~GHz core luminosity). This analysis avoids the double counting problem and accurately separates the individual contributions of SF activity and mAGN activity to the IGRB. Our study also benefits from (1) the analysis of more than twelve years of Fermi-LAT data, (2) a state-of-the art statistical methodology designed to account for systematic errors in the astrophysical background model, and (3) an accurate modeling of gamma-ray propagation to account for spectral changes due to electromagnetic cascades. 

We obtain three primary conclusions: (1) star-formation activity contributes significantly to the IGRB intensity, and this result is robust to our modeling of mAGN activity in these galaxies, (2) the $\gamma$-ray luminosity of mAGN activity has a significant source to source dispersion. In contrast to previous works we show that observed $\gamma$-ray mAGN are likely significant outliers compared to predictions from the radio-$\gamma$-ray correlation, (3) owing to this large source-to-source dispersion, the contribution of mAGN activity can still be significant, but is highly uncertain.

The paper is outlined as follows. In Section~\ref{sec:models} we describe our galaxy catalog, our modeling of Fermi-LAT data, and our statistical methodology. In Section~\ref{sec:contributions} we describe our pipeline for calculating the IGRB contribution from each source class. In Section~\ref{sec:results} we present results for the Radio/Gamma and FIR/Gamma Correlations, along with the modeled contribution of both mAGN activity and SF activity to the IGRB. In Section~\ref{sec:discussion} we discuss our results and conclude.

\section{Methodology}
\label{sec:models}

\subsection{mAGN and Star-Forming Galaxy Selection}
\label{sec:sample}
Our source sample consists of 57 galaxies that have been observed both in the infrared between 8 and 100 $\mu\text{m}$ as well as in the radio at 5GHz. We focus on sources that have 5GHz radio data from their central region, nominally arcseconds in extent. We select only those sources with a viewing angle greater than $\theta_{mAGN}\sim 40^\circ$ with relation to their jet-axis.  The sample is compiled using sources from (1) the mAGN population found in the 1997 Laurent-Muehleisen catalog of radio-loud AGN in the ROSAT All-Sky Survey~(R97, \cite{LaurentMuehleisen1996}), (2) the sample of Di Mauro, et al. (DMa, ~\cite{dimauro:2013xta}), and (3) samples of both radio galaxies and steep-spectrum radio quasars collected by Yuan and Wang~(YW, \cite{Yuan_2011,Yuan2018}). We select sources where those mAGN were also observed by the Infrared Astronomical Satellite (IRAS) between 8 and 100 $\mu\text{m}$~\cite{1988iras....7.....H}. We exclude the giant galaxy at the center of the Perseus cluster, NGC 1275, from our analysis since it is particularly variable in radio as well as in gamma-ray observations~\cite{Ahnen2016,Mukherjee2016,Aleksic2014,Nesterov1995}. Table~\ref{tab:1} compiles our joint catalog of 57 galactic sources. We present the radio catalog from which the 5~GHz core data was taken, the IAU B1950 name, the IRAS name, any common associated name, and the calculated multiwavelength luminosities. 

\begin{table*}[h]
  \centering
\begin{tabular}{L L L L L L L L}
 \text{Catalog} & \text{IAU B1950} & \text{Alt Name} & \text{IRAS name} & \text{L}_\text{ir} (erg/s) & \text{L}_\text{core} (erg/s) & \text{Log}_{10} L_{\gamma }{}^{\text{IR}}(erg/s) & \text{Log}_{10}L_{\gamma }{}^{\text{Rad}}(erg/s)  \\ \hline
 \text{YW} & \text{0045-255} & \text{NGC0253} & \text{IRAS 00450-2533~~~~} & 1.23\times 10^{44}~~ & 1.08\times 10^{37}~~ & 39.9\pm 0.3 & 38.6\pm 0.9 \\
 \text{YW} & \text{0238-084} & \text{NGC1052} & \text{IRAS 02386-0828} & 1.06\times 10^{43} & 3.07\times 10^{39} & 38.8\pm 0.4 & 40.5\pm 0.5 \\
 \text{YW} & \text{0240-002} & \text{3C 71} & \text{IRAS 02401-0013} & 1.05\times 10^{45} & 3.24\times 10^{38} & 40.9\pm 0.2 & 39.8\pm 0.6 \\
 \text{YW} & \text{0244-304} & \text{NGC1097} & \text{IRAS 02441-3029} & 1.95\times 10^{44} & 5.97\times 10^{36} & 40.2\pm 0.2 & 38.4\pm 1.0 \\
 \text{DMa} & \text{None} & \text{IC 310} & \text{IRAS 03135+4108} & 1.56\times 10^{44} & 4.30\times 10^{39} & 40.1\pm 0.3 & 40.6\pm 0.5 \\
 \text{DMa,YW} & \text{0320-373} & \text{For A} & \text{IRAS 03208-3723} & 3.15\times 10^{43} & 1.60\times 10^{38} & 39.3\pm 0.4 & 39.5\pm 0.7 \\
 \text{YW} & \text{0331-363} & \text{NGC 1365} & \text{IRAS 03317-3618} & 5.70\times 10^{44} & 3.36\times 10^{36} & 40.7\pm 0.2 & 38.2\pm 1.0 \\
 \text{DMa,R97} & \text{0433+053B} & \text{3C 120} & \text{IRAS 04305+0514} & 7.13\times 10^{44} & 2.44\times 10^{41} & 40.8\pm 0.2 & 42.0\pm 0.6 \\
 \text{YW} & \text{0505-375} & \text{NGC 1808} & \text{IRAS 05059-3734} & 2.50\times 10^{44} & 1.87\times 10^{37} & 40.3\pm 0.2 & 38.8\pm 0.9 \\
 \text{YW} & \text{0634-205} & \text{None} & \text{IRAS 06343-2032} & 1.82\times 10^{45} & 4.41\times 10^{39} & 41.2\pm 0.2 & 40.7\pm 0.5 \\
 \text{R97} & \text{0652+744} & \text{IC 450} & \text{IRAS 06457+7429} & 1.74\times 10^{44} & 3.99\times 10^{39} & 40.1\pm 0.3 & 40.6\pm 0.5 \\
 \text{YW} & \text{0648+27} & \text{None} & \text{IRAS 06488+2731} & 1.31\times 10^{45} & 1.15\times 10^{40} & 41.1\pm 0.2 & 41.0\pm 0.5 \\
 \text{R97} & \text{0654+070} & \text{None} & \text{IRAS 06518+0707} & 2.25\times 10^{44} & 1.34\times 10^{39} & 40.3\pm 0.2 & 40.2\pm 0.6 \\
 \text{YW} & \text{0722+30} & \text{None} & \text{IRAS 07224+3003} & 2.92\times 10^{44} & 2.11\times 10^{39} & 40.4\pm 0.2 & 40.4\pm 0.5 \\
 \text{R97} & \text{0804+051} & \text{Mrk 1210} & \text{IRAS 08014+0515} & 1.80\times 10^{44} & 1.09\times 10^{39} & 40.1\pm 0.2 & 40.2\pm 0.6 \\
 \text{YW} & \text{0806-103} & \text{3C 195} & \text{IRAS 08065-1018} & 4.63\times 10^{45} & 8.67\times 10^{40} & 41.7\pm 0.2 & 41.7\pm 0.6 \\
 \text{YW} & \text{0836+29B} & \text{4C29.30} & \text{IRAS 08369+2959} & 2.41\times 10^{45} & 3.24\times 10^{40} & 41.4\pm 0.2 & 41.3\pm 0.5 \\
 \text{R97} & \text{0955+696} & \text{M82} & \text{IRAS 09517+6954} & 1.43\times 10^{44} & 2.49\times 10^{36} & 40.0\pm 0.3 & 38.1\pm 1.1 \\
 \text{R97} & \text{1001+556} & \text{NGC 3079} & \text{IRAS 09585+5555} & 1.38\times 10^{44} & 1.03\times 10^{38} & 40.0\pm 0.3 & 39.4\pm 0.7 \\
 \text{R97} & \text{1023+198} & \text{None} & \text{IRAS 10207+2007} & 3.19\times 10^{43} & 4.76\times 10^{37} & 39.3\pm 0.4 & 39.1\pm 0.8 \\
 \text{R97,YW} & \text{1124+387} & \text{NGC 3665} & \text{IRAS 11220+3902} & 3.09\times 10^{43} & 1.00\times 10^{38} & 39.3\pm 0.4 & 39.4\pm 0.7 \\
 \text{YW} & \text{1217+29} & \text{NGC 4278} & \text{IRAS 12175+2933} & 1.69\times 10^{42} & 9.14\times 10^{37} & 37.9\pm 0.6 & 39.3\pm 0.7 \\
 \text{DMa,YW} & \text{1222+131} & \text{3C 272.1} & \text{IRAS 12224+1309} & 5.04\times 10^{42} & 2.15\times 10^{38} & 38.4\pm 0.5 & 39.6\pm 0.7 \\
 \text{DMa,YW} & \text{1228+126} & \text{M 87} & \text{IRAS 12282+1240} & 7.92\times 10^{42} & 6.11\times 10^{39} & 38.7\pm 0.5 & 40.8\pm 0.5 \\
 \text{R97} & \text{1250+411} & \text{M94} & \text{IRAS 12485+4123} & 1.20\times 10^{43} & 4.10\times 10^{35} & 38.9\pm 0.4 & 37.5\pm 1.2 \\
 \text{YW} & \text{1302-491} & \text{NGC4945} & \text{IRAS 13025-4911} & 2.49\times 10^{44} & 1.03\times 10^{37} & 40.3\pm 0.2 & 38.6\pm 0.9 \\
 \text{YW} & \text{1306-241} & \text{PKS 1306-241} & \text{IRAS 13059-2407} & 1.06\times 10^{44} & 4.67\times 10^{39} & 39.9\pm 0.3 & 40.7\pm 0.5 \\
 \text{R97} & \text{1310+370} & \text{None} & \text{IRAS 13086+3719} & 5.38\times 10^{43} & 1.96\times 10^{37} & 39.6\pm 0.3 & 38.8\pm 0.9 \\
 \text{YW} & \text{1318+34} & \text{IC 883} & \text{IRAS 13183+3423} & 1.80\times 10^{45} & 5.14\times 10^{39} & 41.2\pm 0.2 & 40.7\pm 0.5 \\
 \text{DMa,YW} & \text{1322-427} & \text{Centaurus A} & \text{IRAS 13225-4245} & 1.43\times 10^{44} & 2.59\times 10^{39} & 40.0\pm 0.3 & 40.5\pm 0.5 \\
 \text{YW} & \text{1334-296} & \text{M 83} & \text{IRAS 13341-2936} & 8.26\times 10^{43} & 1.34\times 10^{36} & 39.8\pm 0.3 & 37.9\pm 1.1 \\
 \text{R97} & \text{1427+359} & \text{None} & \text{IRAS 14250+3608} & 4.47\times 10^{44} & 3.780\times 10^{38} & 40.6\pm 0.2 & 39.8\pm 0.6 \\
 \text{YW} & \text{1529+242} & \text{3C 321} & \text{IRAS 15295+2414} & 3.96\times 10^{45} & 3.55\times 10^{40} & 41.6\pm 0.2 & 41.4\pm 0.5 \\
 \text{YW} & \text{1559+021} & \text{3C 327} & \text{IRAS 15599+0206} & 4.22\times 10^{45} & 5.18\times 10^{40} & 41.6\pm 0.2 & 41.5\pm 0.5 \\
 \text{R97} & \text{1629+244} & \text{Mrk 883} & \text{IRAS 16277+2433} & 6.93\times 10^{44} & 1.18\times 10^{39} & 40.8\pm 0.2 & 40.2\pm 0.6 \\
 \text{R97} & \text{1652+024} & \text{NGC 6240} & \text{IRAS 16504+0228} & 2.95\times 10^{45} & 2.86\times 10^{39} & 41.5\pm 0.2 & 40.5\pm 0.5 \\
 \text{DMa,YW} & \text{1845+797} & \text{3C 390.3} & \text{IRAS 18456+7943} & 1.15\times 10^{45} & 1.04\times 10^{41} & 41.0\pm 0.2 & 41.7\pm 0.6 \\
 \text{YW} & \text{1957+405} & \text{3C 405/ Cyg A} & \text{IRAS 19577+4035} & 3.37\times 10^{45} & 3.70\times 10^{41} & 41.5\pm 0.2 & 42.2\pm 0.7 \\
 \text{R97} & \text{2034+601} & \text{NGC 6946} & \text{IRAS 20338+5958} & 3.12\times 10^{41} & 2.25\times 10^{34} & 37.1\pm 0.7 & 36.5\pm 1.5 \\
 \text{YW} & \text{2048-572} & \text{IC 5063} & \text{IRAS 20481-5715} & 3.04\times 10^{44} & 4.17\times 10^{38} & 40.4\pm 0.2 & 39.9\pm 0.6 \\
 \text{YW} & \text{2116+26} & \text{NGC 7052} & \text{IRAS 21163+2613} & 7.97\times 10^{43} & 1.34\times 10^{39} & 39.8\pm 0.3 & 40.2\pm 0.6 \\
 \text{R97} & \text{2200+105} & \text{Mrk 520} & \text{IRAS 21582+1018} & 8.36\times 10^{44} & 3.44\times 10^{38} & 40.9\pm 0.2 & 39.8\pm 0.6 \\
 \text{YW} & \text{2254-367} & \text{IC 1459} & \text{IRAS 22544-3643} & 3.05\times 10^{43} & 4.04\times 10^{39} & 39.3\pm 0.4 & 40.6\pm 0.5 \\
 \text{R97} & \text{2303+088} & \text{NGC 7469} & \text{IRAS 23007+0836} & 1.65\times 10^{45} & 5.17\times 10^{38} & 41.2\pm 0.2 & 39.9\pm 0.6 \\
 \text{YW} & \text{0055+30} & \text{NGC 315} & \text{IRAS F00550+3004} & 5.71\times 10^{43} & 2.69\times 10^{40} & 39.6\pm 0.3 & 41.3\pm 0.5 \\
 \text{DMa,YW} & \text{0104+321} & \text{3C 31} & \text{IRAS F01046+3208} & 5.35\times 10^{43} & 3.03\times 10^{39} & 39.6\pm 0.3 & 40.5\pm 0.5 \\
 \text{YW} & \text{0307-305} & \text{None} & \text{IRAS F03079-3031} & 6.90\times 10^{44} & 1.73\times 10^{39} & 40.8\pm 0.2 & 40.3\pm 0.6 \\
 \text{YW} & \text{0326-288} & \text{None} & \text{IRAS F03265-2852} & 2.23\times 10^{45} & 6.19\times 10^{40} & 41.3\pm 0.2 & 41.6\pm 0.6 \\
 \text{YW} & \text{0523-327} & \text{None} & \text{IRAS F05236-3245} & 5.87\times 10^{44} & 1.10\times 10^{40} & 40.7\pm 0.2 & 41.0\pm 0.5 \\
 \text{YW} & \text{0913+38} & \text{None} & \text{IRAS F09137+3831} & 7.46\times 10^{44} & 6.23\times 10^{38} & 40.8\pm 0.2 & 40.0\pm 0.6 \\
 \text{YW} & \text{0958+290} & \text{3C 234} & \text{IRAS 09589+2901} & 8.26\times 10^{45} & 4.41\times 10^{41} & 42.0\pm 0.3 & 42.2\pm 0.7 \\
 \text{DMa,YW} & \text{1350+316} & \text{3C 293} & \text{IRAS F13500+3141} & 2.35\times 10^{44} & 2.43\times 10^{40} & 40.3\pm 0.2 & 41.2\pm 0.5 \\
 \text{DMa,YW} & \text{1448+634} & \text{3C 305} & \text{IRAS F14483+6328} & 2.03\times 10^{44} & 6.06\times 10^{39} & 40.2\pm 0.2 & 40.8\pm 0.5 \\
 \text{DMa,YW} & \text{1833+326} & \text{3C 382} & \text{IRAS F18332+3239} & 4.81\times 10^{44} & 7.70\times 10^{40} & 40.6\pm 0.2 & 41.6\pm 0.6 \\
 \text{YW} & \text{1949+023} & \text{3C 403} & \text{IRAS F19497+0222} & 8.60\times 10^{44} & 4.22\times 10^{39} & 40.9\pm 0.2 & 40.6\pm 0.5 \\
 \text{YW} & \text{2158-380} & \text{None} & \text{IRAS F21583-3800} & 1.64\times 10^{44} & 6.42\times 10^{38} & 40.1\pm 0.3 & 40.0\pm 0.6 \\
 \text{DMa,YW} & \text{2221-023} & \text{3C 445} & \text{IRAS F22212-0221} & 8.78\times 10^{44} & 3.31\times 10^{40} & 40.9\pm 0.2 & 41.3\pm 0.5 \\
\end{tabular}
  \caption{Our sample of selected sources along with (Left-to-Right) radio catalog refs, IAU B1950 designations, alternative/associated names, IRAS identifiers, IR (8-100$\mu$m) luminosities, 5GHz core-radio luminosities, gamma-ray luminosities predicted from our model of mAGN activity, and gamma-ray luminosities predicted from model of SF activity. The catalogs identified in the first column refers to the radio catalog which contains the 5GHz core data as discussed in Sec.~\ref{sec:sample}. Note that the presented gamma-ray luminosities are calculated using the best fit values of $a=1.09\pm0.20$, $b=0.78\pm0.26$, $d=40.81\pm0.19$, $g=40.78\pm0.52$, $\sigma_{SF}=0.20\pm0.17$, and $\sigma_{mAGN}=0.88\pm0.33$. Catalog names refer to: R97 \cite{LaurentMuehleisen1996}, DMa~\cite{dimauro:2013xta}), and YW~\cite{Yuan_2011,Yuan2018}}. 
  \label{tab:1}
\end{table*}

\subsection{Fermi-LAT Data Analysis}
Our analysis depends on the accurate analysis of the $\gamma$-ray flux from our population of SFGs and mAGN. This is made more challenging by the fact that many of these sources are not particularly bright, with expected $\gamma$-ray fluxes that would fall far below the sensitivity of standard Fermi-LAT point source searches~\cite{Ballet:2020hze}. However, we note that analyzing only sources that have bright $\gamma$-ray emission would systematically bias our study. Thus, the accurate determination of the flux and flux uncertainties for a population of sources that is unbiased by their $\gamma$-ray luminosity is paramount.

We utilize a joint-likelihood analysis to accurately include information from the likelihood functions of each source in our analysis. Such techniques have been widely adapted in searches for dim dark matter signals from dwarf spheroidal galaxies~\cite{GeringerSameth:2011iw, Ackermann:2013yva, Fermi-LAT:2016uux, Hoof:2018hyn, Linden:2019soa}, and has been adapted by our group for individual studies of both SFGs~\cite{Linden:2016fdd} and mAGN~\cite{Hooper:2016gjy}.

To calculate the $\gamma$-ray flux for each galaxy, we examine twelve years of Fermi-LAT data (4 August 2008 -- 17 August 2020), including all P8\_R3\_SOURCE class photons with recorded energies that lie between 100~MeV and 100~GeV and which are observed at a zenith angle smaller than 90$^\circ$. We apply standard cuts on the Fermi-LAT livetime.

To calculate the log-likelihood improvement provided by adding a point source at the location of the each galaxy, we produced a binned model of the Fermi-LAT data in a 15.0$^\circ \times$15.0$^\circ$ region of interest with angular bins of size 0.1$^\circ$ and 24 equal logarithmic energy bins spanning 100~MeV to 100~GeV. We utilize the standard Fermi-LAT tools to build a model that includes all 4FGL point and extended sources~\cite{Ballet:2020hze}, as well as the relevant diffuse and isotropic background models. We add a point source to this model at the position of each SFG or mAGN in our sample. In the case that the galaxy in question is also a 4FGL source, we remove the 4FGL source from our sample, and place the new source at the indicated SFG or mAGN position (based on high-resolution optical observations) in order to maintain the consistency of our dataset. This produces only arcsecond-level changes in the position of each Fermi source (which have quoted positions based on the $\gamma$-ray localization) and does not affect our results.

We run our fitting algorithm in two stages. First, we utilize the standard Fermi-LAT implementation of the MINUIT algorithm to find the best-fit flux of each source \emph{within} each individual energy bin. Because we allow the normalization of each source to float independently in each small energy bin, we can fix the spectra of each source to their default Fermi-LAT values without affecting their predicted flux. Second, we fix the normalization of every component besides the source itself, as well as the diffuse and isotropic backgrounds and then scan the likelihood space of source fluxes, allowing the source to take on both negative and positive fluxes (we will discuss our utilization of negative fluxes in the next subsection). 

The end result of this model is a likelihood profile of the source flux in each of the 24 logarithmic energy bins in our analysis. From this analysis, we can quickly calculate the fit of any $\gamma$-ray spectral model to the Fermi data.

\subsection{Statistical Analysis}
If the Fermi-LAT background were perfectly modeled (to the level of Poisson noise), the above treatment would accurately calculate the best-fit flux and spectrum, as well as the uncertainties for every source. The joint-likelihood (obtained from adding the likelihoods of each source) - could then be compared in models with, and without correlations between the radio/FIR and $\gamma$-ray luminosities. 

However, there are important systematic uncertainties in the determination of the astrophysical background, which can affect the statistical significance of any excess at a source location. Importantly, these uncertainties can also result in an oversubtraction of the $\gamma$-ray background, a scenario which results in a best-fit source flux with a negative normalization.

To account for the uncertainty induced by systematic mismodeling, studies have utilized blank sky locations to approximate the flux uncertainties in regions where no signal is expected~(see e.g.~\cite{Ackermann:2013yva}. In most works, the analysis chain is run separately on the true sources and the blank sky locations, and their results are then compared to calculate the statistical significance of any excess (using a bootstrapping technique).

A novel approach, which we first implemented in studies of SFGs~\cite{Linden:2016fdd} and later improved for an analysis of dwarf spheroidal galaxies~\cite{Linden:2019soa}, uses the strong correlation between the statistical significance of a Fermi source and its calculated flux to remove the effect of background mismodeling from the calculated flux at each source location \emph{before} performing the joint-likelihood analysis. While we refer the reader to Refs.~\cite{Linden:2016fdd}~and~\cite{Linden:2019soa} for further details, the basic idea is as follows. The total $\gamma$-ray flux at a given location is modeled as the linear combination of a source flux and a fake flux induced by background mismodeling:

\begin{equation}
    \phi_{{\rm total, i}} = \phi_{{\rm s, i}} + \phi_{{\rm bg}}
\end{equation}

The result is analyzed at each source location $i$. The source term must be non-negative because it represents a real $\gamma$-ray source. The background term is fit to data obtained from blank sky-locations, noting that the probability of obtaining a blank sky location with a given $\gamma$-ray flux is given by:

\begin{equation}
P_{{\rm bg}}(\phi_{\rm{bg}}) = \frac{1}{N} \sum_i A_i \mathcal{L}_i(\phi_{{\rm bg}})
\end{equation}

where $\mathcal{L}$ is the likelihood function of a given source $i$, $A_i$ is a normalization constant set such that the probability distribution integrates to 1, and the sum is taken over all background sources and averaged. Given a likelihood function for the total $\gamma$-ray flux at a source location ($\mathcal{L}_{\rm{total}}$), the likelihood function for the true source flux can be calculated as:

\begin{multline}
\label{eq:theequation}
P(\phi_{\rm{source}}) = \prod_{i}\argmax_{\phi_{s,i}}  \int_{-\infty}^\infty \mathcal{L}(\phi_{s,i} + \phi_{bg}) P_{bg}(\phi_{bg})~ \times \\ \times P_{s, i}\left(\phi_{s,i}\right) {\rm d \phi_{bg}}
\end{multline}

\noindent where the likelihood function is directly measured from Fermi-LAT data. The background term is calculated from the blank sky analysis, and the probability of obtaining a given source flux is based on the multiwavelength correlations discussed in the subsequent section. The result is marginalized over all possible background fluxes, making this a pseudo-Bayesian analysis. In the next section we discuss how to calculate P$_{s,i}(\phi_{s,i})$ using our multiwavelength correlations.

\subsection{Luminosity Relations}
\label{sec:lumRel}
The $\gamma$-ray emission from unresolved galaxies come from a combination of processes relating to star-forming activity and the interactions of particles accelerated near the AGN core. The empirical radio-gamma-ray correlation and IR-gamma-ray correlation are understood to arise from independent mechanisms in each source. As in previous studies, we adopt the following power-law  to fit the correlation between the gamma-ray luminosity produced by star forming activity and the infrared luminosity of the source~\cite{Linden:2016fdd,Tamborra:2014xia,Rojas-Bravo:2016val,2017A&A...602A.123L,2012ApJ...755..164A},
\begin{align}
   \log_{10}{\left( \frac{L_{\gamma,SF}}{\ergs}\right)}=a \log_{10}{\left(\frac{L_{IR}}{10^{45}\ergs}\right)}+g.
   \label{eq:gamma-ir}
\end{align}

We note that this correlation is similar, theoretically, to the long-observed correlation between the far-infrared luminosity and the non-thermal radio synchrotron emission observed from SFGs~\cite{1973A&A....29..231V, 1973A&A....29..249V, 1973A&A....29..263V, 1984ApJ...284..461D, 1985ApJ...298L...7H}. Similarly, as in previous studies, we adopt the following power-law  to fit the relationship between the gamma-ray luminosity powered by a central AGN and the 5~GHz core luminosity of the source~\cite{dimauro:2013xta,Hooper:2016gjy,Inoue:2011bm,Stecker:2019ybn},
\begin{align}
   \log_{10}{\left( \frac{L_{\gamma,5GHz}}{\ergs}\right)}=b \log_{10}{\left(\frac{L_{5GHz}}{10^{40}\ergs}\right)}+d.
   \label{eq:gamma-rad}
\end{align}

Though these correlations are empirically determined, there is significant galaxy-to-galaxy scatter that is parameterized by a spread, $\sigma_{SF/mAGN}$, about the log-linear relationship. Thus, the probability of a galaxy with a measured IR luminosity, producing a given gamma-ray luminosity is given by the following log-gaussian,
\begin{align}
\label{eq:pIR}
    P\left( L_{\gamma,SF}\right)=\frac{1}{\sqrt{2\pi}\sigma_{SF}} \exp{\frac{-\left(\log{\left(\frac{ L_{\gamma,SF}}{L_{IR}^a}  \right)}-g \right)^2}{2\sigma_{SF}^2}},
\end{align}

\noindent where $\sigma_{SF}$ is the intrinsic scatter in the FIR to gamma-ray correlation. Because this correlation has an uncertainty described by a Gaussian distribution in logarithmic space (rather than linear space), we note that the mean $\gamma$-ray emission from a population of galaxies \emph{exceeds} the contribution of the median galaxy. This effect becomes more pronounced as the value of $\sigma$ increases, a result which will have important effects we discuss later. Independently, the probability of a galaxy with a measured 5~GHz core luminosity, producing a given gamma-ray luminosity is given by the following,
\begin{align}
\label{eq:pRad}
    P\left( L_{\gamma,mAGN}\right)=\frac{1}{\sqrt{2\pi}\sigma_{mAGN}} \exp{\frac{-\left(\log{\left(\frac{ L_{\gamma,mAGN}}{L_{5GHz}^b}  \right)}-d \right)^2}{2\sigma_{mAGN}^2}},
\end{align}
where $\sigma_{mAGN}$ is the intrinsic scatter in the radio to gamma-ray correlation.

The total gamma-ray luminosity is the sum of $L_{\gamma,5GHz}$ and $L_{\gamma,SF}$. Therefore, the probability of a galaxy with a measured 5~GHz core luminosity and IR luminosity producing a given total gamma-ray luminosity is given by the convolution of the preceding probability distributions:
\begin{align}
     P\left( L_{\gamma}\right) &= \int  P_{SF}\left( L_{\gamma} - L' \right)  P_{5GHz}\left( L' \right) dL' \\ \nonumber
     &= \int  P_{5GHz}\left( L_{\gamma} - L' \right)  P_{SF}\left( L' \right) dL'.
\end{align}

To compute the joint likelihood, the preceding probability distributions (Eq.~\ref{eq:pIR} and Eq.~\ref{eq:pRad}) are implemented in Eq.~\ref{eq:theequation}.

\subsection{Luminosity Functions}
\label{sec:lunFun}

The luminosity function of a given source class describes the number of sources emitting in a particular band per comoving volume with luminosities in the range between $\log L$ and $\log L + d\log L$, 
\begin{align}
\label{eq:lumi}
    \Phi(L_X,z)=\frac{d^2N}{dVd\log L}.
\end{align}
Since the vast majority of observed SFG and mAGN fall well under Fermi's threshold, it would be impossible to directly describe the redshift and luminosity distribution of the gamma-ray emission of these sources. However, from the multiwavelength correlations in Eq.~\ref{eq:pIR} and Eq.~\ref{eq:pRad}, we can utilize the measured distribution of sources in IR and radio wavelengths and then correlate that emission with its expected Fermi-LAT $\gamma$-ray emission. The gamma-ray luminosity function is then calculated by integrating the multiwavelength correlations described above over the luminosity functions of the corresponding source classes.

\begin{figure*}[t]
\includegraphics[width=0.98\columnwidth]{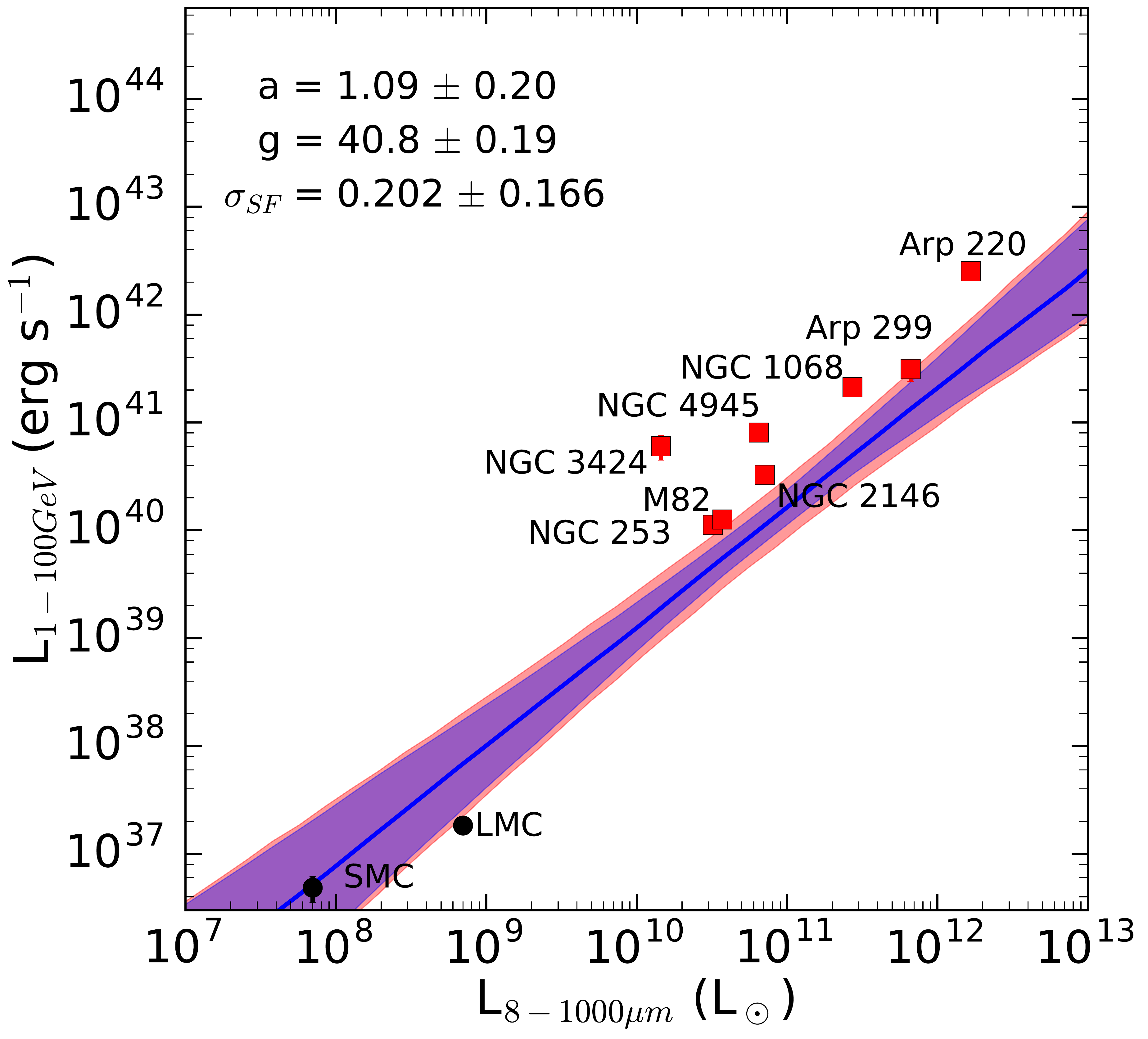}
\includegraphics[width=0.98\columnwidth]{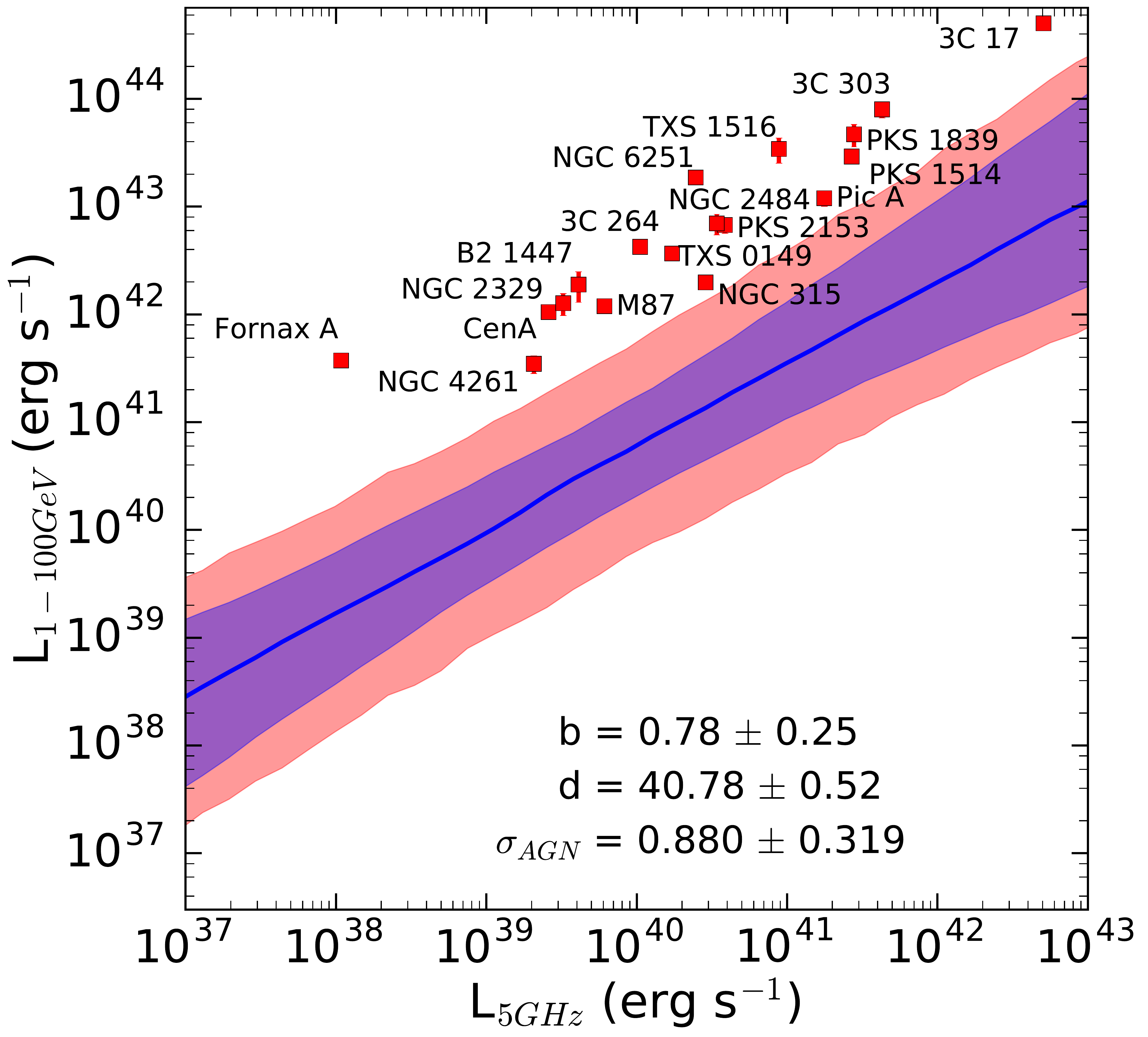}
\caption{The best-fit FIR to $\gamma$-ray correlation (left) and 5~GHz radio to $\gamma$-ray correlation (right), determined by our fitting algorithm, compared to the luminosities of observed and associated SFGs and mAGN detected in the Fermi-LAT point source catalog. The best fit line (blue) represents the median $\gamma$-ray flux of a galaxy in our sample. The blue error band represents the 1$\sigma$ uncertainty in the median flux of a galaxy in our sample, based on the uncertainties in the parameters of the FIR (Radio) to $\gamma$-ray correlations. The pink error bands represent the convolution of this error with the 1$\sigma$ statistical uncertainty in the luminosity of a single SFG (mAGN) galaxy in our sample based on the best-fitting value of the parameter $\sigma$ which describes the source-to-source dispersion. In both cases, we note that the systems observed and positively associated by Fermi observations are systematically brighter than the fluxes predicted by our model, indicating that there exists a significant luminosity threshold that biases the luminosities of observed systems. See text for a detailed discussion.}
\label{fig:fir_radio_correlations_Fermi}
\end{figure*}

The power-law relations between gamma-ray luminosity and radio luminosity, as well as gamma-ray luminosity and IR luminosity, imply the following relationships between their respective luminosity functions,
\begin{align}
    \Phi_{\gamma,mAGN}d\log L_{\gamma,mAGN} &= \Phi_{RG}d\log L_{5GHz} \\
    \Phi_{\gamma,SF}d\log L_{\gamma,SF} &= \Phi_{IR}d\log L_{IR},
\end{align}
where $\Phi_{\gamma,mAGN}$ is the luminosity function of sources emitting gamma-rays from their AGN activity and $\Phi_{\gamma,SF}$ is the number density of sources emitting gamma-rays from their star forming activity. The gamma-ray luminosity function is thus given by the following,
\begin{align}
    &\Phi_{\gamma}(L_\gamma) = \frac{d\log L_{S}}{d\log L_{\gamma}} \int \Phi_{S} (L_{S}) P(L_{\gamma, S}) d\log L_{S},
\end{align}
where $S$ is either the IR or 5~GHz band. Note that when the intrinsic scatter of the probability density approaches zero, the integral over the probability becomes the composition, $\Phi_S (L_S(L_\gamma))$, where $L_S(L_\gamma)$ is given by Eq.~\ref{eq:gamma-ir} and Eq.~\ref{eq:gamma-rad}.

The IR luminosity function is taken to be a modified Schechter function as follows, 
\begin{align}
    \Phi_{IR,X}(L_{IR},z)d\log L_{IR}=\Phi^*_{IR,X}(z) \left( \frac{L_{IR}}{L^*_{X}(z)} \right)^{1-\alpha_{X}}\\ \nonumber
    \times \exp{ \left[ \frac{-1}{2\sigma_{X}} \log^2{ \left( 1-\frac{L_{IR}}{L^*_{X}(z)} \right)}\right] }  d\log{L_{IR}} , 
\end{align}
where the subscript X indexes the source class, i.e. normal galaxies, starburst galaxies, and starburst galaxies with possible active galactic nuclei~\cite{Gruppioni2013}. We adopt the values derived by Ref.~\cite{Linden:2016fdd} for the following parameters, $\log_{10}{(L^*)}=\{ 9.46,11.02.10.57 \}\;\text{L}_{\odot}$ and $\log_{10}{(\Phi^*)}=\{ -2.08,-4.74,-3.25 \}\;\text{Mpc}^{3}$ for the “Normal Galaxy”, “Starburst Galaxy” and “Star-Forming AGN” components respectively.

It should be noted that AGN with viewing angles less than $\theta_c$ are not included in the analysis of mAGN since their observed gamma-ray spectra increasingly comes from beamed anisotropic emission in the jets. These sources are generally classified as on-axis blazars or quasars, whose contribution to the IGRB has been previously accounted for. In recent work, Yuan and Wang find that since mAGN have a large viewing angle between their jets and our line of sight, there is only a negligible contribution to the $\gamma$-ray emission from the on-axis beam. Therefore, the luminosity function of radio cores in the universe follows the radio galaxy (mAGN) luminosity function very closely~\cite{Yuan2018}. The intrinsic core radio luminosity function is intimately related to the core luminosity function of radio galaxies, $\Phi_c^{int} = \kappa \Phi_c^{RG}$, where, $\kappa \sim 1/\cos{\theta_c}\sim 1.03$, is the inverse of the proportion of radio cores with viewing angles below $\theta_c \sim 14^\circ$. The radio core luminosity function, as calculated for radio galaxies, is given by a double-power law as follows,
\begin{align}
    \Phi_{RG,c}(L_{5GHz},z)d\log L_{5GHz}=\Phi^*_{RG}(z) \\ \nonumber
    \left[ \left(\frac{L_{5GHz}}{L^*_{RG}(z)} \right)^{\beta} + \left(\frac{L_{5GHz}}{L^*_{RG}(z)} \right)^{\gamma} \right]^{-1}  d\log{L_{5GHz}} ,
\end{align}

where the redshift evolution of the parameters, $\Phi^*_{RG}$ and $L^*_{RG}$, is parameterized as in Ref.~\cite{Yuan2018},
\begin{align}
    \Phi^*_{RG}(z)= e_{1}(z)\phi_1 \\
    L^*_{RG}(z) = \frac{L^*}{e_2(z)}.
\end{align}
We note that our choice of radio core luminosity function is distinct from the distribution first utilized by Ref.~\cite{dimauro:2013xta} and then later employed by Ref.~\cite{Hooper:2016gjy}. Yuan and Wang construct their semi-parametric  luminosity function using more recent data and discuss the differences between their results and the luminosity function used by Refs.~\cite{Hooper:2016gjy,dimauro:2013xta} in section 6.4 of Ref.~\cite{Yuan2018}. We further explore the effect of our choice of radio-core luminosity function in Sec.~\ref{sec:relativeComp}.

\section{Contributions to the IGRB}
\label{sec:contributions}

In the absence of electromagnetic cascades, the total gamma-ray flux from the unresolved source class, X, is given by:
\begin{align}
\label{eq:contributionEq}
\frac{d^2F}{dE_{\gamma}d\Omega} &= \int_{0}^{z_{max}}dz \frac{d^2V}{dzd\Omega} \int_{L_{\gamma,min}}^{L_\gamma,max} \frac{dN_{\gamma}}{dE_{\gamma}} \\ \nonumber
&\times \Phi(L_{\gamma},z)\left(1-\omega(L_\gamma,z) \right) \exp{(-\tau_{\gamma \gamma})} d\log L_{\gamma},
\end{align}
where $d^2V/dzd\Omega$ is the comoving volume element, and $dN_{\gamma}/dE_{\gamma}$ is the gamma-ray spectrum, which is taken to be a power law with a cutoff at at 100~TeV. For the SF activity contribution, we use a power law spectral index of $\Gamma=2.2\pm 0.04$, while for the mAGN activity contribution we use  $\Gamma=2.25\pm 0.28$~\cite{Hooper:2016gjy,Ambrosone2020}. We note, however, that throughout this paper we use $\Gamma=2.2$ for the source gamma-ray fits which feed into Eq.~\ref{eq:theequation}. The function $\omega$ is the point source detection efficiency, which accounts for the fact that resolved sources do not contribute to the IGRB by definition. We adopt a modified Fermi detection efficiency function as described in Ref~\cite{Hooper:2016gjy}. Note that the gamma-ray luminosity is defined as follows,
\begin{align}
    L_{\gamma} =\int_{0.1\;\text{GeV}}^{1000\;\text{GeV}} E_\gamma \frac{dN_{\gamma}}{dE_{\gamma}} dE_\gamma.
\end{align}
with
\begin{align}
    F_{\gamma} = \frac{L_{\gamma}}{4\pi D_{L}^2(z)},
\end{align}
where $D_{L}(z)$ is the luminosity distance. Finally, the attenuation of the gamma-rays due to pair-production is given by the factor of $\exp{(-\tau_{\gamma \gamma})}$. 

The propagation of gamma rays is complicated by the evolution of electromagnetic cascades via interactions of the gamma-rays with the CMB and the extragalactic background light (EBL) which affects the measured diffuse energy spectrum~\cite{Blanco2017,Blanco2017a}. In order to take into account the effects of electromagnetic cascades and cosmological red-shifting on the observed spectrum, we use the publicly available code, $\gamma$-Cascade~\cite{Blanco2018}.

\begin{figure*}[t!]
\includegraphics[width=0.97\columnwidth]{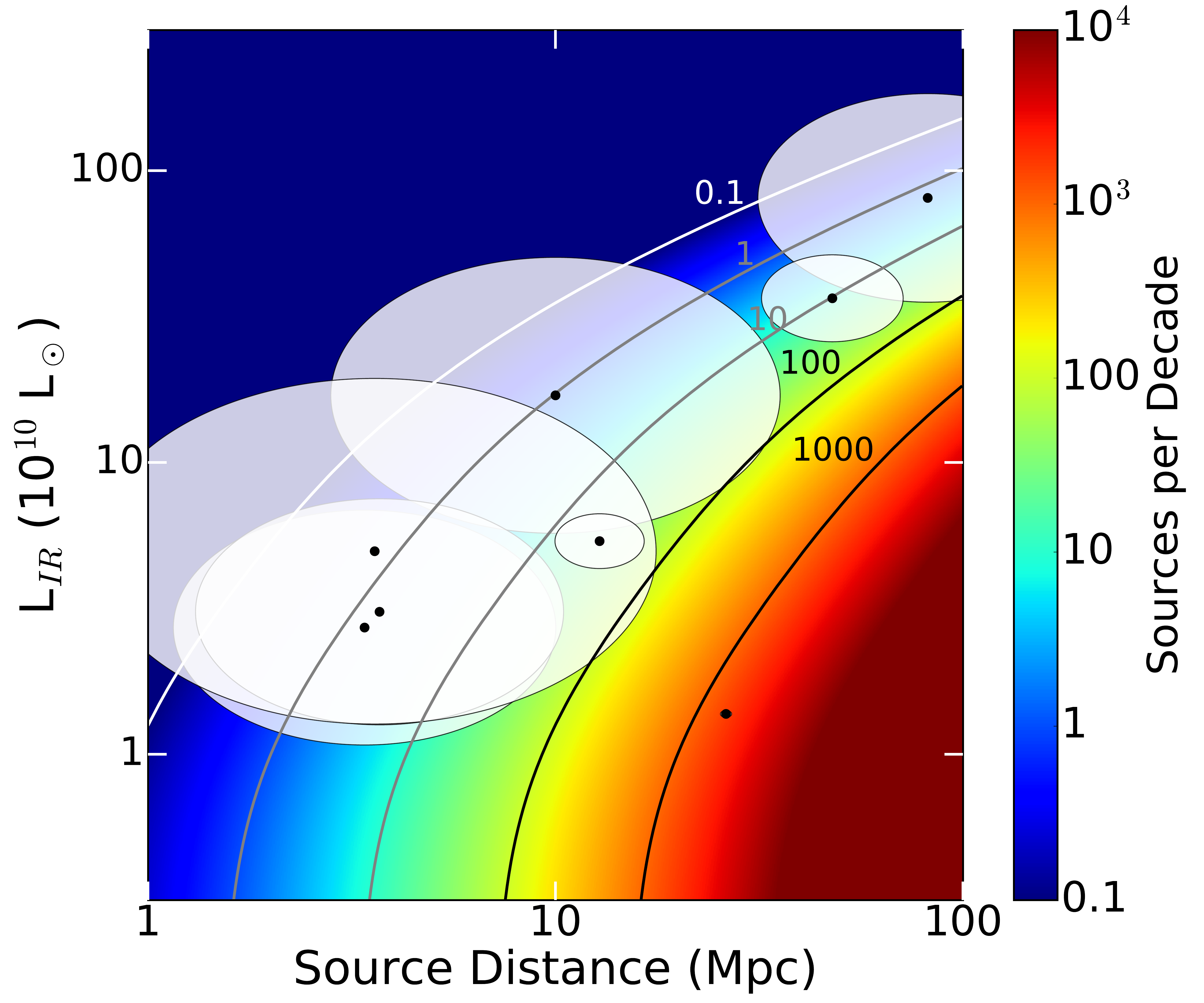}
\includegraphics[width=0.97\columnwidth]{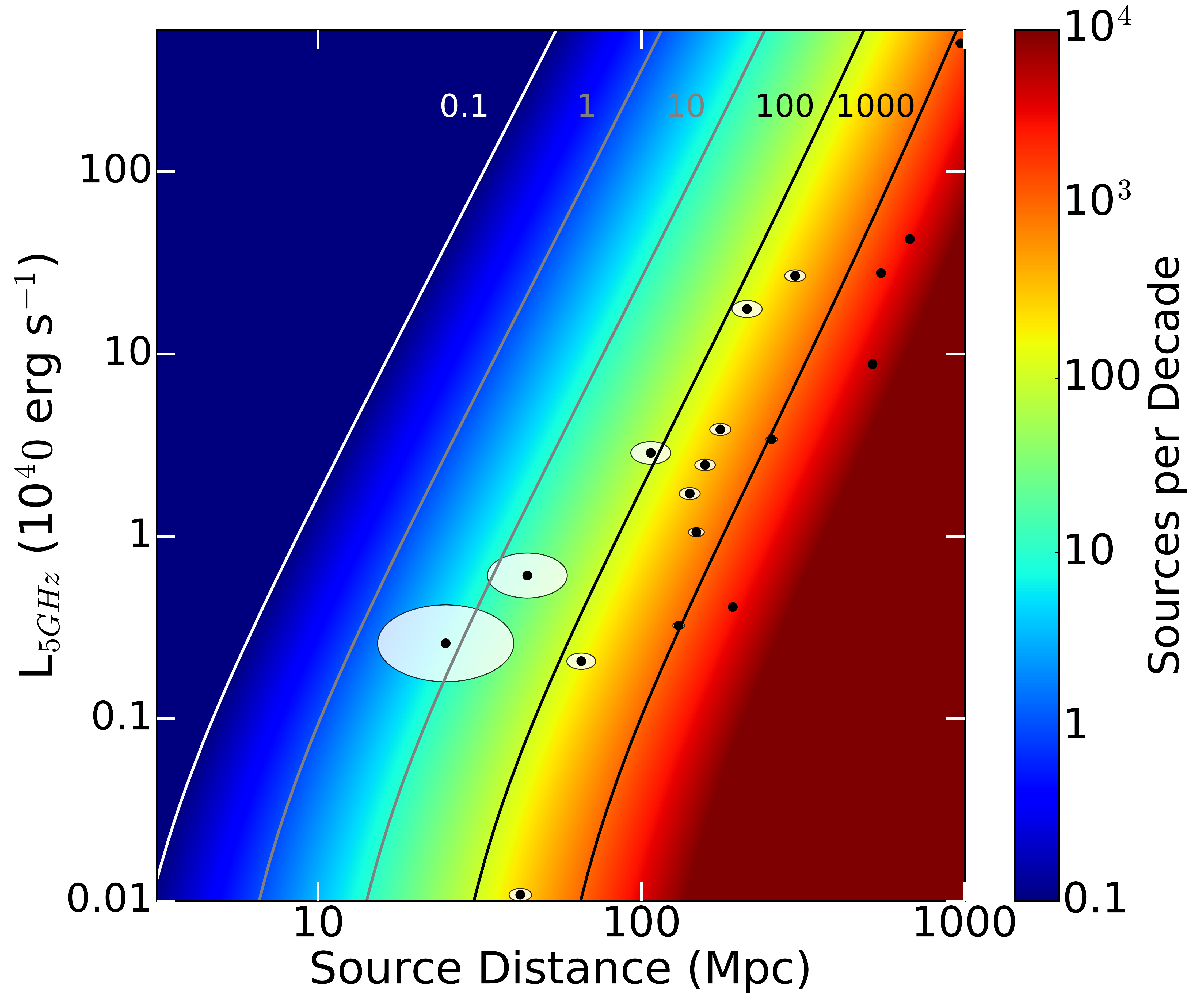}
\caption{The expected density of SFGs (left) and mAGN (right) as a function of the distance to the sources and their IR and 5~GHz radio luminosities, respectively. The source density is the number of expected sources per decade in both IR/radio luminosity and distance, and is evaluated at each individual point. Detected Fermi-LAT sources are shown as black dots, and the white region surrounding that source depicts the closest region surrounding that source where one such galaxy would be expected to exist. If the white region approximately covers the space, then Fermi-LAT observations are likely to be relatively complete, while if the white region covers a small portion of the parameter space, Fermi-LAT observations are likely to be highly incomplete and dominated by upward fluctuations.}
\label{fig:heatmap}
\end{figure*}

Following the prescription derived above, we calculate the contributions of all unresolved mAGN and SF activity to the IGRB and present them in Figs.~\ref{fig:igrb_contributions2} and ~\ref{fig:total_igrb}. Previous works on blazar AGN, i.e. BL Lacerae and Flat Spectrum Radio Quasars (FSRQs) have derived similar contributions to the IGRB~\cite{Fermi2010,Ajello2013,Ajello2011,Cholis2013}. We present these contributions and then combine these along with our calculated diffuse gamma-ray flux from mAGN and SF activity to produce a total calculated IGRB as shown in Fig.~\ref{fig:total_igrb}.

\begin{table*}[t]
  \centering
\[ \begin{pmatrix}
 3.98\times 10^{-2} & -5.76\times 10^{-3} & 2.18\times 10^{-2} & -1.93\times
   10^{-3} & 1.16\times 10^{-2} & 3.46\times 10^{-4} \\
 -5.76\times 10^{-3} & 6.78\times 10^{-2} & 7.47\times 10^{-3} & 1.99\times
   10^{-2} & 1.76\times 10^{-3} & 1.02\times 10^{-2} \\
 2.18\times 10^{-2} & 7.47\times 10^{-3} & 3.78\times 10^{-2} & -1.75\times
   10^{-2} & 3.04\times 10^{-4} & 7.19\times 10^{-3} \\
 -1.93\times 10^{-3} & 1.99\times 10^{-2} & -1.75\times 10^{-2} & 2.75\times
   10^{-1} & -3.2\times 10^{-3} & -1.13\times 10^{-1} \\
 1.16\times 10^{-2} & 1.76\times 10^{-3} & 3.04\times 10^{-4} & -3.2\times
   10^{-3} & 2.93\times 10^{-2} & 2.74\times 10^{-6} \\
 3.46\times 10^{-4} & 1.02\times 10^{-2} & 7.19\times 10^{-3} & -1.13\times
   10^{-1} & 2.74\times 10^{-6} & 1.09\times 10^{-1} \\ 
\end{pmatrix}\]
  \caption{The full covariance matrix, $\text{cov}[X_i,X_j]$ for $i\in\{a,b,d,g,\sigma_{SF},\sigma_{mAGN}\}$, of the best fit parameters in Eq.~\ref{eq:pIR} and Eq.~\ref{eq:pRad}}
  \label{tab:2}
\end{table*}

\section{Results}
\label{sec:results}

Our analysis proceeds as follows. We fit the $\gamma$-ray data from each source as a combination of three terms: a ``fake point source" flux due to background mismodeling, a flux due to SF activity, and a flux due to mAGN activity. The best-fit values of the latter two terms establish FIR-$\gamma$-ray and Radio-$\gamma$-ray correlations, which establish the relationship between standard tracers of SF and mAGN activity and their $\gamma$-ray luminosities. We then extrapolate these relationships down to less luminous and more distant sources, which have undetectable $\gamma$-ray fluxes. Utilizing standard models for galactic FIR and Radio fluxes, we then predict the total $\gamma$-ray luminosity. 

\subsection{Gamma-ray correlations}

In Figure~\ref{fig:fir_radio_correlations_Fermi}, we show the best fit FIR-$\gamma$-ray correlation (left) and Radio-$\gamma$-ray correlation (right) for SF and mAGN activity in our sample, comparing our results to the population of detected Fermi-LAT sources that have been associated with a specific SFG or mAGN source. The best-fit values and approximate 1D uncertainties for each model parameter are shown in the figure insets. However, for all quantitative calculations in the paper, we use the full covariance matrix for our model parameters given in Table~\ref{tab:2}.

\begin{figure*}[t!]
\includegraphics[width=0.97\columnwidth]{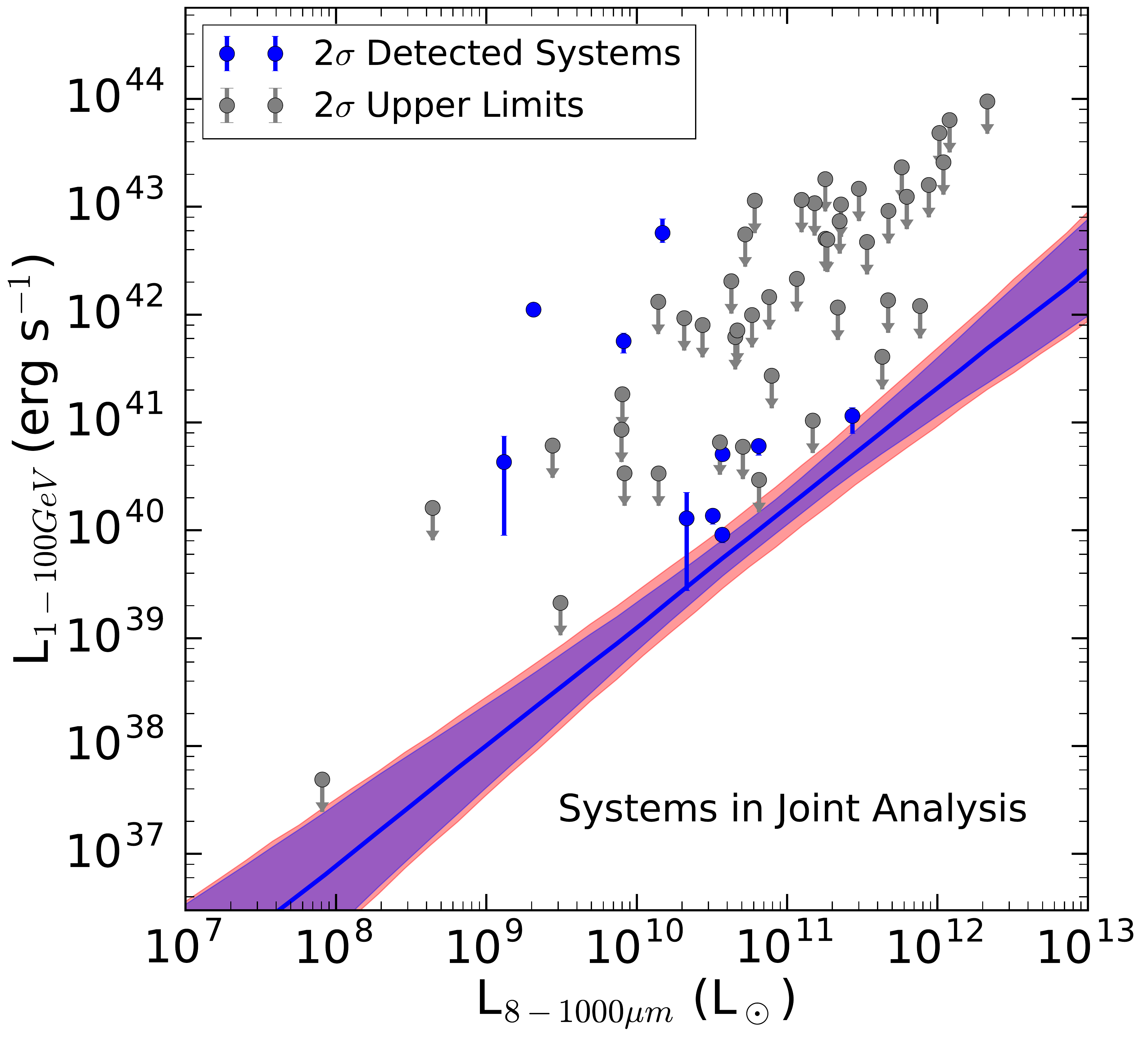}
\includegraphics[width=0.97\columnwidth]{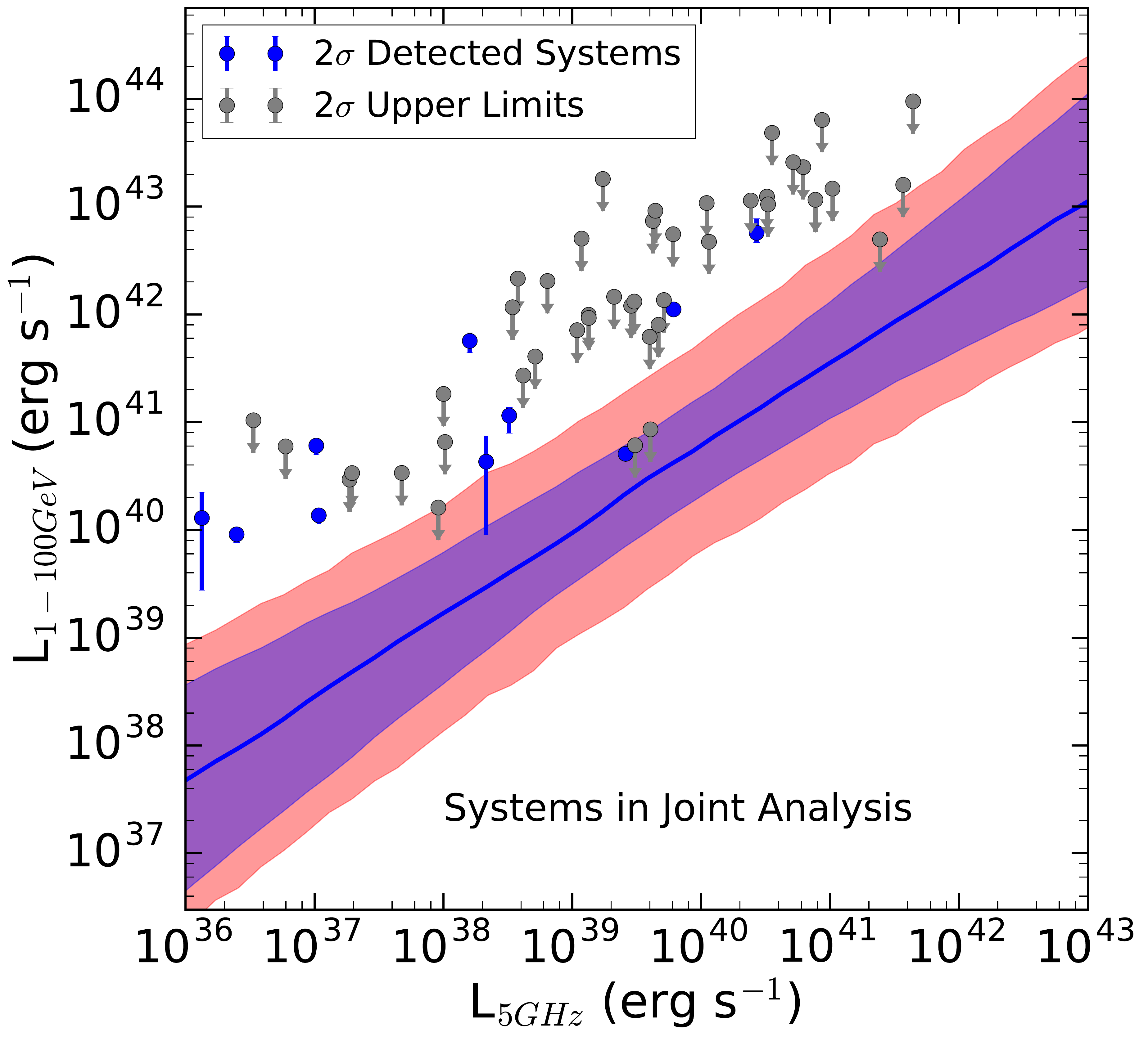}
\caption{Same as Figure~\ref{fig:fir_radio_correlations_Fermi}, except the datapoints represent the calculated $\gamma$-ray luminosities of systems in our joint sample. Systems with a 2$\sigma$ detection (after using our blank sky model to account for background uncertainties) are shown in blue. Systems without a 2$\sigma$ detection are shown with a 2$\sigma$ upper limit in gray. We stress that each $\gamma$-ray luminosity may come either from mAGN or SFG activity, and thus we expect there to be significant upward fluctuations from sources that are $\gamma$-ray bright due to the other (SFG/mAGN) mechanism.}
\label{fig:fir_radio_correlations_catalog}
\end{figure*}

We first examine our physical predictions before comparing with data. Here we note three important trends. First, we observe a statistically significant positive correlation between the FIR/Radio and $\gamma$-ray luminosities in each case. For SFGs, this relation exceeds linearity, indicating that more intense SFGs trap a larger fraction of their proton power and produce an increasingly bright $\gamma$-ray signal. This closely matches previous results by Ref.~\cite{Linden:2016fdd}. For mAGN, this relationship is sub-linear, indicating that the brightest radio AGN produce $\gamma$-ray activity that is dimmer than expected based on the rapidly increasing radio flux. The value of $b$ obtained in our study is slightly smaller than that obtained by Refs.~\cite{dimauro:2013xta, Hooper:2016gjy}, which found best fit values of b=1.008$\pm$0.025 and b=1.158$\pm$ respectively, though we note that the large error bars in these quantities make the results reasonably consistent at around the 1$\sigma$ level. 

Second, we observe very different dispersions for the FIR and Radio correlations. Our best fit value for $\sigma_{SFG}$ is, again, reasonably consistent with the best-fit value obtained in~\cite{Linden:2016fdd}. For mAGN activity, we obtain a best fit value $\sigma_{mAGN}$~=~0.88, which is larger than $\sigma_{mAGN}$~=~0.62 obtained by Ref.~\cite{Hooper:2016gjy}. We stress that our analysis (and all previous studies) obtain much higher variances for mAGN than for SF activity. 

Finally, we note an troubling trend when comparing our best-fit models against Fermi-LAT observed mAGN. While our models for the SFG luminosity fall relatively close to the luminosities of detected SFGs, (though several upward fluctuations -- including the Seyfert galaxy NGC 4945 -- exist), we note that every observed mAGN lies well above the best-fit values predicted by our radio to $\gamma$-ray correlation. 

To understand this mismatch it is important to note three key facts: (1) the source-to-source dispersion in the radio to $\gamma$-ray luminosity is large, (2) the total population of radio galaxies out to $\sim$1~Gpc is much larger than the population observed by the Fermi-LAT, (3) there is a strong sensitivity limit in Fermi-LAT observations, where only the brightest objects are observed. Combining these three features, we find that Fermi-LAT detected mAGN should be dominated by sources that have $\sim$2-$\sigma$ upward fluctuations in their $\gamma$-ray luminosity, compared to the mean of the mAGN population.

In Figure~\ref{fig:heatmap}, we graphically depict this result by plotting the total number density of SFGs (left)~\cite{Gruppioni2013} and mAGN~\cite{Yuan2018} (right) in terms of their FIR or Radio luminosities, respectively. In this parameter space, we also plot the luminosities and distances of detected Fermi-LAT sources. For each source, we add a white ellipse showing the region over which we would expect to detect $\sim$1 source. If Fermi-LAT observations were providing a relatively complete catalog of sources in a given parameter space, the white ovals would approximately fill the image in that region. If they fail to fill the space, then we expect that there are many more sources (correlating roughly to the amount of area that is unfilled) that have similar multiwavelength parameters as the sources that are detected by Fermi, but which are not detected in $\gamma$-rays.

In the case of SFGs, the observed systems fill a significant fraction of the parameter space. In fact some regions are double covered, an effect which is likely due to a combination of small statistics, and the overabundance of SFGs in the local group. In the case of mAGN, however, our observations detect only a handful of sources in regions of parameter space where \emph{hundreds} of sources are known to exist. These observations thus predict that each detected mAGN is approximately a 1 in 100 upward fluctuation, compared to the average $\gamma$-ray flux of sources with similar distances and radio luminosities. Thus, we find that our predicted model is highly consistent with the Fermi-LAT population. In fact, any model that predicts all sources to lie within 1$\sigma$ of the best-fit Radio/$\gamma$-ray correlation would be inconsistent with the underlying data.

\begin{figure}[t]
\includegraphics[width=0.98\columnwidth]{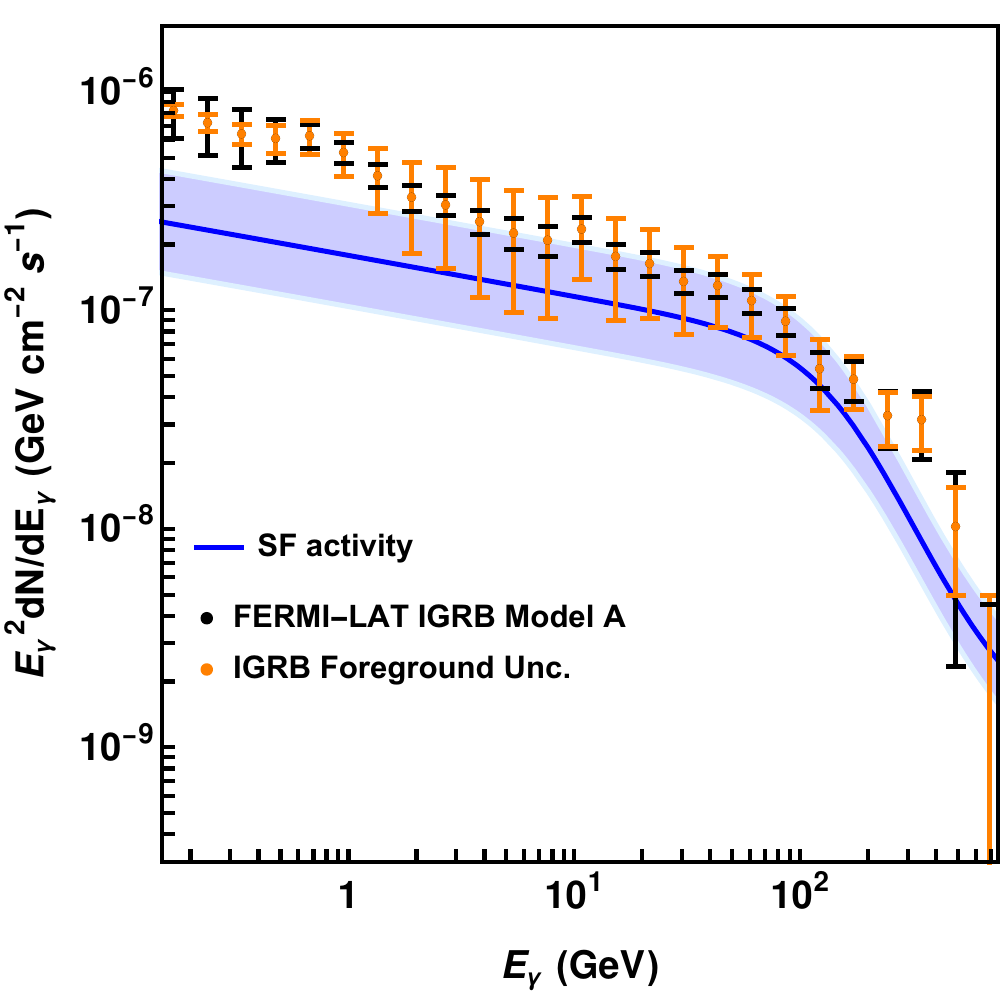}
\includegraphics[width=0.98\columnwidth]{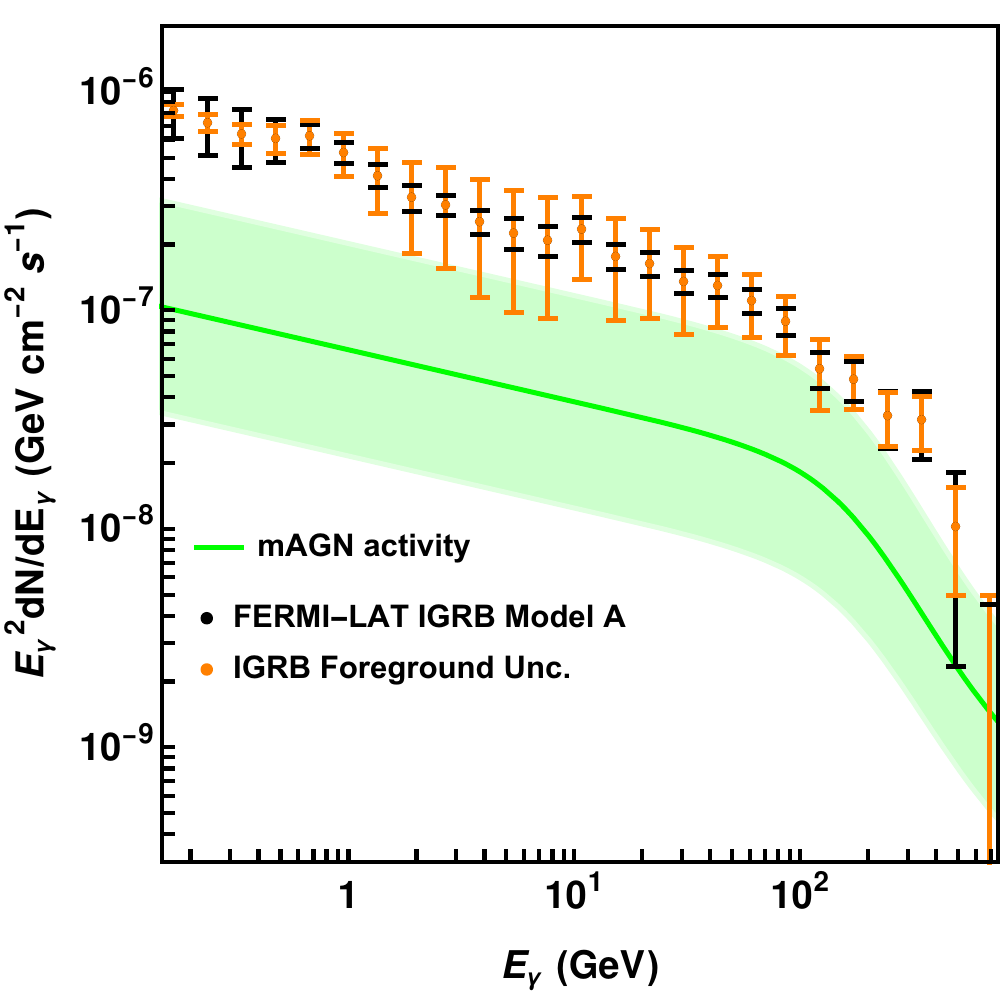}
\caption{{\bf (Top):}  The total contribution of Star-Forming galaxies to the IGRB, based on our best fit models tuned to Fermi-LAT data for nearby galaxies. Shaded regions represent the 1$\sigma$ uncertainties in our model estimate. {\bf (Bottom):} Same, but for Radio Galaxies. Our analysis indicates that SFGs likely produce the majority of the IGRB, though the contribution of Radio Galaxies is highly uncertain.}
\label{fig:igrb_contributions2}
\end{figure}

\begin{figure}[t]
\includegraphics[width=0.98\columnwidth]{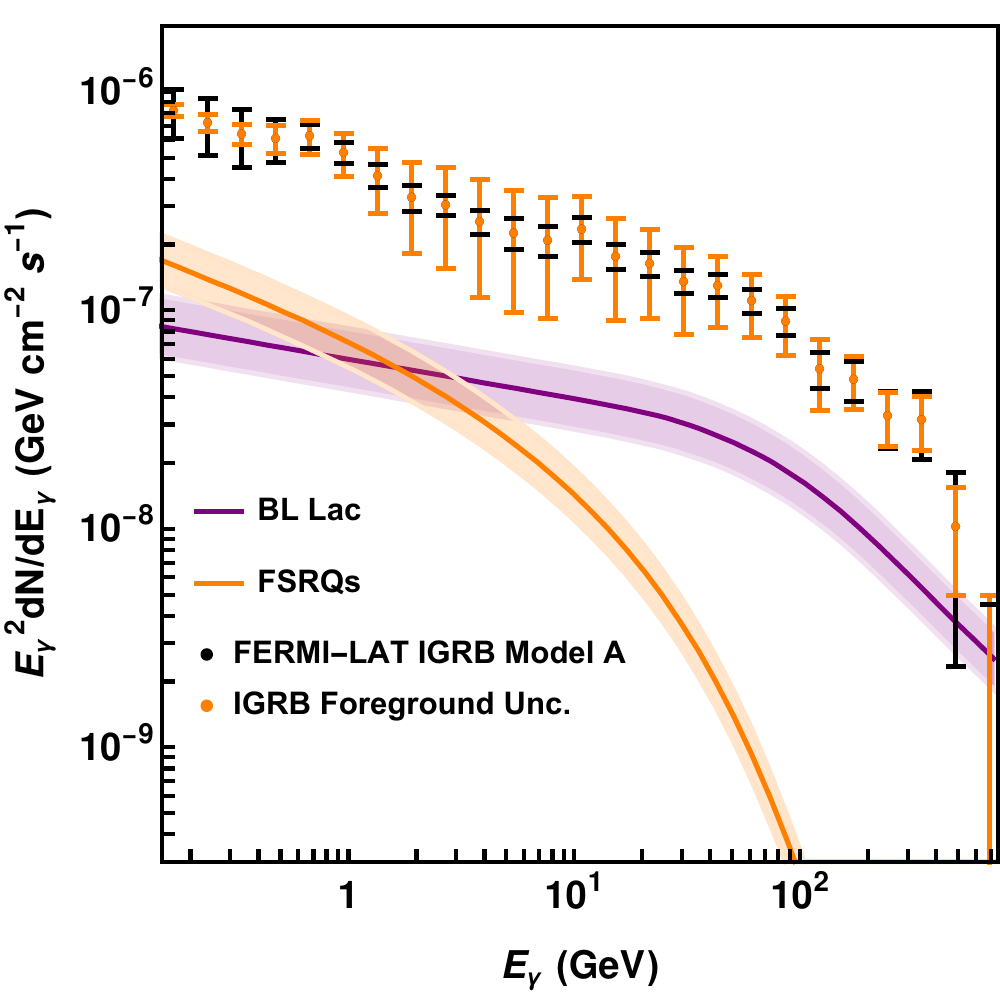}
\includegraphics[width=0.98\columnwidth]{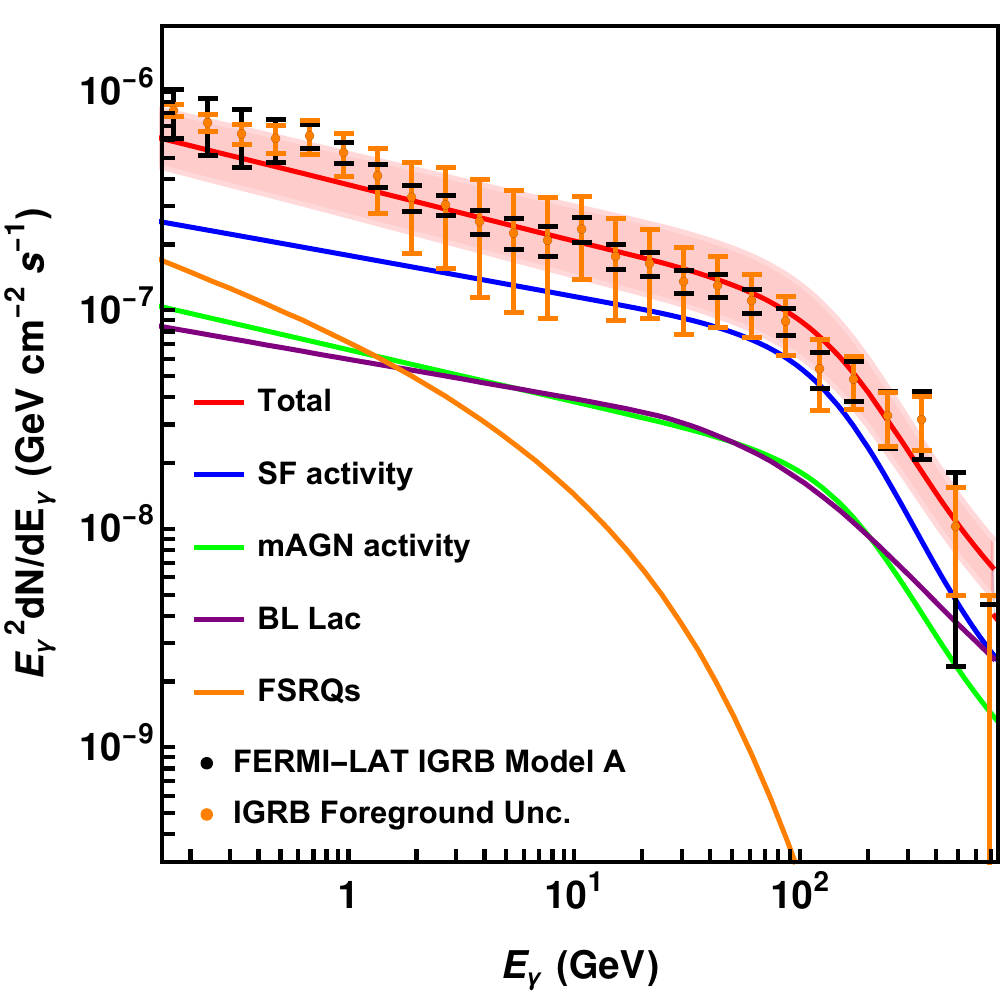}
\caption{{\bf (Top):} The total contribution of blazars to the IGRB from previous calculations~\cite{Fermi2010,Ajello2013,Ajello2011,Cholis2013}. Shaded regions represent the X$\sigma$ uncertainties. {\bf (Bottom):} The total contribution of Star-Forming galaxies, misaligned AGN, and blazars to the IGRB, based on our best fit models tuned to Fermi-LAT data for nearby galaxies. Shaded regions represent the 1$\sigma$ uncertainties in our model estimate.}
\label{fig:total_igrb}
\end{figure}

The key goal of any test of the radio/$\gamma$-ray correlation is to utilize an input population of mAGN that is unbiased by $\gamma$-ray emission (opposed to Fermi detected sources, which are maximally biased by their $\gamma$-ray flux). In Figure~\ref{fig:fir_radio_correlations_catalog}, we plot the best-fit parameters of our FIR and Radio correlations. However, in this plot, we show the best-fitting $\gamma$-ray fluxes (or $\gamma$-ray flux upper limits) for the 57 sources modeled in our analysis.

We note two important results in the comparison to this data. First, we note that we show all 57 sources in both plots -- regardless of whether the source is expected to primarily be a SFG or mAGN. That means that we expect many sources to lie far above the best-fit correlation (e.g., a bright mAGN with a small SFR will lie well above the FIR/$\gamma$-ray correlation). This behavior is directly taken into account in our joint statistical model. The converse is not true. No galaxy should lie below the FIR/radio correlations, unless it is an extreme downward fluctuation compared to our best fit models.

\subsection{IGRB Composition}
In Figure~\ref{fig:igrb_contributions2} we show the calculated contributions to the IGRB from activity related to star formation (SFG) and misaligned actively galactic nuclei (mAGN). We overlay the IGRB measured by Fermi-LAT~\cite{Ackermann:2014usa}. We find that gamma-ray emission from star formation activity is the primary contributor to the IGRB. The shaded regions in the frames of Figure~\ref{fig:igrb_contributions2} are the 1$\sigma$ uncertainties in our model estimates, which come from uncertainties in $a$, $b$, $d$, $g$, $\sigma_{mAGN}$, and $\sigma_{SF}$. Although the SF source class contributes more on average than the mAGN, it should be noted that the mAGN have a significantly larger dispersion. Therefore, a joint model in which the mAGN contribute similarly to SFGs, especially above 100 GeV, is consistent with our analysis. In particular, we find that in 20.5\% of our model simulations, the mAGN contribution is brighter than the contribution from star-forming activity (though never significantly brighter, because the errors on the star-forming contribution are small).

In Figure~\ref{fig:total_igrb} we show the combined contribution to the IGRB from all extragalactic components. In particular, we include contributions from other leading source classes, BL Lacs and FSRQs~\cite{Fermi2010,Ajello2013,Ajello2011,Cholis2013}. We find that the SF source class remains dominant over the entire energy range, though we note that FSRQs are relatively important below a few GeV. Although the average contribution of BL Lacs (from previous studies) and that of mAGN (predicted by our model) is almost identical, anisotropy analyses of BL Lacs have placed significant constraints on their contributions to the IGRB and therefore these sources are not expected to contribute more than about 20\% of the diffuse photons observed by Fermi~\cite{Hooper:2016gjy,Ackermann:2012uf,Cuoco2012,Harding2012}. 

Furthermore, we note that the calculated BL Lac and FSRQ contributions do not take into account the effects of electromagnetic cascades during gamma-ray propagation. Such a calculation is beyond the scope of this study and is left for future work. Until such a calculation is done, an in-depth spectral comparison between the blazar source classes and the mAGN \& SF contributions will have sizable uncertainties. However, for our models of electromagnetic cascades, the unresolved point-source contribution from BL Lacs and FSRQs is expected to outshine the diffuse cascade contribution from their on-axis jet emission. Thus, we do not expect that such an analysis would alter the BL Lac and FSRQ fluxes to the point that they contribute significantly to the IGRB~\cite{Ackermann:2012uf,Cuoco2012,Harding2012}.

\section{Discussion}
\label{sec:discussion}

We have completed the first simultaneous joint-likelihood analysis of $\gamma$-ray data with multiwavelength tracers of both star-forming and AGN activity. Using a robust methodology capable of accurately treating low-significance signals across many sources, we arrive at three robust conclusions:

\begin{itemize}
    \item $\gamma$-rays from star-formation activity significantly contribute to the IGRB, providing 48$^{+33}_{-20}$\% of the total IGRB at~1 GeV and 56$^{+40}_{-23}$\% at 10~GeV. Including mAGN contributions in our analysis does not significantly affect the predicted star-forming flux. 
    
    \item The contribution of mAGN to the IGRB is highly uncertain. In our average model, mAGN produce approximately 18$^{+36}_{-12}$\% of the IGRB flux at 1~GeV. At 10~GeV mAGN produce approximately 18$^{+38}_{-12}$\%. Thus, large mAGN contributions remain viable. 
    
    \item The flux dispersion in the mAGN population is extremely high. Individual mAGN have $\gamma$-ray to radio flux ratios that can differ by nearly an order of magnitude at the 1$\sigma$ level. The analysis of a large, and systematically complete sample of mAGN is paramount to further constrain their $\gamma$-ray properties.
    
\end{itemize}

\subsection{Comparison with Previous SFG Analyses}

\begin{table}[t!]
  \centering
\begin{tabular}{| c | c | c | c |}
\hline
 Model & a & g & $\sigma_{SF}$ \\ \hline \hline
 
 This Paper & ~~1.09 $\pm$ 0.20~~ & ~~40.8 $\pm$ 0.19~~ & ~~0.20 $\pm$ 0.17~~ \\ \hline 
 
 Ref.~\cite{Linden:2016fdd} & 1.18 $\pm$ 0.15 & 40.2 $\pm$ 0.24 & 0.39 $\pm 0.12$ \\ \hline
 
 Ref.~\cite{2012ApJ...755..164A} (All) & 1.17 $\pm$ 0.07 & 40.94 $\pm$ 0.08 & 0.24 \\ \hline
 
 Ref.~\cite{2012ApJ...755..164A} (no AGN) & 1.09 $\pm$ 0.10 & 40.7 $\pm$ 0.10 & 0.25 \\ \hline
 
\end{tabular}
  \caption{Best-Fit parameters for the correlation of $\gamma$-ray emission with star-formation activity in this paper, compared to previous works. We find that the best-fit values of each parameter align very closely with the analysis of Ref.~\cite{2012ApJ...755..164A}. Moreover, the results also closely fit Ref.~\cite{Linden:2016fdd}, after accounting for a degeneracy the best-fit values of $g$ and $\sigma_{SF}$. We note that the best-fit values of $g$ from previous works have been renormalized to a value of 10$^{45}$~erg~s$^{-1}$ rather than 10$^{10}$~L$_\odot$, as used in previous papers. This shift does not significantly produce significant errors in the best fit value of $g$, but may lead to 10\% errors in the quoted uncertainty of $g$.}
  \label{tab:sfg}
\end{table}

Our analyses of the SFG contribution are qualitatively similar to previous work by Refs.~\cite{2012ApJ...755..164A, Linden:2016fdd}. In Table~\ref{tab:sfg} we show the best-fitting values of $a$, $g$ and $\sigma_{SF}$ quoted by each analysis. We find that our best-fit values lie within the 1$\sigma$ uncertainties of Ref.~\cite{2012ApJ...755..164A}, both in models where Swift-BAT detected AGN were included or removed. 

Comparing our results to Ref.~\cite{Linden:2016fdd}, we obtain a slightly higher value for $g$, but a slightly smaller value of $\sigma_{SF}$. However, this is the \emph{expected} impact of adding an mAGN term into the joint-likelihood analysis. mAGN activity in galaxies that are dominantly star-forming will produce additional dispersion in the FIR-$\gamma$-ray correlation. This is likely to significantly boost the $\gamma$-ray luminosity of some SFGs, while leaving the luminosity of the others unchanged. This results in an increase in $\sigma_{SFG}$, which must be compensated for by a decrease in $g$ (which is needed to still fit the majority of galaxies without large mAGN contributions). In Ref.~\cite{Linden:2016fdd} (see Figure 3), it was shown that a decrease of 0.2 dex in $\sigma_{SF}$ correlates with a nearly 0.4 dex shift in $g$. After accounting for this degeneracy, these results are in agreement. 

The similarity of our results to those of previous analyses indicates that the calculated contribution of star-formation activity to the $\gamma$-ray flux of nearby galaxies is robust to contamination from mAGN activity. This is similar to the conclusion of Ref.~\cite{2012ApJ...755..164A}, which found similar results after removing several Swift-BAT detected AGN from their analysis. However, our analysis is significantly more powerful, as it avoids binary classifications and allows for mAGN contributions even from galaxies within extremely bright (BAT-detected) AGN. We furthermore note that the addition of several detected SFGs that were not included in this paper (namely Arp 220, NGC 3424, Arp 299 and NGC 2146), would only strengthen these conclusions. These galaxies were removed from our current analysis only because they lack a measured 5~GHz radio flux. Because these galaxies are in agreement with previous FIR-$\gamma$-ray correlations~\cite{2012ApJ...755..164A, Linden:2016fdd}, they would undoubtedly increase the statistical significance of the analysis here.

\subsection{Comparison with Previous mAGN Analyses}

\begin{table}[t!]
  \centering
\begin{tabular}{| c | c | c | c |}
\hline
 Model & b & d & $\sigma_{mAGN}$ \\ \hline \hline
 
 This Paper & ~~0.78 $\pm$ 0.25~~ & ~~40.78 $\pm$ 0.52~~ & ~~0.88 $\pm$ 0.32~~ \\ \hline 
 
 Ref.~\cite{Hooper:2016gjy} & 1.156 $\pm$ 0.15 & 42.2 $\pm$ * & 0.62 $\pm 0.12$ \\ \hline
 
 Ref.~\cite{dimauro:2013xta} & 1.008 $\pm$ 0.025 & 42.32 $\pm$ * & ? \\ \hline
 
 Ref.~\cite{Stecker:2019ybn} & 0.89 & 42.75 & 0 \\ \hline
 
\end{tabular}
  \caption{Best-Fit parameters for the correlation of $\gamma$-ray emission with mAGN activity in this paper, compared to previous works. Our models predict a smaller mAGN induced $\gamma$-ray flux. We note that values of $d$ in Refs.~\cite{dimauro:2013xta, Hooper:2016gjy} were normalized a flux of 1~erg~s$^{-1}$ which produces large quoted errors in the value, due to significant degeneracies between $d$ and $b$. Ref.~\cite{dimauro:2013xta} shows a statistical uncertainty, but considers it to be due to $\gamma$-ray measurement errors rather than source-to-source dispersion, and does not quantify the value. Ref.~\cite{Stecker:2019ybn} does not provide uncertainties or a source-to-source dispersion term.}
  \label{tab:mAGN}
\end{table}

Unlike the case of SFGs, we find that our mAGN analysis differs significantly from previous studies by Refs.\cite{dimauro:2013xta, Hooper:2016gjy, Stecker:2019ybn}. In particular, our analysis predicts a much smaller value of $d$, but a much larger value of $\sigma_{mAGN}$. Our best fit value of $b$ is similar to Ref.~\cite{Stecker:2019ybn}, but differs from Refs.~\cite{dimauro:2013xta, Hooper:2016gjy}.

We first stress that Refs.~\cite{dimauro:2013xta, Stecker:2019ybn} make several simplifying assumptions that systematically skew their results. Most importantly, these studies both choose to fit the Radio/$\gamma$-ray correlation \emph{only} using galaxies that are detected in $\gamma$-rays. Their choice to remove an mAGN from the sample \emph{because} it is not a bright $\gamma$-ray emission source causes the radio-$\gamma$-ray correlation to be highly biased towards bright sources. Secondly, these papers do not consider any source-to-source dispersion in the radio-$\gamma$-ray correlation, causing their results to be dominated by the brightest (most statistically significant) sources. We note that Ref.~\cite{dimauro:2013xta} does include an uncertainty band, but they calculate this number based only on the statistical variations in the $\gamma$-ray flux of detected sources. 

The comparison with Ref.~\cite{Hooper:2016gjy} is more nuanced, as this paper does include sub-threshold mAGN and source-to-source dispersion. Our analysis finds a significantly smaller value of $b$ and $d$, but with a higher value of $\sigma_{mAGN}$ (which as noted previously, is degenerate with $d$). We do note that the catalogs differ markedly, most importantly due to our choices to more strongly cut on quasar-like activity, as well our constraint to include only sources with measured LIR fluxes. Importantly, our catalog contains no sources with 5~GHz fluxes exceeding 10$^{43}$~erg~s$^{-1}$, while Ref.~\cite{Hooper:2016gjy} includes several. Additionally, we note that Ref.~\cite{Hooper:2016gjy} borrowed heavily from Ref.~\cite{dimauro:2013xta} in constructing their catalog, meaning that there may be some bias towards catalog detected sources. Still, there is no obvious reason for the difference between these results.

The evidence supporting our result stems from the interpretation of Figure~\ref{fig:heatmap}, which does not rely on the sources included in either analysis. We find that most Fermi-LAT \emph{detected} mAGN exist in portions of the Radio Luminosity/Distance parameter space where several hundred sources are expected. Because this parameter space does not depend on the $\gamma$-ray luminosity -- the fact that only a few dozen mAGN are detected by Fermi (and not several hundred), implies that the mAGN that are observed have much higher $\gamma$-ray luminosities than the average source in our population. However, the best-fit analysis of Ref.~\cite{Hooper:2016gjy} includes only one source that (barely) exceeds the 1$\sigma$ source-to-source dispersion, while Figure~\ref{fig:heatmap} indicates that we should see dozens of such sources. 

On the contrary, our analysis is consistent with the hypothesis that there is a strong selection effect on Fermi-LAT detections. If the mAGN population has a large dispersion in the radio/$\gamma$-ray correlation, Fermi-LAT observations should detect only the tip of the iceberg, finding sources that have $\gamma$-ray luminosities far above the best-fit correlation. This same argument applies (even more strongly) to the analyses of Refs.~\cite{dimauro:2013xta, Stecker:2019ybn}. An analysis that includes only detected Fermi-LAT sources is likely to highly bias the underlying correlation.

Finally, we note that, compared to previous works, our joint-likelihood analysis correctly assigns some fraction of the $\gamma$-ray emission from mAGN to star-formation within those same galaxies, decreasing the contribution of mAGN to the IGRB. This effect, which is theoretically expected as most mAGN galaxies also host star-formation, could not be tested by previous analyses.

To conclude, we note an important cross-check of our results. As mentioned previously, comparing the Fermi mAGN population with the full population (shown in Figure~\ref{fig:heatmap}) clearly implies that all detected mAGN should be $\sim$2$\sigma$ upward fluctuations in the Radio to $\gamma$-ray correlation. This is consistent with the radio-$\gamma$-ray correlation obtained by our fit (and shown in Figure~\ref{fig:fir_radio_correlations_Fermi}. However, we stress that our model \emph{does not know anything} about the total number of mAGN that exist in the local universe. It only knows about the relative radio and $\gamma$-ray luminosities (and their uncertainties) in the systems that we chose to analyze. Thus, the fact that our fit (in contrast with previous results) predicts that the observed mAGN population is brighter than the average mAGN population, actually serves as a strong implicit cross-check of our analysis. 

\subsection{Comparison of Luminosity functions and multiwavelength Correlations}
\label{sec:relativeComp}

Here we try to estimate how the contribution to the IGRB would change in light of a new luminosity function and/or a new radio-gamma correlation. We can get a rough measure of how bright the flux is predicted to be given an mAGN luminosity function $\phi_{RG,i}$ using the following integral,
\begin{align}
   f_i &= \int \frac{L_{\gamma}}{D^2_l(z)} \Phi_{\gamma,i}(L_{\gamma},z)  d\log{L}_{\gamma}\frac{dV}{d\Omega dz}dz \\ \nonumber
    &\approx  \int \frac{\overline{L}_{\gamma}}{D^2_l(z)} \Phi_{RG,i}(L_{5GHz},z)  d\log{L_{5GHz}} \frac{dV}{d\Omega dz}dz,
\end{align}
Where in the second line we assume a weak luminosity evolution as is the case with mAGNs (see the discussion of luminosity evolution in Ref~\cite{Yuan2018}). Since there is a linear correlation between $L_\gamma$ and $L_{5GHz}$, the mean value of $L_\gamma$ can be estimated as the mean value of $L_\gamma$ at the central value of $L_{5GHz} = 10^{40}\;erg/s$. However, since the distribution in Eq.~\ref{eq:pRad} is log-normal with a SD of $\sigma_{mAGN}$, the mean value is given by $\overline{L}_{\gamma}\approx10^{\frac{\sigma^2}{2}+d}10^{40}\;erg/s$. Intuitively, this means that the galaxy-to-galaxy spread in the $\gamma$-ray-radio correlation serves to increase the mean predicted luminosity of a source class. Finally, we can estimate the relative change in the flux contribution to the IGRB comparing two choices of $\Phi$ and $\overline{L}_{\gamma}$,
\begin{align}
\label{eq:q}
    q &= \frac{f_i}{f_j} \\ \nonumber
     &\approx \left(\frac{\overline{L}_{\gamma,i}}{\overline{L}_{\gamma,j}} \right) \left( \frac{\int \frac{1}{D^2_l(z)} \Phi_{RG,i}(L_{5},z)  d\log{L_{5}} \frac{dV}{d\Omega dz}dz} {\int \frac{1}{D^2_l(z)} \Phi_{RG,j}(L_{5},z)  d\log{L_{5}} \frac{dV}{d\Omega dz}dz} \right) \\ \nonumber
     &\approx \left(\frac{10^{\frac{\sigma_i^2}{2}+d_i}}{10^{\frac{\sigma_j^2}{2}+d_j}} \right) \left( \frac{\int \frac{1}{D^2_l(z)} \Phi_{RG,i}(L_{5},z)  d\log{L_{5}} \frac{dV}{d\Omega dz}dz} {\int \frac{1}{D^2_l(z)} \Phi_{RG,j}(L_{5},z)  d\log{L_{5}} \frac{dV}{d\Omega dz}dz} \right) \\ \nonumber
     &\approx q_{fit}(d_i,d_j,\sigma_i,\sigma_j)q_{\Phi}(\Phi_{RG,i},\Phi_{RG,j}),
\end{align}
where we've abbreviated $L_{5GHz}$ into $L_{5}$. The separation in Eq.~\ref{eq:q} allows us to estimate the relative change in $\gamma$-ray flux due to a change in luminosity functions independently from changes in the radio-gamma correlation while still taking into account the galaxy-to-galaxy scatter.

\begin{figure}[t]
\includegraphics[width=\columnwidth]{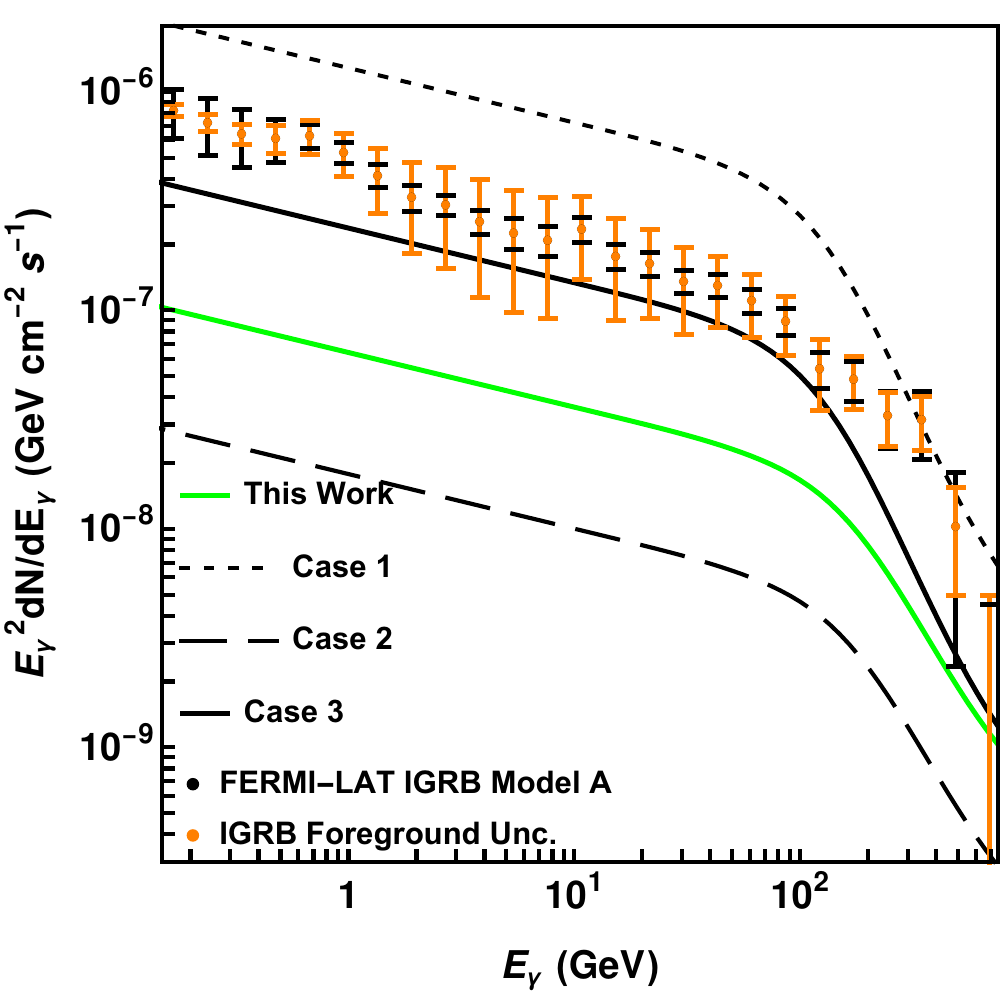}
\caption{The mean IGRB contribution from mAGN using different Luminosity functions (LF) and fitting parameters. This work used the parametric LF from Yuan and Wang~\cite{Yuan2018} and our best fit parameters (Table~\ref{tab:sfg}). Case 1 used the LF from Di Mauro~\cite{dimauro:2013xta} and our fitting parameters. Case 2 used the LF from Yuan and Wang and the fitting parameters from Hooper et al.~\cite{Hooper:2016gjy}. Case 3 used the LF from Di Mauro and the fitting parameters from Hooper et al.}
\label{fig:compared}
\end{figure}

Applying this analysis to Ref.~\cite{Hooper:2016gjy}, we find that the ratio of the predicted gamma-ray luminosity found by Hooper et al. $\overline{L}_{\gamma,HLL}$ to that found in this work $\bar{L}_{\gamma,BL}$ to be $q_{fit} \sim 16.8$. Similarly, the ratio in the luminosity function-dependent integrals is calculated from the luminosity function $\Phi_{RG,i}(L_{5GHz},z)$ used in this work, as laid out in Ref.~\cite{Yuan2018} while that of Hooper et al. is taken from that used by Di Mauro~\cite{dimauro:2013xta} and found to be $q_{\Phi}\sim 0.17$. Therefore, one would expect that the IGRB contribution from mAGN would be $q \sim 2.9$ larger. Indeed we note that in the lower energy bins, $E_{\gamma}<10\; \text{GeV}$, where relative differences due to the effects of propagation are mild, Hooper et al. find that mAGN contribute on average about 3 times more than what we predict. To show the validity on these heuristic arguments, we compare the mean IGRB contribution of mAGN using these different luminosity functions and fitting parameters after running them through our propagation analysis in Fig.~\ref{fig:compared}.

\subsection{Implications for Extragalactic Neutrinos}
Understanding the origin of the extragalactic IceCube neutrino flux~\cite{Aartsen:2013bka, Aartsen:2015knd, Aartsen:2016xlq} is critical to understanding natures most powerful accelerators. Studies that consider spatial and temporal associations between IceCube neutrinos and gamma-ray sources have strongly constrained the contribution of gamma-ray bursts~\cite{2012Natur.484..351I, Aartsen:2018fpd} and on-axis blazar jets~\cite{Aartsen:2016lir, Hooper:2018wyk}, respectively, to the diffuse neutrino background. Due to these constraints, many models have examined the possibility that SFGs~\cite{Loeb:2006tw, Murase:2013rfa, Anchordoqui:2014yva} or mAGN~\cite{Tjus:2014dna, Giacinti:2015pya, Murase:2015ndr, Hooper:2016jls,Fang:2017zjf}, dominate the IceCube signal. However, it should be noted that the sources that dominate the high-energy neutrino flux do not necessarily need to be bright $\gamma$-ray sources, for example in scenarios where ``hidden" accelerators dominate the neutrino flux~\cite{Murase:2015xka, Senno:2015tsn}.

Unlike blazars, for which it is (impressively) possible to associate a neutrino flux with a specific source~\cite{IceCube:2018cha}, the low-luminosity of individual mAGN and SFGs makes strong associations unlikely (though see~\cite{Inoue:2019yfs, Anchordoqui:2021vms,Murase:2019vdl,Kheirandish:2021wkm}). Thus, the composition of the IGRB at GeV energies shines light into the relative importance of mAGN and SFGs~\cite{Bechtol:2015uqb, Linden:2016fdd, Hooper:2016gjy}. This is well-motivated, because the same hadronic interactions that produce $\gamma$-ray emission also produce high-energy neutrinos. 

In this paper, we show that SF activity is likely to be the primary contributor to the IGRB at GeV energies-- while mAGN are likely subdominant. This raises the possibility that SF activity may be a brighter neutrino source than mAGN. However, significant caution is warranted for two reasons. First, the uncertainties (especially on the mAGN contribution) are large, making it difficult to rule out sizable mAGN contributions to the IGRB at high statistical significance.

Second, and more importantly, our model is constrained primarily by the relative luminosities of SF activity and mAGN at energies between 1-10~GeV, where Fermi-LAT observations are most constraining. Notably, the spectral indices of the the SF and mAGN components are not constrained by our model -- as we adopt a universal spectrum for each component (this provided a good fit to our data). Small changes in this spectral index could produce large changes in the SF and mAGN fluxes at PeV-energies, where the IceCube neutrinos are observed. Additionally, even if this spectrum is an appropriate average for each source class, a source-to-source dispersion in the spectral index may significantly affect the total luminosity of the population at PeV energies.

In the case of SF activity, these spectral distortions may occur in scenarios where different classes of supernovae dominate cosmic-ray injection at different energy scales~\cite{Xiao:2016rvd}. Depending on the dispersion in spectral indices, the neutrino contribution from SF activity can vary significantly. Models with mixed spectral indices can contribute significantly below 100~TeV and less so above that energy~\cite{Ambrosone2020}.  However, in models where the SF spectrum is relatively soft, it may be difficult to explain the intensity of the neutrino flux within any SFG model~\cite{Bechtol:2015uqb}, particularly at energies between 10-100~TeV~\cite{Palladino:2018bqf}. These results suggest that while models of the SF component can be constructed to explain the IceCube data above \emph{or} below 100~TeV, at least two source classes are needed to explain the whole energy range.

Thus, while our model would hint towards an SF activity explanation for the IceCube neutrino flux, we strongly caution that this is far from a definitive picture. In particular, significant work in fitting the TeV $\gamma$-ray fluxes from SF activity and mAGN is required to better understand the spectral features of both emission sources. Moreover, a full model --- including not only SF activity and mAGN, but also contributions from on-axis AGN (blazars), GRBs, and hidden neutrino sources, may be necessary to explain the totality of the IceCube signal.

\subsection{Uncertainties above 100~GeV}
The uncertainties related to the injected spectra of our source classes suggest caution should be taken when analysing these results above 100 GeV. This is particularly true, as cascades and $\gamma$-ray attenuation begin to significantly distort the injected $\gamma$-ray spectrum at high energies. Since the magnitude of electromagnetic cascade production depends on the injected gamma-rays above 1 TeV, uncertainties in spectral cutoffs as well as the distribution of spectral indices for mAGN translate into uncertainties in the cascade contribution and spectral distortion above 100 GeV. Additionally, the choice of models for the extra-galactic background light (EBL) and extra-galactic magnetic field (EGM) also affects the spectral shape of the resulting diffuse flux, particularly around the break at around 100~GeV.

We note, in particular, that this implies that the high-energy contribution from mAGN (in this paper) should not be compared with the contribution of BL Lac objects in the high energy range (as cascades were not taken into account in previous studies). Finally, we point out that the intrinsic spectral shape above 100 GeV would also make a significant difference in the relative contribution of different source classes to the diffuse high-energy neutrino flux (see also ref.~\cite{Ambrosone2020}).




\section{Conclusions}

Previous work has found that mAGN and SFGs are the two most likely source classes capable of explaining the IGRB. However, these studies have come to different conclusions regarding which is more important. This was primarily due to two weaknesses in previous work: (1) each only examined one source class at a time, mistreating systems which produced $\gamma$-ray emission from both central black holes and SF activity, (2) most studies utilized source samples that were systematically biased by the presence of bright $\gamma$-ray emission. 

In this paper, we have completed a systematic joint-likelihood analysis that simultaneously accounts for both the $\gamma$-ray contributions of SF and mAGN activity. Using this methodology, we arrive at consistent predictions for the correlations between $\gamma$-ray luminosity and FIR \& 5GHz-core radio luminosity. We find that SF activity contributes a significant fraction of the observed IGRB, and its contribution is independent of contamination from potential mAGN activity (for example in Seyfert galaxies). On average, the contribution of SF activity is three times higher than that of mAGN.

While mAGN activity is usually subdominant to that of SF activity -- and often similar to the contribution of on-axis blazars -- we note that the uncertainty in the mAGN contribution is extremely large. This is due to significant source-to-source dispersion in the radio/$\gamma$-ray correlation among the class of mAGN. We note that a similar uncertainty does not apply to on-axis blazars, which have a total contribution that is well-constrained by anisotropy analyses. 

The difference between our IGRB predictions and those of previous studies stems from changes in both the multiwavelength correlations, as well as multiwavelength luminosity functions. In both cases, our FIR-$\gamma$-ray correlation is consistent with previous results. However, our mAGN models are starkly different. Our analysis of the radio-$\gamma$-ray correlation finds a significantly smaller median $\gamma$-ray luminosity but a much larger galaxy-to-galaxy scatter than previous models. Averaging over the mAGN population, we find that an average mAGNs $\gamma$-ray luminosity is approximately 18 times smaller than previous models predict. However, our model also uses an updated mAGN luminosity function, which predicts that (for a given radio-$\gamma$-ray correlation) the contribution of mAGN to the IGRB would be six-times larger than in previous studies. Combining these two factors, we find that the total contribution of mAGN activity to the IGRB is, on average, three times smaller than previous studies.

We note that the critical systematic uncertainty affecting our analysis (and all analyses of the IGRB) is the existence of a complete local catalog of FIR and Radio luminosities for nearby galaxies that is unbiased by the presence of any associated $\gamma$-ray emission. Additionally, observations of the diffuse $\gamma$-ray flux above 1~TeV would significantly constrain the spectral composition of our source classes, providing more accurate determinations of the very high-energy $\gamma$-ray and neutrino emission from these sources.

\newpage 
\section*{Acknowledgements}
We thank John Beacom, Dan Hooper and Kohta Murase for helpful comments. The work of CB was supported in part by NASA through the NASA Hubble Fellowship Program grant HST-HF2-51451.001-A awarded by the Space Telescope Science Institute, which is operated by the Association of Universities for Research in Astronomy, Inc., for NASA, under contract NAS5-26555 as well as by the European Research Council under grant 742104. TL is partially supported by the Swedish Research Council under contract 2019-05135, the Swedish National Space Agency under contract 117/19 and the European Research Council under grant 742104. This project made use of resources provided by the Swedish National Infrastructure for Computing (SNIC) under the project No. 2020/5-463 partially funded by the Swedish Research Council through grant agreement no. 2018-05973.

\newpage
\bibliography{main.bib}

\begin{thebibliography}{84}%
\makeatletter
\providecommand \@ifxundefined [1]{%
 \@ifx{#1\undefined}
}%
\providecommand \@ifnum [1]{%
 \ifnum #1\expandafter \@firstoftwo
 \else \expandafter \@secondoftwo
 \fi
}%
\providecommand \@ifx [1]{%
 \ifx #1\expandafter \@firstoftwo
 \else \expandafter \@secondoftwo
 \fi
}%
\providecommand \natexlab [1]{#1}%
\providecommand \enquote  [1]{``#1''}%
\providecommand \bibnamefont  [1]{#1}%
\providecommand \bibfnamefont [1]{#1}%
\providecommand \citenamefont [1]{#1}%
\providecommand \href@noop [0]{\@secondoftwo}%
\providecommand \href [0]{\begingroup \@sanitize@url \@href}%
\providecommand \@href[1]{\@@startlink{#1}\@@href}%
\providecommand \@@href[1]{\endgroup#1\@@endlink}%
\providecommand \@sanitize@url [0]{\catcode `\\12\catcode `\$12\catcode
  `\&12\catcode `\#12\catcode `\^12\catcode `\_12\catcode `\%12\relax}%
\providecommand \@@startlink[1]{}%
\providecommand \@@endlink[0]{}%
\providecommand \url  [0]{\begingroup\@sanitize@url \@url }%
\providecommand \@url [1]{\endgroup\@href {#1}{\urlprefix }}%
\providecommand \urlprefix  [0]{URL }%
\providecommand \Eprint [0]{\href }%
\providecommand \doibase [0]{http://dx.doi.org/}%
\providecommand \selectlanguage [0]{\@gobble}%
\providecommand \bibinfo  [0]{\@secondoftwo}%
\providecommand \bibfield  [0]{\@secondoftwo}%
\providecommand \translation [1]{[#1]}%
\providecommand \BibitemOpen [0]{}%
\providecommand \bibitemStop [0]{}%
\providecommand \bibitemNoStop [0]{.\EOS\space}%
\providecommand \EOS [0]{\spacefactor3000\relax}%
\providecommand \BibitemShut  [1]{\csname bibitem#1\endcsname}%
\let\auto@bib@innerbib\@empty
\bibitem [{\citenamefont {Ajello}\ \emph {et~al.}(2015)\citenamefont {Ajello}
  \emph {et~al.}}]{Ajello:2015mfa}%
  \BibitemOpen
  \bibfield  {author} {\bibinfo {author} {\bibfnamefont {M.}~\bibnamefont
  {Ajello}} \emph {et~al.},\ }\href {\doibase 10.1088/2041-8205/800/2/L27}
  {\bibfield  {journal} {\bibinfo  {journal} {Astrophys. J. Lett.}\ }\textbf
  {\bibinfo {volume} {800}},\ \bibinfo {pages} {L27} (\bibinfo {year}
  {2015})},\ \Eprint {http://arxiv.org/abs/1501.05301} {arXiv:1501.05301
  [astro-ph.HE]} \BibitemShut {NoStop}%
\bibitem [{\citenamefont {Lisanti}\ \emph {et~al.}(2016)\citenamefont
  {Lisanti}, \citenamefont {Mishra-Sharma}, \citenamefont {Necib},\ and\
  \citenamefont {Safdi}}]{Lisanti:2016jub}%
  \BibitemOpen
  \bibfield  {author} {\bibinfo {author} {\bibfnamefont {M.}~\bibnamefont
  {Lisanti}}, \bibinfo {author} {\bibfnamefont {S.}~\bibnamefont
  {Mishra-Sharma}}, \bibinfo {author} {\bibfnamefont {L.}~\bibnamefont
  {Necib}}, \ and\ \bibinfo {author} {\bibfnamefont {B.~R.}\ \bibnamefont
  {Safdi}},\ }\href {\doibase 10.3847/0004-637X/832/2/117} {\bibfield
  {journal} {\bibinfo  {journal} {Astrophys. J.}\ }\textbf {\bibinfo {volume}
  {832}},\ \bibinfo {pages} {117} (\bibinfo {year} {2016})},\ \Eprint
  {http://arxiv.org/abs/1606.04101} {arXiv:1606.04101 [astro-ph.HE]}
  \BibitemShut {NoStop}%
\bibitem [{\citenamefont {Di~Mauro}\ \emph {et~al.}(2018)\citenamefont
  {Di~Mauro}, \citenamefont {Manconi}, \citenamefont {Zechlin}, \citenamefont
  {Ajello}, \citenamefont {Charles},\ and\ \citenamefont
  {Donato}}]{DiMauro:2017ing}%
  \BibitemOpen
  \bibfield  {author} {\bibinfo {author} {\bibfnamefont {M.}~\bibnamefont
  {Di~Mauro}}, \bibinfo {author} {\bibfnamefont {S.}~\bibnamefont {Manconi}},
  \bibinfo {author} {\bibfnamefont {H.~S.}\ \bibnamefont {Zechlin}}, \bibinfo
  {author} {\bibfnamefont {M.}~\bibnamefont {Ajello}}, \bibinfo {author}
  {\bibfnamefont {E.}~\bibnamefont {Charles}}, \ and\ \bibinfo {author}
  {\bibfnamefont {F.}~\bibnamefont {Donato}},\ }\href {\doibase
  10.3847/1538-4357/aab3e5} {\bibfield  {journal} {\bibinfo  {journal}
  {Astrophys. J.}\ }\textbf {\bibinfo {volume} {856}},\ \bibinfo {pages} {106}
  (\bibinfo {year} {2018})},\ \Eprint {http://arxiv.org/abs/1711.03111}
  {arXiv:1711.03111 [astro-ph.HE]} \BibitemShut {NoStop}%
\bibitem [{\citenamefont {Manconi}\ \emph {et~al.}(2020)\citenamefont
  {Manconi}, \citenamefont {Korsmeier}, \citenamefont {Donato}, \citenamefont
  {Fornengo}, \citenamefont {Regis},\ and\ \citenamefont
  {Zechlin}}]{Manconi:2019ynl}%
  \BibitemOpen
  \bibfield  {author} {\bibinfo {author} {\bibfnamefont {S.}~\bibnamefont
  {Manconi}}, \bibinfo {author} {\bibfnamefont {M.}~\bibnamefont {Korsmeier}},
  \bibinfo {author} {\bibfnamefont {F.}~\bibnamefont {Donato}}, \bibinfo
  {author} {\bibfnamefont {N.}~\bibnamefont {Fornengo}}, \bibinfo {author}
  {\bibfnamefont {M.}~\bibnamefont {Regis}}, \ and\ \bibinfo {author}
  {\bibfnamefont {H.}~\bibnamefont {Zechlin}},\ }\href {\doibase
  10.1103/PhysRevD.101.103026} {\bibfield  {journal} {\bibinfo  {journal}
  {Phys. Rev. D}\ }\textbf {\bibinfo {volume} {101}},\ \bibinfo {pages}
  {103026} (\bibinfo {year} {2020})},\ \Eprint
  {http://arxiv.org/abs/1912.01622} {arXiv:1912.01622 [astro-ph.HE]}
  \BibitemShut {NoStop}%
\bibitem [{\citenamefont {{Marcotulli}}\ \emph {et~al.}(2020)\citenamefont
  {{Marcotulli}}, \citenamefont {{Di Mauro}},\ and\ \citenamefont
  {{Ajello}}}]{2020ApJ...896....6M}%
  \BibitemOpen
  \bibfield  {author} {\bibinfo {author} {\bibfnamefont {L.}~\bibnamefont
  {{Marcotulli}}}, \bibinfo {author} {\bibfnamefont {M.}~\bibnamefont {{Di
  Mauro}}}, \ and\ \bibinfo {author} {\bibfnamefont {M.}~\bibnamefont
  {{Ajello}}},\ }\href {\doibase 10.3847/1538-4357/ab8cbd} {\bibfield
  {journal} {\bibinfo  {journal} {\apj}\ }\textbf {\bibinfo {volume} {896}},\
  \bibinfo {eid} {6} (\bibinfo {year} {2020})},\ \Eprint
  {http://arxiv.org/abs/2006.04703} {arXiv:2006.04703 [astro-ph.HE]}
  \BibitemShut {NoStop}%
\bibitem [{\citenamefont {Ackermann}\ \emph
  {et~al.}(2012{\natexlab{a}})\citenamefont {Ackermann} \emph
  {et~al.}}]{Ackermann:2012uf}%
  \BibitemOpen
  \bibfield  {author} {\bibinfo {author} {\bibfnamefont {M.}~\bibnamefont
  {Ackermann}} \emph {et~al.} (\bibinfo {collaboration} {Fermi-LAT}),\ }\href
  {\doibase 10.1103/PhysRevD.85.083007} {\bibfield  {journal} {\bibinfo
  {journal} {Phys. Rev. D}\ }\textbf {\bibinfo {volume} {85}},\ \bibinfo
  {pages} {083007} (\bibinfo {year} {2012}{\natexlab{a}})},\ \Eprint
  {http://arxiv.org/abs/1202.2856} {arXiv:1202.2856 [astro-ph.HE]} \BibitemShut
  {NoStop}%
\bibitem [{\citenamefont {Di~Mauro}\ \emph {et~al.}(2014)\citenamefont
  {Di~Mauro}, \citenamefont {Calore}, \citenamefont {Donato}, \citenamefont
  {Ajello},\ and\ \citenamefont {Latronico}}]{dimauro:2013xta}%
  \BibitemOpen
  \bibfield  {author} {\bibinfo {author} {\bibfnamefont {M.}~\bibnamefont
  {Di~Mauro}}, \bibinfo {author} {\bibfnamefont {F.}~\bibnamefont {Calore}},
  \bibinfo {author} {\bibfnamefont {F.}~\bibnamefont {Donato}}, \bibinfo
  {author} {\bibfnamefont {M.}~\bibnamefont {Ajello}}, \ and\ \bibinfo {author}
  {\bibfnamefont {L.}~\bibnamefont {Latronico}},\ }\href {\doibase
  10.1088/0004-637X/780/2/161} {\bibfield  {journal} {\bibinfo  {journal}
  {Astrophys. J.}\ }\textbf {\bibinfo {volume} {780}},\ \bibinfo {pages} {161}
  (\bibinfo {year} {2014})},\ \Eprint {http://arxiv.org/abs/1304.0908}
  {arXiv:1304.0908 [astro-ph.HE]} \BibitemShut {NoStop}%
\bibitem [{\citenamefont {Ackermann}\ \emph {et~al.}(2015)\citenamefont
  {Ackermann} \emph {et~al.}}]{Ackermann:2014usa}%
  \BibitemOpen
  \bibfield  {author} {\bibinfo {author} {\bibfnamefont {M.}~\bibnamefont
  {Ackermann}} \emph {et~al.} (\bibinfo {collaboration} {Fermi-LAT}),\ }\href
  {\doibase 10.1088/0004-637X/799/1/86} {\bibfield  {journal} {\bibinfo
  {journal} {Astrophys. J.}\ }\textbf {\bibinfo {volume} {799}},\ \bibinfo
  {pages} {86} (\bibinfo {year} {2015})},\ \Eprint
  {http://arxiv.org/abs/1410.3696} {arXiv:1410.3696 [astro-ph.HE]} \BibitemShut
  {NoStop}%
\bibitem [{\citenamefont {Hooper}\ \emph {et~al.}(2016)\citenamefont {Hooper},
  \citenamefont {Linden},\ and\ \citenamefont {Lopez}}]{Hooper:2016gjy}%
  \BibitemOpen
  \bibfield  {author} {\bibinfo {author} {\bibfnamefont {D.}~\bibnamefont
  {Hooper}}, \bibinfo {author} {\bibfnamefont {T.}~\bibnamefont {Linden}}, \
  and\ \bibinfo {author} {\bibfnamefont {A.}~\bibnamefont {Lopez}},\ }\href
  {\doibase 10.1088/1475-7516/2016/08/019} {\bibfield  {journal} {\bibinfo
  {journal} {JCAP}\ }\textbf {\bibinfo {volume} {08}},\ \bibinfo {pages} {019}
  (\bibinfo {year} {2016})},\ \Eprint {http://arxiv.org/abs/1604.08505}
  {arXiv:1604.08505 [astro-ph.HE]} \BibitemShut {NoStop}%
\bibitem [{\citenamefont {Linden}(2017)}]{Linden:2016fdd}%
  \BibitemOpen
  \bibfield  {author} {\bibinfo {author} {\bibfnamefont {T.}~\bibnamefont
  {Linden}},\ }\href {\doibase 10.1103/PhysRevD.96.083001} {\bibfield
  {journal} {\bibinfo  {journal} {Phys. Rev. D}\ }\textbf {\bibinfo {volume}
  {96}},\ \bibinfo {pages} {083001} (\bibinfo {year} {2017})},\ \Eprint
  {http://arxiv.org/abs/1612.03175} {arXiv:1612.03175 [astro-ph.HE]}
  \BibitemShut {NoStop}%
\bibitem [{\citenamefont {Blanco}\ and\ \citenamefont
  {Hooper}(2017)}]{Blanco2017}%
  \BibitemOpen
  \bibfield  {author} {\bibinfo {author} {\bibfnamefont {C.}~\bibnamefont
  {Blanco}}\ and\ \bibinfo {author} {\bibfnamefont {D.}~\bibnamefont
  {Hooper}},\ }\href {\doibase 10.1088/1475-7516/2017/12/017} {\  (\bibinfo
  {year} {2017}),\ 10.1088/1475-7516/2017/12/017},\ \Eprint
  {http://arxiv.org/abs/1706.07047} {arXiv:1706.07047 [astro-ph.HE]}
  \BibitemShut {NoStop}%
\bibitem [{\citenamefont {Komis}\ \emph {et~al.}(2019)\citenamefont {Komis},
  \citenamefont {Pavlidou},\ and\ \citenamefont {Zezas}}]{Komis:2017jta}%
  \BibitemOpen
  \bibfield  {author} {\bibinfo {author} {\bibfnamefont {I.}~\bibnamefont
  {Komis}}, \bibinfo {author} {\bibfnamefont {V.}~\bibnamefont {Pavlidou}}, \
  and\ \bibinfo {author} {\bibfnamefont {A.}~\bibnamefont {Zezas}},\ }\href
  {\doibase 10.1093/mnras/sty3354} {\bibfield  {journal} {\bibinfo  {journal}
  {Mon. Not. Roy. Astron. Soc.}\ }\textbf {\bibinfo {volume} {483}},\ \bibinfo
  {pages} {4020} (\bibinfo {year} {2019})},\ \Eprint
  {http://arxiv.org/abs/1711.11046} {arXiv:1711.11046 [astro-ph.HE]}
  \BibitemShut {NoStop}%
\bibitem [{\citenamefont {Stecker}\ \emph {et~al.}(2019)\citenamefont
  {Stecker}, \citenamefont {Shrader},\ and\ \citenamefont
  {Malkan}}]{Stecker:2019ybn}%
  \BibitemOpen
  \bibfield  {author} {\bibinfo {author} {\bibfnamefont {F.~W.}\ \bibnamefont
  {Stecker}}, \bibinfo {author} {\bibfnamefont {C.~R.}\ \bibnamefont
  {Shrader}}, \ and\ \bibinfo {author} {\bibfnamefont {M.~A.}\ \bibnamefont
  {Malkan}},\ }\href {\doibase 10.3847/1538-4357/ab23ee} {\bibfield  {journal}
  {\bibinfo  {journal} {Astrophys. J.}\ }\textbf {\bibinfo {volume} {879}},\
  \bibinfo {pages} {68} (\bibinfo {year} {2019})},\ \Eprint
  {http://arxiv.org/abs/1903.06544} {arXiv:1903.06544 [astro-ph.GA]}
  \BibitemShut {NoStop}%
\bibitem [{\citenamefont {Zechlin}\ \emph
  {et~al.}(2016{\natexlab{a}})\citenamefont {Zechlin}, \citenamefont {Cuoco},
  \citenamefont {Donato}, \citenamefont {Fornengo},\ and\ \citenamefont
  {Vittino}}]{Zechlin:2015wdz}%
  \BibitemOpen
  \bibfield  {author} {\bibinfo {author} {\bibfnamefont {H.-S.}\ \bibnamefont
  {Zechlin}}, \bibinfo {author} {\bibfnamefont {A.}~\bibnamefont {Cuoco}},
  \bibinfo {author} {\bibfnamefont {F.}~\bibnamefont {Donato}}, \bibinfo
  {author} {\bibfnamefont {N.}~\bibnamefont {Fornengo}}, \ and\ \bibinfo
  {author} {\bibfnamefont {A.}~\bibnamefont {Vittino}},\ }\href {\doibase
  10.3847/0067-0049/225/2/18} {\bibfield  {journal} {\bibinfo  {journal}
  {Astrophys. J. Suppl.}\ }\textbf {\bibinfo {volume} {225}},\ \bibinfo {pages}
  {18} (\bibinfo {year} {2016}{\natexlab{a}})},\ \Eprint
  {http://arxiv.org/abs/1512.07190} {arXiv:1512.07190 [astro-ph.HE]}
  \BibitemShut {NoStop}%
\bibitem [{\citenamefont {Fornasa}\ \emph {et~al.}(2016)\citenamefont {Fornasa}
  \emph {et~al.}}]{Fornasa:2016ohl}%
  \BibitemOpen
  \bibfield  {author} {\bibinfo {author} {\bibfnamefont {M.}~\bibnamefont
  {Fornasa}} \emph {et~al.},\ }\href {\doibase 10.1103/PhysRevD.94.123005}
  {\bibfield  {journal} {\bibinfo  {journal} {Phys. Rev. D}\ }\textbf {\bibinfo
  {volume} {94}},\ \bibinfo {pages} {123005} (\bibinfo {year} {2016})},\
  \Eprint {http://arxiv.org/abs/1608.07289} {arXiv:1608.07289 [astro-ph.HE]}
  \BibitemShut {NoStop}%
\bibitem [{\citenamefont {Zechlin}\ \emph
  {et~al.}(2016{\natexlab{b}})\citenamefont {Zechlin}, \citenamefont {Cuoco},
  \citenamefont {Donato}, \citenamefont {Fornengo},\ and\ \citenamefont
  {Regis}}]{Zechlin:2016pme}%
  \BibitemOpen
  \bibfield  {author} {\bibinfo {author} {\bibfnamefont {H.-S.}\ \bibnamefont
  {Zechlin}}, \bibinfo {author} {\bibfnamefont {A.}~\bibnamefont {Cuoco}},
  \bibinfo {author} {\bibfnamefont {F.}~\bibnamefont {Donato}}, \bibinfo
  {author} {\bibfnamefont {N.}~\bibnamefont {Fornengo}}, \ and\ \bibinfo
  {author} {\bibfnamefont {M.}~\bibnamefont {Regis}},\ }\href {\doibase
  10.3847/2041-8205/826/2/L31} {\bibfield  {journal} {\bibinfo  {journal}
  {Astrophys. J. Lett.}\ }\textbf {\bibinfo {volume} {826}},\ \bibinfo {pages}
  {L31} (\bibinfo {year} {2016}{\natexlab{b}})},\ \Eprint
  {http://arxiv.org/abs/1605.04256} {arXiv:1605.04256 [astro-ph.HE]}
  \BibitemShut {NoStop}%
\bibitem [{\citenamefont {Ando}\ \emph {et~al.}(2017)\citenamefont {Ando},
  \citenamefont {Fornasa}, \citenamefont {Fornengo}, \citenamefont {Regis},\
  and\ \citenamefont {Zechlin}}]{Ando:2017alx}%
  \BibitemOpen
  \bibfield  {author} {\bibinfo {author} {\bibfnamefont {S.}~\bibnamefont
  {Ando}}, \bibinfo {author} {\bibfnamefont {M.}~\bibnamefont {Fornasa}},
  \bibinfo {author} {\bibfnamefont {N.}~\bibnamefont {Fornengo}}, \bibinfo
  {author} {\bibfnamefont {M.}~\bibnamefont {Regis}}, \ and\ \bibinfo {author}
  {\bibfnamefont {H.-S.}\ \bibnamefont {Zechlin}},\ }\href {\doibase
  10.1103/PhysRevD.95.123006} {\bibfield  {journal} {\bibinfo  {journal} {Phys.
  Rev. D}\ }\textbf {\bibinfo {volume} {95}},\ \bibinfo {pages} {123006}
  (\bibinfo {year} {2017})},\ \Eprint {http://arxiv.org/abs/1701.06988}
  {arXiv:1701.06988 [astro-ph.HE]} \BibitemShut {NoStop}%
\bibitem [{\citenamefont {Ackermann}\ \emph {et~al.}(2018)\citenamefont
  {Ackermann} \emph {et~al.}}]{Ackermann:2018wlo}%
  \BibitemOpen
  \bibfield  {author} {\bibinfo {author} {\bibfnamefont {M.}~\bibnamefont
  {Ackermann}} \emph {et~al.} (\bibinfo {collaboration} {Fermi-LAT}),\ }\href
  {\doibase 10.1103/PhysRevLett.121.241101} {\bibfield  {journal} {\bibinfo
  {journal} {Phys. Rev. Lett.}\ }\textbf {\bibinfo {volume} {121}},\ \bibinfo
  {pages} {241101} (\bibinfo {year} {2018})},\ \Eprint
  {http://arxiv.org/abs/1812.02079} {arXiv:1812.02079 [astro-ph.HE]}
  \BibitemShut {NoStop}%
\bibitem [{\citenamefont {Socrates}\ \emph {et~al.}(2008)\citenamefont
  {Socrates}, \citenamefont {Davis},\ and\ \citenamefont
  {Ramirez-Ruiz}}]{Socrates:2006dv}%
  \BibitemOpen
  \bibfield  {author} {\bibinfo {author} {\bibfnamefont {A.}~\bibnamefont
  {Socrates}}, \bibinfo {author} {\bibfnamefont {S.~W.}\ \bibnamefont {Davis}},
  \ and\ \bibinfo {author} {\bibfnamefont {E.}~\bibnamefont {Ramirez-Ruiz}},\
  }\href {\doibase 10.1086/590046} {\bibfield  {journal} {\bibinfo  {journal}
  {Astrophys. J.}\ }\textbf {\bibinfo {volume} {687}},\ \bibinfo {pages} {202}
  (\bibinfo {year} {2008})},\ \Eprint {http://arxiv.org/abs/astro-ph/0609796}
  {arXiv:astro-ph/0609796} \BibitemShut {NoStop}%
\bibitem [{\citenamefont {Kormendy}\ and\ \citenamefont
  {Ho}(2013)}]{Kormendy:2013dxa}%
  \BibitemOpen
  \bibfield  {author} {\bibinfo {author} {\bibfnamefont {J.}~\bibnamefont
  {Kormendy}}\ and\ \bibinfo {author} {\bibfnamefont {L.~C.}\ \bibnamefont
  {Ho}},\ }\href {\doibase 10.1146/annurev-astro-082708-101811} {\bibfield
  {journal} {\bibinfo  {journal} {Ann. Rev. Astron. Astrophys.}\ }\textbf
  {\bibinfo {volume} {51}},\ \bibinfo {pages} {511} (\bibinfo {year} {2013})},\
  \Eprint {http://arxiv.org/abs/1304.7762} {arXiv:1304.7762 [astro-ph.CO]}
  \BibitemShut {NoStop}%
\bibitem [{\citenamefont {{Sijacki}}\ \emph {et~al.}(2015)\citenamefont
  {{Sijacki}}, \citenamefont {{Vogelsberger}}, \citenamefont {{Genel}},
  \citenamefont {{Springel}}, \citenamefont {{Torrey}}, \citenamefont
  {{Snyder}}, \citenamefont {{Nelson}},\ and\ \citenamefont
  {{Hernquist}}}]{2015MNRAS.452..575S}%
  \BibitemOpen
  \bibfield  {author} {\bibinfo {author} {\bibfnamefont {D.}~\bibnamefont
  {{Sijacki}}}, \bibinfo {author} {\bibfnamefont {M.}~\bibnamefont
  {{Vogelsberger}}}, \bibinfo {author} {\bibfnamefont {S.}~\bibnamefont
  {{Genel}}}, \bibinfo {author} {\bibfnamefont {V.}~\bibnamefont {{Springel}}},
  \bibinfo {author} {\bibfnamefont {P.}~\bibnamefont {{Torrey}}}, \bibinfo
  {author} {\bibfnamefont {G.~F.}\ \bibnamefont {{Snyder}}}, \bibinfo {author}
  {\bibfnamefont {D.}~\bibnamefont {{Nelson}}}, \ and\ \bibinfo {author}
  {\bibfnamefont {L.}~\bibnamefont {{Hernquist}}},\ }\href {\doibase
  10.1093/mnras/stv1340} {\bibfield  {journal} {\bibinfo  {journal} {\mnras}\
  }\textbf {\bibinfo {volume} {452}},\ \bibinfo {pages} {575} (\bibinfo {year}
  {2015})},\ \Eprint {http://arxiv.org/abs/1408.6842} {arXiv:1408.6842
  [astro-ph.GA]} \BibitemShut {NoStop}%
\bibitem [{\citenamefont {Pakmor}\ \emph {et~al.}(2016)\citenamefont {Pakmor},
  \citenamefont {Pfrommer}, \citenamefont {Simpson},\ and\ \citenamefont
  {Springel}}]{Pakmor:2016zbu}%
  \BibitemOpen
  \bibfield  {author} {\bibinfo {author} {\bibfnamefont {R.}~\bibnamefont
  {Pakmor}}, \bibinfo {author} {\bibfnamefont {C.}~\bibnamefont {Pfrommer}},
  \bibinfo {author} {\bibfnamefont {C.~M.}\ \bibnamefont {Simpson}}, \ and\
  \bibinfo {author} {\bibfnamefont {V.}~\bibnamefont {Springel}},\ }\href
  {\doibase 10.3847/2041-8205/824/2/L30} {\bibfield  {journal} {\bibinfo
  {journal} {Astrophys. J. Lett.}\ }\textbf {\bibinfo {volume} {824}},\
  \bibinfo {pages} {L30} (\bibinfo {year} {2016})},\ \Eprint
  {http://arxiv.org/abs/1605.00643} {arXiv:1605.00643 [astro-ph.GA]}
  \BibitemShut {NoStop}%
\bibitem [{\citenamefont {Wiener}\ \emph {et~al.}(2017)\citenamefont {Wiener},
  \citenamefont {Pfrommer},\ and\ \citenamefont {Oh}}]{Wiener:2016zcr}%
  \BibitemOpen
  \bibfield  {author} {\bibinfo {author} {\bibfnamefont {J.}~\bibnamefont
  {Wiener}}, \bibinfo {author} {\bibfnamefont {C.}~\bibnamefont {Pfrommer}}, \
  and\ \bibinfo {author} {\bibfnamefont {P.}~\bibnamefont {Oh}},\ }\href
  {\doibase 10.1093/mnras/stx127} {\bibfield  {journal} {\bibinfo  {journal}
  {Mon. Not. Roy. Astron. Soc.}\ }\textbf {\bibinfo {volume} {467}},\ \bibinfo
  {pages} {906} (\bibinfo {year} {2017})},\ \Eprint
  {http://arxiv.org/abs/1608.02585} {arXiv:1608.02585 [astro-ph.GA]}
  \BibitemShut {NoStop}%
\bibitem [{\citenamefont {{Lanzuisi}}\ \emph {et~al.}(2017)\citenamefont
  {{Lanzuisi}}, \citenamefont {{Delvecchio}}, \citenamefont {{Berta}},
  \citenamefont {{Brusa}}, \citenamefont {{Comastri}}, \citenamefont {{Gilli}},
  \citenamefont {{Gruppioni}}, \citenamefont {{Marchesi}}, \citenamefont
  {{Perna}}, \citenamefont {{Pozzi}}, \citenamefont {{Salvato}}, \citenamefont
  {{Symeonidis}}, \citenamefont {{Vignali}}, \citenamefont {{Vito}},
  \citenamefont {{Volonteri}},\ and\ \citenamefont
  {{Zamorani}}}]{2017A&A...602A.123L}%
  \BibitemOpen
  \bibfield  {author} {\bibinfo {author} {\bibfnamefont {G.}~\bibnamefont
  {{Lanzuisi}}}, \bibinfo {author} {\bibfnamefont {I.}~\bibnamefont
  {{Delvecchio}}}, \bibinfo {author} {\bibfnamefont {S.}~\bibnamefont
  {{Berta}}}, \bibinfo {author} {\bibfnamefont {M.}~\bibnamefont {{Brusa}}},
  \bibinfo {author} {\bibfnamefont {A.}~\bibnamefont {{Comastri}}}, \bibinfo
  {author} {\bibfnamefont {R.}~\bibnamefont {{Gilli}}}, \bibinfo {author}
  {\bibfnamefont {C.}~\bibnamefont {{Gruppioni}}}, \bibinfo {author}
  {\bibfnamefont {S.}~\bibnamefont {{Marchesi}}}, \bibinfo {author}
  {\bibfnamefont {M.}~\bibnamefont {{Perna}}}, \bibinfo {author} {\bibfnamefont
  {F.}~\bibnamefont {{Pozzi}}}, \bibinfo {author} {\bibfnamefont
  {M.}~\bibnamefont {{Salvato}}}, \bibinfo {author} {\bibfnamefont
  {M.}~\bibnamefont {{Symeonidis}}}, \bibinfo {author} {\bibfnamefont
  {C.}~\bibnamefont {{Vignali}}}, \bibinfo {author} {\bibfnamefont
  {F.}~\bibnamefont {{Vito}}}, \bibinfo {author} {\bibfnamefont
  {M.}~\bibnamefont {{Volonteri}}}, \ and\ \bibinfo {author} {\bibfnamefont
  {G.}~\bibnamefont {{Zamorani}}},\ }\href {\doibase
  10.1051/0004-6361/201629955} {\bibfield  {journal} {\bibinfo  {journal}
  {\aap}\ }\textbf {\bibinfo {volume} {602}},\ \bibinfo {eid} {A123} (\bibinfo
  {year} {2017})},\ \Eprint {http://arxiv.org/abs/1702.07357} {arXiv:1702.07357
  [astro-ph.GA]} \BibitemShut {NoStop}%
\bibitem [{\citenamefont {Ackermann}\ \emph
  {et~al.}(2012{\natexlab{b}})\citenamefont {Ackermann}, \citenamefont
  {Ajello}, \citenamefont {Allafort}, \citenamefont {Baldini}, \citenamefont
  {Ballet}, \citenamefont {Bastieri}, \citenamefont {Bechtol}, \citenamefont
  {Bellazzini},\ and\ \citenamefont {et~al.}}]{2012ApJ...755..164A}%
  \BibitemOpen
  \bibfield  {author} {\bibinfo {author} {\bibfnamefont {M.}~\bibnamefont
  {Ackermann}}, \bibinfo {author} {\bibfnamefont {M.}~\bibnamefont {Ajello}},
  \bibinfo {author} {\bibfnamefont {A.}~\bibnamefont {Allafort}}, \bibinfo
  {author} {\bibfnamefont {L.}~\bibnamefont {Baldini}}, \bibinfo {author}
  {\bibfnamefont {J.}~\bibnamefont {Ballet}}, \bibinfo {author} {\bibfnamefont
  {D.}~\bibnamefont {Bastieri}}, \bibinfo {author} {\bibfnamefont
  {K.}~\bibnamefont {Bechtol}}, \bibinfo {author} {\bibfnamefont
  {R.}~\bibnamefont {Bellazzini}}, \ and\ \bibinfo {author} {\bibnamefont
  {et~al.}},\ }\href {\doibase 10.1088/0004-637X/755/2/164} {\bibfield
  {journal} {\bibinfo  {journal} {\apj}\ }\textbf {\bibinfo {volume} {755}},\
  \bibinfo {eid} {164} (\bibinfo {year} {2012}{\natexlab{b}})},\ \Eprint
  {http://arxiv.org/abs/1206.1346} {arXiv:1206.1346 [astro-ph.HE]} \BibitemShut
  {NoStop}%
\bibitem [{\citenamefont {{Kornecki}}\ \emph {et~al.}(2020)\citenamefont
  {{Kornecki}}, \citenamefont {{Pellizza}}, \citenamefont {{del Palacio}},
  \citenamefont {{M{\"u}ller}}, \citenamefont {{Albacete-Colombo}},\ and\
  \citenamefont {{Romero}}}]{2020A&A...641A.147K}%
  \BibitemOpen
  \bibfield  {author} {\bibinfo {author} {\bibfnamefont {P.}~\bibnamefont
  {{Kornecki}}}, \bibinfo {author} {\bibfnamefont {L.~J.}\ \bibnamefont
  {{Pellizza}}}, \bibinfo {author} {\bibfnamefont {S.}~\bibnamefont {{del
  Palacio}}}, \bibinfo {author} {\bibfnamefont {A.~L.}\ \bibnamefont
  {{M{\"u}ller}}}, \bibinfo {author} {\bibfnamefont {J.~F.}\ \bibnamefont
  {{Albacete-Colombo}}}, \ and\ \bibinfo {author} {\bibfnamefont {G.~E.}\
  \bibnamefont {{Romero}}},\ }\href {\doibase 10.1051/0004-6361/202038428}
  {\bibfield  {journal} {\bibinfo  {journal} {\aap}\ }\textbf {\bibinfo
  {volume} {641}},\ \bibinfo {eid} {A147} (\bibinfo {year} {2020})},\ \Eprint
  {http://arxiv.org/abs/2007.07430} {arXiv:2007.07430 [astro-ph.HE]}
  \BibitemShut {NoStop}%
\bibitem [{\citenamefont {Ajello}\ \emph {et~al.}(2020)\citenamefont {Ajello},
  \citenamefont {Di~Mauro}, \citenamefont {Paliya},\ and\ \citenamefont
  {Garrappa}}]{Ajello:2020zna}%
  \BibitemOpen
  \bibfield  {author} {\bibinfo {author} {\bibfnamefont {M.}~\bibnamefont
  {Ajello}}, \bibinfo {author} {\bibfnamefont {M.}~\bibnamefont {Di~Mauro}},
  \bibinfo {author} {\bibfnamefont {V.~S.}\ \bibnamefont {Paliya}}, \ and\
  \bibinfo {author} {\bibfnamefont {S.}~\bibnamefont {Garrappa}},\ }\href
  {\doibase 10.3847/1538-4357/ab86a6} {\bibfield  {journal} {\bibinfo
  {journal} {Astrophys. J.}\ }\textbf {\bibinfo {volume} {894}},\ \bibinfo
  {pages} {88} (\bibinfo {year} {2020})},\ \Eprint
  {http://arxiv.org/abs/2003.05493} {arXiv:2003.05493 [astro-ph.GA]}
  \BibitemShut {NoStop}%
\bibitem [{\citenamefont {Laurent-Muehleisen}\ \emph
  {et~al.}(1996)\citenamefont {Laurent-Muehleisen}, \citenamefont {Kollgaard},
  \citenamefont {Ryan}, \citenamefont {Feigelson}, \citenamefont {Brinkmann},\
  and\ \citenamefont {Siebert}}]{LaurentMuehleisen1996}%
  \BibitemOpen
  \bibfield  {author} {\bibinfo {author} {\bibfnamefont {S.~A.}\ \bibnamefont
  {Laurent-Muehleisen}}, \bibinfo {author} {\bibfnamefont {R.~I.}\ \bibnamefont
  {Kollgaard}}, \bibinfo {author} {\bibfnamefont {P.~J.}\ \bibnamefont {Ryan}},
  \bibinfo {author} {\bibfnamefont {E.~D.}\ \bibnamefont {Feigelson}}, \bibinfo
  {author} {\bibfnamefont {W.}~\bibnamefont {Brinkmann}}, \ and\ \bibinfo
  {author} {\bibfnamefont {J.}~\bibnamefont {Siebert}},\ }\href {\doibase
  10.1051/aas:1997331} {\  (\bibinfo {year} {1996}),\ 10.1051/aas:1997331},\
  \Eprint {http://arxiv.org/abs/astro-ph/9607058} {arXiv:astro-ph/9607058
  [astro-ph]} \BibitemShut {NoStop}%
\bibitem [{\citenamefont {Yuan}\ and\ \citenamefont {Wang}(2011)}]{Yuan_2011}%
  \BibitemOpen
  \bibfield  {author} {\bibinfo {author} {\bibfnamefont {Z.}~\bibnamefont
  {Yuan}}\ and\ \bibinfo {author} {\bibfnamefont {J.}~\bibnamefont {Wang}},\
  }\href {\doibase 10.1088/0004-637x/744/2/84} {\bibfield  {journal} {\bibinfo
  {journal} {The Astrophysical Journal}\ }\textbf {\bibinfo {volume} {744}},\
  \bibinfo {pages} {84} (\bibinfo {year} {2011})}\BibitemShut {NoStop}%
\bibitem [{\citenamefont {Yuan}\ \emph {et~al.}(2018)\citenamefont {Yuan},
  \citenamefont {Wang}, \citenamefont {Worrall}, \citenamefont {Zhang},\ and\
  \citenamefont {Mao}}]{Yuan2018}%
  \BibitemOpen
  \bibfield  {author} {\bibinfo {author} {\bibfnamefont {Z.}~\bibnamefont
  {Yuan}}, \bibinfo {author} {\bibfnamefont {J.}~\bibnamefont {Wang}}, \bibinfo
  {author} {\bibfnamefont {D.~M.}\ \bibnamefont {Worrall}}, \bibinfo {author}
  {\bibfnamefont {B.-B.}\ \bibnamefont {Zhang}}, \ and\ \bibinfo {author}
  {\bibfnamefont {J.}~\bibnamefont {Mao}},\ }\href {\doibase
  10.3847/1538-4365/aaed3b} {\bibfield  {journal} {\bibinfo  {journal} {The
  Astrophysical Journal Supplement Series}\ }\textbf {\bibinfo {volume}
  {239}},\ \bibinfo {pages} {33} (\bibinfo {year} {2018})}\BibitemShut
  {NoStop}%
\bibitem [{198(1988)}]{1988iras....7.....H}%
  \BibitemOpen
  \href@noop {} {\emph {\bibinfo {title} {Infrared astronomical satellite
  (IRAS) catalogs and atlases. Volume 7}}},\ Vol.~\bibinfo {volume} {7}\
  (\bibinfo {year} {1988})\BibitemShut {NoStop}%
\bibitem [{\citenamefont {Ahnen}\ \emph {et~al.}(2016)\citenamefont {Ahnen},
  \citenamefont {Ansoldi}, \citenamefont {Antonelli}, \citenamefont {Antoranz},
  \citenamefont {Babic}, \citenamefont {Banerjee}, \citenamefont {Bangale},
  \citenamefont {de~Almeida}, \citenamefont {Barrio}, \citenamefont
  {Gonz{\'{a}}lez} \emph {et~al.}}]{Ahnen2016}%
  \BibitemOpen
  \bibfield  {author} {\bibinfo {author} {\bibfnamefont {M.~L.}\ \bibnamefont
  {Ahnen}}, \bibinfo {author} {\bibfnamefont {S.}~\bibnamefont {Ansoldi}},
  \bibinfo {author} {\bibfnamefont {L.~A.}\ \bibnamefont {Antonelli}}, \bibinfo
  {author} {\bibfnamefont {P.}~\bibnamefont {Antoranz}}, \bibinfo {author}
  {\bibfnamefont {A.}~\bibnamefont {Babic}}, \bibinfo {author} {\bibfnamefont
  {B.}~\bibnamefont {Banerjee}}, \bibinfo {author} {\bibfnamefont
  {P.}~\bibnamefont {Bangale}}, \bibinfo {author} {\bibfnamefont {U.~B.}\
  \bibnamefont {de~Almeida}}, \bibinfo {author} {\bibfnamefont {J.~A.}\
  \bibnamefont {Barrio}}, \bibinfo {author} {\bibfnamefont {J.~B.}\
  \bibnamefont {Gonz{\'{a}}lez}},  \emph {et~al.},\ }\href {\doibase
  10.1051/0004-6361/201527846} {\bibfield  {journal} {\bibinfo  {journal}
  {Astronomy {\&} Astrophysics}\ }\textbf {\bibinfo {volume} {589}},\ \bibinfo
  {pages} {A33} (\bibinfo {year} {2016})}\BibitemShut {NoStop}%
\bibitem [{\citenamefont {Mukherjee}\ and\ \citenamefont
  {Collaboration}(2016)}]{Mukherjee2016}%
  \BibitemOpen
  \bibfield  {author} {\bibinfo {author} {\bibfnamefont {R.}~\bibnamefont
  {Mukherjee}}\ and\ \bibinfo {author} {\bibfnamefont {V.}~\bibnamefont
  {Collaboration}},\ }\href@noop {} {\bibfield  {journal} {\bibinfo  {journal}
  {The Astronomer's Telegram}\ }\textbf {\bibinfo {volume} {9690}},\ \bibinfo
  {pages} {1} (\bibinfo {year} {2016})}\BibitemShut {NoStop}%
\bibitem [{\citenamefont {Aleksi{\'{c}}}\ \emph {et~al.}(2014)\citenamefont
  {Aleksi{\'{c}}}, \citenamefont {Ansoldi}, \citenamefont {Antonelli},
  \citenamefont {Antoranz}, \citenamefont {Babic}, \citenamefont {Bangale},
  \citenamefont {de~Almeida}, \citenamefont {Barrio}, \citenamefont
  {Gonz{\'{a}}lez} \emph {et~al.}}]{Aleksic2014}%
  \BibitemOpen
  \bibfield  {author} {\bibinfo {author} {\bibfnamefont {J.}~\bibnamefont
  {Aleksi{\'{c}}}}, \bibinfo {author} {\bibfnamefont {S.}~\bibnamefont
  {Ansoldi}}, \bibinfo {author} {\bibfnamefont {L.~A.}\ \bibnamefont
  {Antonelli}}, \bibinfo {author} {\bibfnamefont {P.}~\bibnamefont {Antoranz}},
  \bibinfo {author} {\bibfnamefont {A.}~\bibnamefont {Babic}}, \bibinfo
  {author} {\bibfnamefont {P.}~\bibnamefont {Bangale}}, \bibinfo {author}
  {\bibfnamefont {U.~B.}\ \bibnamefont {de~Almeida}}, \bibinfo {author}
  {\bibfnamefont {J.~A.}\ \bibnamefont {Barrio}}, \bibinfo {author}
  {\bibfnamefont {J.~B.}\ \bibnamefont {Gonz{\'{a}}lez}},  \emph {et~al.},\
  }\href {\doibase 10.1051/0004-6361/201322951} {\bibfield  {journal} {\bibinfo
   {journal} {Astronomy {\&} Astrophysics}\ }\textbf {\bibinfo {volume}
  {564}},\ \bibinfo {pages} {A5} (\bibinfo {year} {2014})}\BibitemShut
  {NoStop}%
\bibitem [{\citenamefont {Nesterov}\ \emph {et~al.}(1995)\citenamefont
  {Nesterov}, \citenamefont {Lyuty},\ and\ \citenamefont
  {Valtaoja}}]{Nesterov1995}%
  \BibitemOpen
  \bibfield  {author} {\bibinfo {author} {\bibfnamefont {N.~S.}\ \bibnamefont
  {Nesterov}}, \bibinfo {author} {\bibfnamefont {V.~M.}\ \bibnamefont {Lyuty}},
  \ and\ \bibinfo {author} {\bibfnamefont {E.}~\bibnamefont {Valtaoja}},\
  }\href@noop {} {\bibfield  {journal} {\bibinfo  {journal} {\aap}\ }\textbf
  {\bibinfo {volume} {296}},\ \bibinfo {pages} {628} (\bibinfo {year}
  {1995})}\BibitemShut {NoStop}%
\bibitem [{\citenamefont {Ballet}\ \emph {et~al.}(2020)\citenamefont {Ballet},
  \citenamefont {Burnett}, \citenamefont {Digel},\ and\ \citenamefont
  {Lott}}]{Ballet:2020hze}%
  \BibitemOpen
  \bibfield  {author} {\bibinfo {author} {\bibfnamefont {J.}~\bibnamefont
  {Ballet}}, \bibinfo {author} {\bibfnamefont {T.~H.}\ \bibnamefont {Burnett}},
  \bibinfo {author} {\bibfnamefont {S.~W.}\ \bibnamefont {Digel}}, \ and\
  \bibinfo {author} {\bibfnamefont {B.}~\bibnamefont {Lott}} (\bibinfo
  {collaboration} {Fermi-LAT}),\ }\href@noop {} {\  (\bibinfo {year} {2020})},\
  \Eprint {http://arxiv.org/abs/2005.11208} {arXiv:2005.11208 [astro-ph.HE]}
  \BibitemShut {NoStop}%
\bibitem [{\citenamefont {Geringer-Sameth}\ and\ \citenamefont
  {Koushiappas}(2011)}]{GeringerSameth:2011iw}%
  \BibitemOpen
  \bibfield  {author} {\bibinfo {author} {\bibfnamefont {A.}~\bibnamefont
  {Geringer-Sameth}}\ and\ \bibinfo {author} {\bibfnamefont {S.~M.}\
  \bibnamefont {Koushiappas}},\ }\href {\doibase
  10.1103/PhysRevLett.107.241303} {\bibfield  {journal} {\bibinfo  {journal}
  {Phys. Rev. Lett.}\ }\textbf {\bibinfo {volume} {107}},\ \bibinfo {pages}
  {241303} (\bibinfo {year} {2011})},\ \Eprint {http://arxiv.org/abs/1108.2914}
  {arXiv:1108.2914 [astro-ph.CO]} \BibitemShut {NoStop}%
\bibitem [{\citenamefont {Ackermann}\ \emph {et~al.}(2014)\citenamefont
  {Ackermann} \emph {et~al.}}]{Ackermann:2013yva}%
  \BibitemOpen
  \bibfield  {author} {\bibinfo {author} {\bibfnamefont {M.}~\bibnamefont
  {Ackermann}} \emph {et~al.} (\bibinfo {collaboration} {Fermi-LAT}),\ }\href
  {\doibase 10.1103/PhysRevD.89.042001} {\bibfield  {journal} {\bibinfo
  {journal} {Phys. Rev. D}\ }\textbf {\bibinfo {volume} {89}},\ \bibinfo
  {pages} {042001} (\bibinfo {year} {2014})},\ \Eprint
  {http://arxiv.org/abs/1310.0828} {arXiv:1310.0828 [astro-ph.HE]} \BibitemShut
  {NoStop}%
\bibitem [{\citenamefont {Albert}\ \emph {et~al.}(2017)\citenamefont {Albert}
  \emph {et~al.}}]{Fermi-LAT:2016uux}%
  \BibitemOpen
  \bibfield  {author} {\bibinfo {author} {\bibfnamefont {A.}~\bibnamefont
  {Albert}} \emph {et~al.} (\bibinfo {collaboration} {Fermi-LAT, DES}),\ }\href
  {\doibase 10.3847/1538-4357/834/2/110} {\bibfield  {journal} {\bibinfo
  {journal} {Astrophys. J.}\ }\textbf {\bibinfo {volume} {834}},\ \bibinfo
  {pages} {110} (\bibinfo {year} {2017})},\ \Eprint
  {http://arxiv.org/abs/1611.03184} {arXiv:1611.03184 [astro-ph.HE]}
  \BibitemShut {NoStop}%
\bibitem [{\citenamefont {Hoof}\ \emph {et~al.}(2020)\citenamefont {Hoof},
  \citenamefont {Geringer-Sameth},\ and\ \citenamefont
  {Trotta}}]{Hoof:2018hyn}%
  \BibitemOpen
  \bibfield  {author} {\bibinfo {author} {\bibfnamefont {S.}~\bibnamefont
  {Hoof}}, \bibinfo {author} {\bibfnamefont {A.}~\bibnamefont
  {Geringer-Sameth}}, \ and\ \bibinfo {author} {\bibfnamefont {R.}~\bibnamefont
  {Trotta}},\ }\href {\doibase 10.1088/1475-7516/2020/02/012} {\bibfield
  {journal} {\bibinfo  {journal} {JCAP}\ }\textbf {\bibinfo {volume} {02}},\
  \bibinfo {pages} {012} (\bibinfo {year} {2020})},\ \Eprint
  {http://arxiv.org/abs/1812.06986} {arXiv:1812.06986 [astro-ph.CO]}
  \BibitemShut {NoStop}%
\bibitem [{\citenamefont {Linden}(2020)}]{Linden:2019soa}%
  \BibitemOpen
  \bibfield  {author} {\bibinfo {author} {\bibfnamefont {T.}~\bibnamefont
  {Linden}},\ }\href {\doibase 10.1103/PhysRevD.101.043017} {\bibfield
  {journal} {\bibinfo  {journal} {Phys. Rev. D}\ }\textbf {\bibinfo {volume}
  {101}},\ \bibinfo {pages} {043017} (\bibinfo {year} {2020})},\ \Eprint
  {http://arxiv.org/abs/1905.11992} {arXiv:1905.11992 [astro-ph.HE]}
  \BibitemShut {NoStop}%
\bibitem [{\citenamefont {Tamborra}\ \emph {et~al.}(2014)\citenamefont
  {Tamborra}, \citenamefont {Ando},\ and\ \citenamefont
  {Murase}}]{Tamborra:2014xia}%
  \BibitemOpen
  \bibfield  {author} {\bibinfo {author} {\bibfnamefont {I.}~\bibnamefont
  {Tamborra}}, \bibinfo {author} {\bibfnamefont {S.}~\bibnamefont {Ando}}, \
  and\ \bibinfo {author} {\bibfnamefont {K.}~\bibnamefont {Murase}},\ }\href
  {\doibase 10.1088/1475-7516/2014/09/043} {\bibfield  {journal} {\bibinfo
  {journal} {JCAP}\ }\textbf {\bibinfo {volume} {09}},\ \bibinfo {pages} {043}
  (\bibinfo {year} {2014})},\ \Eprint {http://arxiv.org/abs/1404.1189}
  {arXiv:1404.1189 [astro-ph.HE]} \BibitemShut {NoStop}%
\bibitem [{\citenamefont {Rojas-Bravo}\ and\ \citenamefont
  {Araya}(2016)}]{Rojas-Bravo:2016val}%
  \BibitemOpen
  \bibfield  {author} {\bibinfo {author} {\bibfnamefont {C.}~\bibnamefont
  {Rojas-Bravo}}\ and\ \bibinfo {author} {\bibfnamefont {M.}~\bibnamefont
  {Araya}},\ }\href {\doibase 10.1093/mnras/stw2059} {\bibfield  {journal}
  {\bibinfo  {journal} {Mon. Not. Roy. Astron. Soc.}\ }\textbf {\bibinfo
  {volume} {463}},\ \bibinfo {pages} {1068} (\bibinfo {year} {2016})},\ \Eprint
  {http://arxiv.org/abs/1608.04413} {arXiv:1608.04413 [astro-ph.HE]}
  \BibitemShut {NoStop}%
\bibitem [{\citenamefont {{van der
  Kruit}}(1973{\natexlab{a}})}]{1973A&A....29..231V}%
  \BibitemOpen
  \bibfield  {author} {\bibinfo {author} {\bibfnamefont {P.~C.}\ \bibnamefont
  {{van der Kruit}}},\ }\href@noop {} {\bibfield  {journal} {\bibinfo
  {journal} {\aap}\ }\textbf {\bibinfo {volume} {29}},\ \bibinfo {pages} {231}
  (\bibinfo {year} {1973}{\natexlab{a}})}\BibitemShut {NoStop}%
\bibitem [{\citenamefont {{van der
  Kruit}}(1973{\natexlab{b}})}]{1973A&A....29..249V}%
  \BibitemOpen
  \bibfield  {author} {\bibinfo {author} {\bibfnamefont {P.~C.}\ \bibnamefont
  {{van der Kruit}}},\ }\href@noop {} {\bibfield  {journal} {\bibinfo
  {journal} {\aap}\ }\textbf {\bibinfo {volume} {29}},\ \bibinfo {pages} {249}
  (\bibinfo {year} {1973}{\natexlab{b}})}\BibitemShut {NoStop}%
\bibitem [{\citenamefont {{van der
  Kruit}}(1973{\natexlab{c}})}]{1973A&A....29..263V}%
  \BibitemOpen
  \bibfield  {author} {\bibinfo {author} {\bibfnamefont {P.~C.}\ \bibnamefont
  {{van der Kruit}}},\ }\href@noop {} {\bibfield  {journal} {\bibinfo
  {journal} {\aap}\ }\textbf {\bibinfo {volume} {29}},\ \bibinfo {pages} {263}
  (\bibinfo {year} {1973}{\natexlab{c}})}\BibitemShut {NoStop}%
\bibitem [{\citenamefont {{Dickey}}\ and\ \citenamefont
  {{Salpeter}}(1984)}]{1984ApJ...284..461D}%
  \BibitemOpen
  \bibfield  {author} {\bibinfo {author} {\bibfnamefont {J.~M.}\ \bibnamefont
  {{Dickey}}}\ and\ \bibinfo {author} {\bibfnamefont {E.~E.}\ \bibnamefont
  {{Salpeter}}},\ }\href {\doibase 10.1086/162428} {\bibfield  {journal}
  {\bibinfo  {journal} {\apj}\ }\textbf {\bibinfo {volume} {284}},\ \bibinfo
  {pages} {461} (\bibinfo {year} {1984})}\BibitemShut {NoStop}%
\bibitem [{\citenamefont {{Helou}}\ \emph {et~al.}(1985)\citenamefont
  {{Helou}}, \citenamefont {{Soifer}},\ and\ \citenamefont
  {{Rowan-Robinson}}}]{1985ApJ...298L...7H}%
  \BibitemOpen
  \bibfield  {author} {\bibinfo {author} {\bibfnamefont {G.}~\bibnamefont
  {{Helou}}}, \bibinfo {author} {\bibfnamefont {B.~T.}\ \bibnamefont
  {{Soifer}}}, \ and\ \bibinfo {author} {\bibfnamefont {M.}~\bibnamefont
  {{Rowan-Robinson}}},\ }\href {\doibase 10.1086/184556} {\bibfield  {journal}
  {\bibinfo  {journal} {\apjl}\ }\textbf {\bibinfo {volume} {298}},\ \bibinfo
  {pages} {L7} (\bibinfo {year} {1985})}\BibitemShut {NoStop}%
\bibitem [{\citenamefont {Inoue}(2011)}]{Inoue:2011bm}%
  \BibitemOpen
  \bibfield  {author} {\bibinfo {author} {\bibfnamefont {Y.}~\bibnamefont
  {Inoue}},\ }\href {\doibase 10.1088/0004-637X/733/1/66} {\bibfield  {journal}
  {\bibinfo  {journal} {Astrophys. J.}\ }\textbf {\bibinfo {volume} {733}},\
  \bibinfo {pages} {66} (\bibinfo {year} {2011})},\ \Eprint
  {http://arxiv.org/abs/1103.3946} {arXiv:1103.3946 [astro-ph.HE]} \BibitemShut
  {NoStop}%
\bibitem [{\citenamefont {Gruppioni}\ \emph {et~al.}(2013)\citenamefont
  {Gruppioni}, \citenamefont {Pozzi}, \citenamefont {Rodighiero}, \citenamefont
  {Delvecchio}, \citenamefont {Berta}, \citenamefont {Pozzetti}, \citenamefont
  {Zamorani}, \citenamefont {Andreani}, \citenamefont {Cimatti} \emph
  {et~al.}}]{Gruppioni2013}%
  \BibitemOpen
  \bibfield  {author} {\bibinfo {author} {\bibfnamefont {C.}~\bibnamefont
  {Gruppioni}}, \bibinfo {author} {\bibfnamefont {F.}~\bibnamefont {Pozzi}},
  \bibinfo {author} {\bibfnamefont {G.}~\bibnamefont {Rodighiero}}, \bibinfo
  {author} {\bibfnamefont {I.}~\bibnamefont {Delvecchio}}, \bibinfo {author}
  {\bibfnamefont {S.}~\bibnamefont {Berta}}, \bibinfo {author} {\bibfnamefont
  {L.}~\bibnamefont {Pozzetti}}, \bibinfo {author} {\bibfnamefont
  {G.}~\bibnamefont {Zamorani}}, \bibinfo {author} {\bibfnamefont
  {P.}~\bibnamefont {Andreani}}, \bibinfo {author} {\bibfnamefont
  {A.}~\bibnamefont {Cimatti}},  \emph {et~al.},\ }\href {\doibase
  10.1093/mnras/stt308} {\  (\bibinfo {year} {2013}),\ 10.1093/mnras/stt308},\
  \Eprint {http://arxiv.org/abs/1302.5209} {arXiv:1302.5209 [astro-ph.CO]}
  \BibitemShut {NoStop}%
\bibitem [{\citenamefont {Ambrosone}\ \emph {et~al.}(2020)\citenamefont
  {Ambrosone}, \citenamefont {Chianese}, \citenamefont {Fiorillo},
  \citenamefont {Marinelli}, \citenamefont {Miele},\ and\ \citenamefont
  {Pisanti}}]{Ambrosone2020}%
  \BibitemOpen
  \bibfield  {author} {\bibinfo {author} {\bibfnamefont {A.}~\bibnamefont
  {Ambrosone}}, \bibinfo {author} {\bibfnamefont {M.}~\bibnamefont {Chianese}},
  \bibinfo {author} {\bibfnamefont {D.~F.~G.}\ \bibnamefont {Fiorillo}},
  \bibinfo {author} {\bibfnamefont {A.}~\bibnamefont {Marinelli}}, \bibinfo
  {author} {\bibfnamefont {G.}~\bibnamefont {Miele}}, \ and\ \bibinfo {author}
  {\bibfnamefont {O.}~\bibnamefont {Pisanti}},\ }\href@noop {} {\  (\bibinfo
  {year} {2020})},\ \Eprint {http://arxiv.org/abs/2011.02483} {arXiv:2011.02483
  [astro-ph.HE]} \BibitemShut {NoStop}%
\bibitem [{\citenamefont {Blanco}\ \emph {et~al.}(2017)\citenamefont {Blanco},
  \citenamefont {Harding},\ and\ \citenamefont {Hooper}}]{Blanco2017a}%
  \BibitemOpen
  \bibfield  {author} {\bibinfo {author} {\bibfnamefont {C.}~\bibnamefont
  {Blanco}}, \bibinfo {author} {\bibfnamefont {J.~P.}\ \bibnamefont {Harding}},
  \ and\ \bibinfo {author} {\bibfnamefont {D.}~\bibnamefont {Hooper}},\ }\href
  {\doibase 10.1088/1475-7516/2018/04/060} {\  (\bibinfo {year} {2017}),\
  10.1088/1475-7516/2018/04/060},\ \Eprint {http://arxiv.org/abs/1712.02805}
  {arXiv:1712.02805 [hep-ph]} \BibitemShut {NoStop}%
\bibitem [{\citenamefont {Blanco}(2018)}]{Blanco2018}%
  \BibitemOpen
  \bibfield  {author} {\bibinfo {author} {\bibfnamefont {C.}~\bibnamefont
  {Blanco}},\ }\href {\doibase 10.1088/1475-7516/2019/01/013} {\bibfield
  {journal} {\bibinfo  {journal} {JCAP 01 (2019) 013}\ } (\bibinfo {year}
  {2018}),\ 10.1088/1475-7516/2019/01/013},\ \Eprint
  {http://arxiv.org/abs/1804.00005} {arXiv:1804.00005 [astro-ph.HE]}
  \BibitemShut {NoStop}%
\bibitem [{\citenamefont {Collaboration}(2010)}]{Fermi2010}%
  \BibitemOpen
  \bibfield  {author} {\bibinfo {author} {\bibfnamefont {T.~F.-L. A.~T.}\
  \bibnamefont {Collaboration}},\ }\href {\doibase 10.1088/0004-637X/720/1/435}
  {\  (\bibinfo {year} {2010}),\ 10.1088/0004-637X/720/1/435},\ \Eprint
  {http://arxiv.org/abs/1003.0895} {arXiv:1003.0895 [astro-ph.CO]} \BibitemShut
  {NoStop}%
\bibitem [{\citenamefont {Ajello}\ \emph {et~al.}(2013)\citenamefont {Ajello},
  \citenamefont {Romani}, \citenamefont {Gasparrini}, \citenamefont {Shaw},
  \citenamefont {Bolmer}, \citenamefont {Cotter}, \citenamefont {Finke},
  \citenamefont {Greiner}, \citenamefont {Healey} \emph {et~al.}}]{Ajello2013}%
  \BibitemOpen
  \bibfield  {author} {\bibinfo {author} {\bibfnamefont {M.}~\bibnamefont
  {Ajello}}, \bibinfo {author} {\bibfnamefont {R.~W.}\ \bibnamefont {Romani}},
  \bibinfo {author} {\bibfnamefont {D.}~\bibnamefont {Gasparrini}}, \bibinfo
  {author} {\bibfnamefont {M.~S.}\ \bibnamefont {Shaw}}, \bibinfo {author}
  {\bibfnamefont {J.}~\bibnamefont {Bolmer}}, \bibinfo {author} {\bibfnamefont
  {G.}~\bibnamefont {Cotter}}, \bibinfo {author} {\bibfnamefont
  {J.}~\bibnamefont {Finke}}, \bibinfo {author} {\bibfnamefont
  {J.}~\bibnamefont {Greiner}}, \bibinfo {author} {\bibfnamefont {S.~E.}\
  \bibnamefont {Healey}},  \emph {et~al.},\ }\href {\doibase
  10.1088/0004-637X/780/1/73} {\  (\bibinfo {year} {2013}),\
  10.1088/0004-637X/780/1/73},\ \Eprint {http://arxiv.org/abs/1310.0006}
  {arXiv:1310.0006 [astro-ph.CO]} \BibitemShut {NoStop}%
\bibitem [{\citenamefont {Ajello}\ \emph {et~al.}(2011)\citenamefont {Ajello},
  \citenamefont {Shaw}, \citenamefont {Romani}, \citenamefont {Dermer},
  \citenamefont {Costamante}, \citenamefont {King}, \citenamefont
  {Max-Moerbeck}, \citenamefont {Readhead}, \citenamefont {Reimer},
  \citenamefont {Richards},\ and\ \citenamefont {Stevenson}}]{Ajello2011}%
  \BibitemOpen
  \bibfield  {author} {\bibinfo {author} {\bibfnamefont {M.}~\bibnamefont
  {Ajello}}, \bibinfo {author} {\bibfnamefont {M.~S.}\ \bibnamefont {Shaw}},
  \bibinfo {author} {\bibfnamefont {R.~W.}\ \bibnamefont {Romani}}, \bibinfo
  {author} {\bibfnamefont {C.~D.}\ \bibnamefont {Dermer}}, \bibinfo {author}
  {\bibfnamefont {L.}~\bibnamefont {Costamante}}, \bibinfo {author}
  {\bibfnamefont {O.~G.}\ \bibnamefont {King}}, \bibinfo {author}
  {\bibfnamefont {W.}~\bibnamefont {Max-Moerbeck}}, \bibinfo {author}
  {\bibfnamefont {A.}~\bibnamefont {Readhead}}, \bibinfo {author}
  {\bibfnamefont {A.}~\bibnamefont {Reimer}}, \bibinfo {author} {\bibfnamefont
  {J.~L.}\ \bibnamefont {Richards}}, \ and\ \bibinfo {author} {\bibfnamefont
  {M.}~\bibnamefont {Stevenson}},\ }\href {\doibase
  10.1088/0004-637X/751/2/108} {\  (\bibinfo {year} {2011}),\
  10.1088/0004-637X/751/2/108},\ \Eprint {http://arxiv.org/abs/1110.3787}
  {arXiv:1110.3787 [astro-ph.CO]} \BibitemShut {NoStop}%
\bibitem [{\citenamefont {Cholis}\ \emph {et~al.}(2013)\citenamefont {Cholis},
  \citenamefont {Hooper},\ and\ \citenamefont {McDermott}}]{Cholis2013}%
  \BibitemOpen
  \bibfield  {author} {\bibinfo {author} {\bibfnamefont {I.}~\bibnamefont
  {Cholis}}, \bibinfo {author} {\bibfnamefont {D.}~\bibnamefont {Hooper}}, \
  and\ \bibinfo {author} {\bibfnamefont {S.~D.}\ \bibnamefont {McDermott}},\
  }\href {\doibase 10.1088/1475-7516/2014/02/014} {\  (\bibinfo {year}
  {2013}),\ 10.1088/1475-7516/2014/02/014},\ \Eprint
  {http://arxiv.org/abs/1312.0608} {arXiv:1312.0608 [astro-ph.CO]} \BibitemShut
  {NoStop}%
\bibitem [{\citenamefont {Cuoco}\ \emph {et~al.}(2012)\citenamefont {Cuoco},
  \citenamefont {Komatsu},\ and\ \citenamefont {Siegal-Gaskins}}]{Cuoco2012}%
  \BibitemOpen
  \bibfield  {author} {\bibinfo {author} {\bibfnamefont {A.}~\bibnamefont
  {Cuoco}}, \bibinfo {author} {\bibfnamefont {E.}~\bibnamefont {Komatsu}}, \
  and\ \bibinfo {author} {\bibfnamefont {J.}~\bibnamefont {Siegal-Gaskins}},\
  }\href {\doibase 10.1103/PhysRevD.86.063004} {\bibfield  {journal} {\bibinfo
  {journal} {Phys.Rev. D86 (2012) 063004}\ } (\bibinfo {year} {2012}),\
  10.1103/PhysRevD.86.063004},\ \Eprint {http://arxiv.org/abs/1202.5309}
  {arXiv:1202.5309 [astro-ph.CO]} \BibitemShut {NoStop}%
\bibitem [{\citenamefont {Harding}\ and\ \citenamefont
  {Abazajian}(2012)}]{Harding2012}%
  \BibitemOpen
  \bibfield  {author} {\bibinfo {author} {\bibfnamefont {J.~P.}\ \bibnamefont
  {Harding}}\ and\ \bibinfo {author} {\bibfnamefont {K.~N.}\ \bibnamefont
  {Abazajian}},\ }\href {\doibase 10.1088/1475-7516/2012/11/026} {\  (\bibinfo
  {year} {2012}),\ 10.1088/1475-7516/2012/11/026},\ \Eprint
  {http://arxiv.org/abs/1206.4734} {arXiv:1206.4734 [astro-ph.HE]} \BibitemShut
  {NoStop}%
\bibitem [{\citenamefont {Aartsen}\ \emph {et~al.}(2013)\citenamefont {Aartsen}
  \emph {et~al.}}]{Aartsen:2013bka}%
  \BibitemOpen
  \bibfield  {author} {\bibinfo {author} {\bibfnamefont {M.~G.}\ \bibnamefont
  {Aartsen}} \emph {et~al.} (\bibinfo {collaboration} {IceCube}),\ }\href
  {\doibase 10.1103/PhysRevLett.111.021103} {\bibfield  {journal} {\bibinfo
  {journal} {Phys. Rev. Lett.}\ }\textbf {\bibinfo {volume} {111}},\ \bibinfo
  {pages} {021103} (\bibinfo {year} {2013})},\ \Eprint
  {http://arxiv.org/abs/1304.5356} {arXiv:1304.5356 [astro-ph.HE]} \BibitemShut
  {NoStop}%
\bibitem [{\citenamefont {Aartsen}\ \emph {et~al.}(2015)\citenamefont {Aartsen}
  \emph {et~al.}}]{Aartsen:2015knd}%
  \BibitemOpen
  \bibfield  {author} {\bibinfo {author} {\bibfnamefont {M.~G.}\ \bibnamefont
  {Aartsen}} \emph {et~al.} (\bibinfo {collaboration} {IceCube}),\ }\href
  {\doibase 10.1088/0004-637X/809/1/98} {\bibfield  {journal} {\bibinfo
  {journal} {Astrophys. J.}\ }\textbf {\bibinfo {volume} {809}},\ \bibinfo
  {pages} {98} (\bibinfo {year} {2015})},\ \Eprint
  {http://arxiv.org/abs/1507.03991} {arXiv:1507.03991 [astro-ph.HE]}
  \BibitemShut {NoStop}%
\bibitem [{\citenamefont {Aartsen}\ \emph {et~al.}(2016)\citenamefont {Aartsen}
  \emph {et~al.}}]{Aartsen:2016xlq}%
  \BibitemOpen
  \bibfield  {author} {\bibinfo {author} {\bibfnamefont {M.~G.}\ \bibnamefont
  {Aartsen}} \emph {et~al.} (\bibinfo {collaboration} {IceCube}),\ }\href
  {\doibase 10.3847/0004-637X/833/1/3} {\bibfield  {journal} {\bibinfo
  {journal} {Astrophys. J.}\ }\textbf {\bibinfo {volume} {833}},\ \bibinfo
  {pages} {3} (\bibinfo {year} {2016})},\ \Eprint
  {http://arxiv.org/abs/1607.08006} {arXiv:1607.08006 [astro-ph.HE]}
  \BibitemShut {NoStop}%
\bibitem [{\citenamefont {{Icecube Collaboration}}\ \emph
  {et~al.}(2012)\citenamefont {{Icecube Collaboration}}, \citenamefont
  {{Abbasi}}, \citenamefont {{Abdou}}, \citenamefont {{Abu-Zayyad}},
  \citenamefont {{Ackermann}}, \citenamefont {{Adams}}, \citenamefont
  {{Aguilar}}, \citenamefont {{Ahlers}}, \citenamefont {{Altmann}},\ and\
  \citenamefont {{et al.}}}]{2012Natur.484..351I}%
  \BibitemOpen
  \bibfield  {author} {\bibinfo {author} {\bibnamefont {{Icecube
  Collaboration}}}, \bibinfo {author} {\bibfnamefont {R.}~\bibnamefont
  {{Abbasi}}}, \bibinfo {author} {\bibfnamefont {Y.}~\bibnamefont {{Abdou}}},
  \bibinfo {author} {\bibfnamefont {T.}~\bibnamefont {{Abu-Zayyad}}}, \bibinfo
  {author} {\bibfnamefont {M.}~\bibnamefont {{Ackermann}}}, \bibinfo {author}
  {\bibfnamefont {J.}~\bibnamefont {{Adams}}}, \bibinfo {author} {\bibfnamefont
  {J.~A.}\ \bibnamefont {{Aguilar}}}, \bibinfo {author} {\bibfnamefont
  {M.}~\bibnamefont {{Ahlers}}}, \bibinfo {author} {\bibfnamefont
  {D.}~\bibnamefont {{Altmann}}}, \ and\ \bibinfo {author} {\bibnamefont {{et
  al.}}},\ }\href {\doibase 10.1038/nature11068} {\bibfield  {journal}
  {\bibinfo  {journal} {\nat}\ }\textbf {\bibinfo {volume} {484}},\ \bibinfo
  {pages} {351} (\bibinfo {year} {2012})},\ \Eprint
  {http://arxiv.org/abs/1204.4219} {arXiv:1204.4219 [astro-ph.HE]} \BibitemShut
  {NoStop}%
\bibitem [{\citenamefont {Aartsen}\ \emph {et~al.}(2019)\citenamefont {Aartsen}
  \emph {et~al.}}]{Aartsen:2018fpd}%
  \BibitemOpen
  \bibfield  {author} {\bibinfo {author} {\bibfnamefont {M.~G.}\ \bibnamefont
  {Aartsen}} \emph {et~al.} (\bibinfo {collaboration} {IceCube}),\ }\href
  {\doibase 10.1103/PhysRevLett.122.051102} {\bibfield  {journal} {\bibinfo
  {journal} {Phys. Rev. Lett.}\ }\textbf {\bibinfo {volume} {122}},\ \bibinfo
  {pages} {051102} (\bibinfo {year} {2019})},\ \Eprint
  {http://arxiv.org/abs/1807.11492} {arXiv:1807.11492 [astro-ph.HE]}
  \BibitemShut {NoStop}%
\bibitem [{\citenamefont {Aartsen}\ \emph {et~al.}(2017)\citenamefont {Aartsen}
  \emph {et~al.}}]{Aartsen:2016lir}%
  \BibitemOpen
  \bibfield  {author} {\bibinfo {author} {\bibfnamefont {M.~G.}\ \bibnamefont
  {Aartsen}} \emph {et~al.} (\bibinfo {collaboration} {IceCube}),\ }\href
  {\doibase 10.3847/1538-4357/835/1/45} {\bibfield  {journal} {\bibinfo
  {journal} {Astrophys. J.}\ }\textbf {\bibinfo {volume} {835}},\ \bibinfo
  {pages} {45} (\bibinfo {year} {2017})},\ \Eprint
  {http://arxiv.org/abs/1611.03874} {arXiv:1611.03874 [astro-ph.HE]}
  \BibitemShut {NoStop}%
\bibitem [{\citenamefont {Hooper}\ \emph {et~al.}(2019)\citenamefont {Hooper},
  \citenamefont {Linden},\ and\ \citenamefont {Vieregg}}]{Hooper:2018wyk}%
  \BibitemOpen
  \bibfield  {author} {\bibinfo {author} {\bibfnamefont {D.}~\bibnamefont
  {Hooper}}, \bibinfo {author} {\bibfnamefont {T.}~\bibnamefont {Linden}}, \
  and\ \bibinfo {author} {\bibfnamefont {A.}~\bibnamefont {Vieregg}},\ }\href
  {\doibase 10.1088/1475-7516/2019/02/012} {\bibfield  {journal} {\bibinfo
  {journal} {JCAP}\ }\textbf {\bibinfo {volume} {02}},\ \bibinfo {pages} {012}
  (\bibinfo {year} {2019})},\ \Eprint {http://arxiv.org/abs/1810.02823}
  {arXiv:1810.02823 [astro-ph.HE]} \BibitemShut {NoStop}%
\bibitem [{\citenamefont {Loeb}\ and\ \citenamefont
  {Waxman}(2006)}]{Loeb:2006tw}%
  \BibitemOpen
  \bibfield  {author} {\bibinfo {author} {\bibfnamefont {A.}~\bibnamefont
  {Loeb}}\ and\ \bibinfo {author} {\bibfnamefont {E.}~\bibnamefont {Waxman}},\
  }\href {\doibase 10.1088/1475-7516/2006/05/003} {\bibfield  {journal}
  {\bibinfo  {journal} {JCAP}\ }\textbf {\bibinfo {volume} {05}},\ \bibinfo
  {pages} {003} (\bibinfo {year} {2006})},\ \Eprint
  {http://arxiv.org/abs/astro-ph/0601695} {arXiv:astro-ph/0601695} \BibitemShut
  {NoStop}%
\bibitem [{\citenamefont {Murase}\ \emph {et~al.}(2013)\citenamefont {Murase},
  \citenamefont {Ahlers},\ and\ \citenamefont {Lacki}}]{Murase:2013rfa}%
  \BibitemOpen
  \bibfield  {author} {\bibinfo {author} {\bibfnamefont {K.}~\bibnamefont
  {Murase}}, \bibinfo {author} {\bibfnamefont {M.}~\bibnamefont {Ahlers}}, \
  and\ \bibinfo {author} {\bibfnamefont {B.~C.}\ \bibnamefont {Lacki}},\ }\href
  {\doibase 10.1103/PhysRevD.88.121301} {\bibfield  {journal} {\bibinfo
  {journal} {Phys. Rev. D}\ }\textbf {\bibinfo {volume} {88}},\ \bibinfo
  {pages} {121301} (\bibinfo {year} {2013})},\ \Eprint
  {http://arxiv.org/abs/1306.3417} {arXiv:1306.3417 [astro-ph.HE]} \BibitemShut
  {NoStop}%
\bibitem [{\citenamefont {Anchordoqui}\ \emph {et~al.}(2014)\citenamefont
  {Anchordoqui}, \citenamefont {Paul}, \citenamefont {da~Silva}, \citenamefont
  {Torres},\ and\ \citenamefont {Vlcek}}]{Anchordoqui:2014yva}%
  \BibitemOpen
  \bibfield  {author} {\bibinfo {author} {\bibfnamefont {L.~A.}\ \bibnamefont
  {Anchordoqui}}, \bibinfo {author} {\bibfnamefont {T.~C.}\ \bibnamefont
  {Paul}}, \bibinfo {author} {\bibfnamefont {L.~H.~M.}\ \bibnamefont
  {da~Silva}}, \bibinfo {author} {\bibfnamefont {D.~F.}\ \bibnamefont
  {Torres}}, \ and\ \bibinfo {author} {\bibfnamefont {B.~J.}\ \bibnamefont
  {Vlcek}},\ }\href {\doibase 10.1103/PhysRevD.89.127304} {\bibfield  {journal}
  {\bibinfo  {journal} {Phys. Rev. D}\ }\textbf {\bibinfo {volume} {89}},\
  \bibinfo {pages} {127304} (\bibinfo {year} {2014})},\ \Eprint
  {http://arxiv.org/abs/1405.7648} {arXiv:1405.7648 [astro-ph.HE]} \BibitemShut
  {NoStop}%
\bibitem [{\citenamefont {Becker~Tjus}\ \emph {et~al.}(2014)\citenamefont
  {Becker~Tjus}, \citenamefont {Eichmann}, \citenamefont {Halzen},
  \citenamefont {Kheirandish},\ and\ \citenamefont {Saba}}]{Tjus:2014dna}%
  \BibitemOpen
  \bibfield  {author} {\bibinfo {author} {\bibfnamefont {J.}~\bibnamefont
  {Becker~Tjus}}, \bibinfo {author} {\bibfnamefont {B.}~\bibnamefont
  {Eichmann}}, \bibinfo {author} {\bibfnamefont {F.}~\bibnamefont {Halzen}},
  \bibinfo {author} {\bibfnamefont {A.}~\bibnamefont {Kheirandish}}, \ and\
  \bibinfo {author} {\bibfnamefont {S.~M.}\ \bibnamefont {Saba}},\ }\href
  {\doibase 10.1103/PhysRevD.89.123005} {\bibfield  {journal} {\bibinfo
  {journal} {Phys. Rev. D}\ }\textbf {\bibinfo {volume} {89}},\ \bibinfo
  {pages} {123005} (\bibinfo {year} {2014})},\ \Eprint
  {http://arxiv.org/abs/1406.0506} {arXiv:1406.0506 [astro-ph.HE]} \BibitemShut
  {NoStop}%
\bibitem [{\citenamefont {Giacinti}\ \emph {et~al.}(2015)\citenamefont
  {Giacinti}, \citenamefont {Kachelrie\ss{}}, \citenamefont {Kalashev},
  \citenamefont {Neronov},\ and\ \citenamefont {Semikoz}}]{Giacinti:2015pya}%
  \BibitemOpen
  \bibfield  {author} {\bibinfo {author} {\bibfnamefont {G.}~\bibnamefont
  {Giacinti}}, \bibinfo {author} {\bibfnamefont {M.}~\bibnamefont
  {Kachelrie\ss{}}}, \bibinfo {author} {\bibfnamefont {O.}~\bibnamefont
  {Kalashev}}, \bibinfo {author} {\bibfnamefont {A.}~\bibnamefont {Neronov}}, \
  and\ \bibinfo {author} {\bibfnamefont {D.~V.}\ \bibnamefont {Semikoz}},\
  }\href {\doibase 10.1103/PhysRevD.92.083016} {\bibfield  {journal} {\bibinfo
  {journal} {Phys. Rev. D}\ }\textbf {\bibinfo {volume} {92}},\ \bibinfo
  {pages} {083016} (\bibinfo {year} {2015})},\ \Eprint
  {http://arxiv.org/abs/1507.07534} {arXiv:1507.07534 [astro-ph.HE]}
  \BibitemShut {NoStop}%
\bibitem [{\citenamefont {Murase}(2017)}]{Murase:2015ndr}%
  \BibitemOpen
  \bibfield  {author} {\bibinfo {author} {\bibfnamefont {K.}~\bibnamefont
  {Murase}},\ }\enquote {\bibinfo {title} {{Active Galactic Nuclei as
  High-Energy Neutrino Sources}},}\ in\ \href {\doibase
  10.1142/9789814759410_0002} {\emph {\bibinfo {booktitle} {{Neutrino
  Astronomy}: {Current Status, Future Prospects}}}},\ \bibinfo {editor} {edited
  by\ \bibinfo {editor} {\bibfnamefont {T.}~\bibnamefont {Gaisser}}\ and\
  \bibinfo {editor} {\bibfnamefont {A.}~\bibnamefont {Karle}}}\ (\bibinfo
  {year} {2017})\ \Eprint {http://arxiv.org/abs/1511.01590} {arXiv:1511.01590
  [astro-ph.HE]} \BibitemShut {NoStop}%
\bibitem [{\citenamefont {Hooper}(2016)}]{Hooper:2016jls}%
  \BibitemOpen
  \bibfield  {author} {\bibinfo {author} {\bibfnamefont {D.}~\bibnamefont
  {Hooper}},\ }\href {\doibase 10.1088/1475-7516/2016/09/002} {\bibfield
  {journal} {\bibinfo  {journal} {JCAP}\ }\textbf {\bibinfo {volume} {09}},\
  \bibinfo {pages} {002} (\bibinfo {year} {2016})},\ \Eprint
  {http://arxiv.org/abs/1605.06504} {arXiv:1605.06504 [astro-ph.HE]}
  \BibitemShut {NoStop}%
\bibitem [{\citenamefont {Fang}\ and\ \citenamefont
  {Murase}(2018)}]{Fang:2017zjf}%
  \BibitemOpen
  \bibfield  {author} {\bibinfo {author} {\bibfnamefont {K.}~\bibnamefont
  {Fang}}\ and\ \bibinfo {author} {\bibfnamefont {K.}~\bibnamefont {Murase}},\
  }\href {\doibase 10.1038/s41567-017-0025-4} {\bibfield  {journal} {\bibinfo
  {journal} {Nature Phys.}\ }\textbf {\bibinfo {volume} {14}},\ \bibinfo
  {pages} {396} (\bibinfo {year} {2018})},\ \Eprint
  {http://arxiv.org/abs/1704.00015} {arXiv:1704.00015 [astro-ph.HE]}
  \BibitemShut {NoStop}%
\bibitem [{\citenamefont {Murase}\ \emph {et~al.}(2016)\citenamefont {Murase},
  \citenamefont {Guetta},\ and\ \citenamefont {Ahlers}}]{Murase:2015xka}%
  \BibitemOpen
  \bibfield  {author} {\bibinfo {author} {\bibfnamefont {K.}~\bibnamefont
  {Murase}}, \bibinfo {author} {\bibfnamefont {D.}~\bibnamefont {Guetta}}, \
  and\ \bibinfo {author} {\bibfnamefont {M.}~\bibnamefont {Ahlers}},\ }\href
  {\doibase 10.1103/PhysRevLett.116.071101} {\bibfield  {journal} {\bibinfo
  {journal} {Phys. Rev. Lett.}\ }\textbf {\bibinfo {volume} {116}},\ \bibinfo
  {pages} {071101} (\bibinfo {year} {2016})},\ \Eprint
  {http://arxiv.org/abs/1509.00805} {arXiv:1509.00805 [astro-ph.HE]}
  \BibitemShut {NoStop}%
\bibitem [{\citenamefont {Senno}\ \emph {et~al.}(2016)\citenamefont {Senno},
  \citenamefont {Murase},\ and\ \citenamefont {Meszaros}}]{Senno:2015tsn}%
  \BibitemOpen
  \bibfield  {author} {\bibinfo {author} {\bibfnamefont {N.}~\bibnamefont
  {Senno}}, \bibinfo {author} {\bibfnamefont {K.}~\bibnamefont {Murase}}, \
  and\ \bibinfo {author} {\bibfnamefont {P.}~\bibnamefont {Meszaros}},\ }\href
  {\doibase 10.1103/PhysRevD.93.083003} {\bibfield  {journal} {\bibinfo
  {journal} {Phys. Rev. D}\ }\textbf {\bibinfo {volume} {93}},\ \bibinfo
  {pages} {083003} (\bibinfo {year} {2016})},\ \Eprint
  {http://arxiv.org/abs/1512.08513} {arXiv:1512.08513 [astro-ph.HE]}
  \BibitemShut {NoStop}%
\bibitem [{\citenamefont {Aartsen}\ \emph {et~al.}(2018)\citenamefont {Aartsen}
  \emph {et~al.}}]{IceCube:2018cha}%
  \BibitemOpen
  \bibfield  {author} {\bibinfo {author} {\bibfnamefont {M.~G.}\ \bibnamefont
  {Aartsen}} \emph {et~al.} (\bibinfo {collaboration} {IceCube}),\ }\href
  {\doibase 10.1126/science.aat2890} {\bibfield  {journal} {\bibinfo  {journal}
  {Science}\ }\textbf {\bibinfo {volume} {361}},\ \bibinfo {pages} {147}
  (\bibinfo {year} {2018})},\ \Eprint {http://arxiv.org/abs/1807.08794}
  {arXiv:1807.08794 [astro-ph.HE]} \BibitemShut {NoStop}%
\bibitem [{\citenamefont {Inoue}\ \emph {et~al.}(2020)\citenamefont {Inoue},
  \citenamefont {Khangulyan},\ and\ \citenamefont {Doi}}]{Inoue:2019yfs}%
  \BibitemOpen
  \bibfield  {author} {\bibinfo {author} {\bibfnamefont {Y.}~\bibnamefont
  {Inoue}}, \bibinfo {author} {\bibfnamefont {D.}~\bibnamefont {Khangulyan}}, \
  and\ \bibinfo {author} {\bibfnamefont {A.}~\bibnamefont {Doi}},\ }\href
  {\doibase 10.3847/2041-8213/ab7661} {\bibfield  {journal} {\bibinfo
  {journal} {Astrophys. J. Lett.}\ }\textbf {\bibinfo {volume} {891}},\
  \bibinfo {pages} {L33} (\bibinfo {year} {2020})},\ \Eprint
  {http://arxiv.org/abs/1909.02239} {arXiv:1909.02239 [astro-ph.HE]}
  \BibitemShut {NoStop}%
\bibitem [{\citenamefont {Anchordoqui}\ \emph {et~al.}(2021)\citenamefont
  {Anchordoqui}, \citenamefont {Krizmanic},\ and\ \citenamefont
  {Stecker}}]{Anchordoqui:2021vms}%
  \BibitemOpen
  \bibfield  {author} {\bibinfo {author} {\bibfnamefont {L.~A.}\ \bibnamefont
  {Anchordoqui}}, \bibinfo {author} {\bibfnamefont {J.~F.}\ \bibnamefont
  {Krizmanic}}, \ and\ \bibinfo {author} {\bibfnamefont {F.~W.}\ \bibnamefont
  {Stecker}},\ }\href@noop {} {\  (\bibinfo {year} {2021})},\ \Eprint
  {http://arxiv.org/abs/2102.12409} {arXiv:2102.12409 [astro-ph.HE]}
  \BibitemShut {NoStop}%
\bibitem [{\citenamefont {Murase}\ \emph {et~al.}(2020)\citenamefont {Murase},
  \citenamefont {Kimura},\ and\ \citenamefont {Meszaros}}]{Murase:2019vdl}%
  \BibitemOpen
  \bibfield  {author} {\bibinfo {author} {\bibfnamefont {K.}~\bibnamefont
  {Murase}}, \bibinfo {author} {\bibfnamefont {S.~S.}\ \bibnamefont {Kimura}},
  \ and\ \bibinfo {author} {\bibfnamefont {P.}~\bibnamefont {Meszaros}},\
  }\href {\doibase 10.1103/PhysRevLett.125.011101} {\bibfield  {journal}
  {\bibinfo  {journal} {Phys. Rev. Lett.}\ }\textbf {\bibinfo {volume} {125}},\
  \bibinfo {pages} {011101} (\bibinfo {year} {2020})},\ \Eprint
  {http://arxiv.org/abs/1904.04226} {arXiv:1904.04226 [astro-ph.HE]}
  \BibitemShut {NoStop}%
\bibitem [{\citenamefont {Kheirandish}\ \emph {et~al.}(2021)\citenamefont
  {Kheirandish}, \citenamefont {Murase},\ and\ \citenamefont
  {Kimura}}]{Kheirandish:2021wkm}%
  \BibitemOpen
  \bibfield  {author} {\bibinfo {author} {\bibfnamefont {A.}~\bibnamefont
  {Kheirandish}}, \bibinfo {author} {\bibfnamefont {K.}~\bibnamefont {Murase}},
  \ and\ \bibinfo {author} {\bibfnamefont {S.~S.}\ \bibnamefont {Kimura}},\
  }\href@noop {} {\  (\bibinfo {year} {2021})},\ \Eprint
  {http://arxiv.org/abs/2102.04475} {arXiv:2102.04475 [astro-ph.HE]}
  \BibitemShut {NoStop}%
\bibitem [{\citenamefont {Bechtol}\ \emph {et~al.}(2017)\citenamefont
  {Bechtol}, \citenamefont {Ahlers}, \citenamefont {Di~Mauro}, \citenamefont
  {Ajello},\ and\ \citenamefont {Vandenbroucke}}]{Bechtol:2015uqb}%
  \BibitemOpen
  \bibfield  {author} {\bibinfo {author} {\bibfnamefont {K.}~\bibnamefont
  {Bechtol}}, \bibinfo {author} {\bibfnamefont {M.}~\bibnamefont {Ahlers}},
  \bibinfo {author} {\bibfnamefont {M.}~\bibnamefont {Di~Mauro}}, \bibinfo
  {author} {\bibfnamefont {M.}~\bibnamefont {Ajello}}, \ and\ \bibinfo {author}
  {\bibfnamefont {J.}~\bibnamefont {Vandenbroucke}},\ }\href {\doibase
  10.3847/1538-4357/836/1/47} {\bibfield  {journal} {\bibinfo  {journal}
  {Astrophys. J.}\ }\textbf {\bibinfo {volume} {836}},\ \bibinfo {pages} {47}
  (\bibinfo {year} {2017})},\ \Eprint {http://arxiv.org/abs/1511.00688}
  {arXiv:1511.00688 [astro-ph.HE]} \BibitemShut {NoStop}%
\bibitem [{\citenamefont {Xiao}\ \emph {et~al.}(2016)\citenamefont {Xiao},
  \citenamefont {M\'esz\'aros}, \citenamefont {Murase},\ and\ \citenamefont
  {Dai}}]{Xiao:2016rvd}%
  \BibitemOpen
  \bibfield  {author} {\bibinfo {author} {\bibfnamefont {D.}~\bibnamefont
  {Xiao}}, \bibinfo {author} {\bibfnamefont {P.}~\bibnamefont {M\'esz\'aros}},
  \bibinfo {author} {\bibfnamefont {K.}~\bibnamefont {Murase}}, \ and\ \bibinfo
  {author} {\bibfnamefont {Z.-g.}\ \bibnamefont {Dai}},\ }\href {\doibase
  10.3847/0004-637X/826/2/133} {\bibfield  {journal} {\bibinfo  {journal}
  {Astrophys. J.}\ }\textbf {\bibinfo {volume} {826}},\ \bibinfo {pages} {133}
  (\bibinfo {year} {2016})},\ \Eprint {http://arxiv.org/abs/1604.08131}
  {arXiv:1604.08131 [astro-ph.HE]} \BibitemShut {NoStop}%
\bibitem [{\citenamefont {Palladino}\ \emph {et~al.}(2019)\citenamefont
  {Palladino}, \citenamefont {Fedynitch}, \citenamefont {Rasmussen},\ and\
  \citenamefont {Taylor}}]{Palladino:2018bqf}%
  \BibitemOpen
  \bibfield  {author} {\bibinfo {author} {\bibfnamefont {A.}~\bibnamefont
  {Palladino}}, \bibinfo {author} {\bibfnamefont {A.}~\bibnamefont
  {Fedynitch}}, \bibinfo {author} {\bibfnamefont {R.~W.}\ \bibnamefont
  {Rasmussen}}, \ and\ \bibinfo {author} {\bibfnamefont {A.~M.}\ \bibnamefont
  {Taylor}},\ }\href {\doibase 10.1088/1475-7516/2019/09/004} {\bibfield
  {journal} {\bibinfo  {journal} {JCAP}\ }\textbf {\bibinfo {volume} {09}},\
  \bibinfo {pages} {004} (\bibinfo {year} {2019})},\ \Eprint
  {http://arxiv.org/abs/1812.04685} {arXiv:1812.04685 [astro-ph.HE]}
  \BibitemShut {NoStop}%
\end{thebibliography}%

\end{document}